\documentclass[twocolumn, times]{aastex7}

\newcommand{\water}{H$_2$O}
\newcommand{\coo}{CO$_2$}

\newcommand{\ocn}{OCN$^-$}

\newcommand{\hydrogenmol}{H$_2$}
\newcommand{\hhnu}{H$_2 ~v=0-0$ S}
\newcommand{\Eup}{$E_{\rm u}/k$}

\newcommand{\lbol}{$L_{\mathrm{bol}}$}

\def\msun{\rm\,M_\odot}

\newcommand{\msunyr}{M$_\odot$ yr$^{-1}$}

\newcommand{\lsun}{{$\rm\,L_\odot$}}

\newcommand{\kms}{$\rm km\ s^{-1}$}

\newcommand{\e}{${\rm e^{-}}$}

\definecolor{orange}{rgb}{.8,0.4,0}

\newcommand\submm{submillimeter}

\accepted{August 7, 2026}
\shorttitle{IPA: Morphology and Kinematics of \hydrogenmol\ winds}
\shortauthors{The IPA Team}

\usepackage{amsmath}

\usepackage{booktabs}
\usepackage{subcaption} 
\usepackage{siunitx}

\begin{document}
 
\title{
IPA: Morphology and Kinematics of Molecular Hydrogen Winds in Five Young Protostars across the Mass Spectrum Observed with JWST
}

\author[0000-0002-9497-8856]{Himanshu Tyagi}
\affiliation{Department of Astronomy and Astrophysics, Tata Institute of Fundamental Research, Homi Bhabha Road, Colaba, Mumbai 400005, IN}
\email[show]{tyagihimanshu027@gmail.com}

\author[0000-0002-3530-304X]{P. Manoj}
\affiliation{Department of Astronomy and Astrophysics, Tata Institute of Fundamental Research, Homi Bhabha Road, Colaba, Mumbai 400005, IN}
\email{mpuravankara@gmail.com}

\author[0000-0002-0554-1151]{Mayank Narang}
\affiliation{Jet Propulsion Laboratory, California Institute of Technology, 4800 Oak Grove Drive, Pasadena, CA 91109, USA}
\email{mayank.narang@jpl.nasa.gov}

\author[0000-0001-7629-3573]{S. Thomas Megeath}
\affiliation{Ritter Astrophysical Research Center, Department of Physics and Astronomy, University of Toledo, Toledo, OH, US}
\email{tommegeath@gmail.com}

\author[0000-0002-6447-899X]{Robert Gutermuth}
\affiliation{University of Massachusetts Amherst, Amherst, MA, US}
\email{rob.gutermuth@gmail.com}

\author[0000-0003-1430-8519]{Lee Hartmann}
\affiliation{University of Michigan, Ann Arbor, MI, US}
\email{lhartm@umich.edu}

\author[0000-0001-8876-6614]{Alessio Caratti o Garatti}
\affiliation{INAF-Osservatorio Astronomico di Capodimonte, IT}
\email{alessio.caratti@inaf.it}

\author[0000-0001-8302-0530]{Dan M. Watson}
\affiliation{University of Rochester, Rochester, NY, US}
\email{dmw@pas.rochester.edu}

\author[0000-0001-8341-1646]{David A. Neufeld}
\affiliation{William H. Miller III Department of Physics and Astronomy, The Johns Hopkins University, Baltimore, MD, USA}
\email{neufeld@jhu.edu}

\author[0000-0001-7591-1907]{Ewine F. Van Dishoeck}
\affiliation{Leiden Observatory, Universiteit Leiden, Leiden, Zuid-Holland, NL}
\affiliation{Max-Planck Institut f\"ur Extraterrestrische Physik, Garching bei München, DE}
\email{ewine@strw.leidenuniv.nl}

\author[0000-0001-5175-1777]{Neal J. Evans II}
\affiliation{Department of Astronomy, The University of Texas at Austin, 2515 Speedway, Stop C1400, Austin, Texas 78712-1205, USA}
\email{nje@astro.as.utexas.edu}

\author[0009-0006-3123-053X]{Vinod Chandra Pathak}
\affiliation{Department of Astronomy and Astrophysics, Tata Institute of Fundamental Research, Homi Bhabha Road, Colaba, Mumbai 400005, IN}
\email{vinodchandrapathak139@gmail.com}

\author[0000-0002-6136-5578]{Samuel A. Federman}
\affiliation{Ritter Astrophysical Research Center, Department of Physics and Astronomy, University of Toledo, Toledo, OH, US}
\affiliation{INAF-Osservatorio Astronomico di Capodimonte, IT}
\email{sam.federman@gmail.com}

\author[0000-0001-7491-0048]{Tyler L. Bourke}
\affiliation{SKA Observatory, Jodrell Bank, Lower Withington, Macclesfield SK11 9FT, UK}
\email{tyler.bourke@skao.int}

\author[0000-0001-8227-2816]{Yao-Lun Yang}
\affiliation{Star and Planet Formation Laboratory, RIKEN Pioneering Research Institute, Wako-shi, Saitama, 351-0106, Japan}
\email{yaolunyang.astro@gmail.com}

\author[0000-0002-7506-5429]{Guillem Anglada}
\affiliation{Instituto de Astrof{\'i}sica de Andaluc{\'i}a, CSIC, Glorieta de la Astronom{\'i}a s/n, E-18008 Granada, ES}
\email{guillem@iaa.es}

\author[0000-0002-1700-090X]{Henrik Beuther}
\affiliation{Max Planck Institute for Astronomy, Heidelberg, Baden Wuerttemberg, DE}
\email{beuther@mpia-hd.mpg.de}

\author[0000-0002-4540-6587]{Leslie W. Looney}
\affiliation{Department of Astronomy, University of Illinois, 1002 West Green St, Urbana, IL 61801, US}
\email{lwl@illinois.edu}

\author[0000-0003-2309-8963]{Rolf Kuiper}
\affiliation{Faculty of Physics, University of Duisburg-Essen, Lotharstra{\ss}e 1, D-47057 Duisburg, Germany}
\email{rolf.kuiper@uni-due.de}

\author[0000-0001-9443-0463]{Pamela Klaassen}
\affiliation{United Kingdom Astronomy Technology Centre, Edinburgh, GB}
\email{pamela.klaassen@stfc.ac.uk}

\author[0000-0002-4448-3871]{Pooneh Nazari}
\affiliation{European Southern Observatory, Karl-Schwarzschild-Strasse 2, 85748 Garching, Germany}
\email{Pooneh.Nazari@eso.org}


\author[0000-0001-8075-3819]{Bihan Banerjee}
\affiliation{Department of Astronomy and Astrophysics, Tata Institute of Fundamental Research, Homi Bhabha Road, Colaba, Mumbai 400005, IN}
\email{banerjeebihan@gmail.com}

\author[0000-0003-1665-5709]{Joel Green}
\affiliation{Space Telescope Science Institute, 3700 San Martin Drive, Baltimore, MD 21218, US}
\email{jgreen@stsci.edu}

\author[0009-0007-8087-5657]{Sujay Vijay Jadhav}
\affiliation{Department of Astronomy and Astrophysics, Tata Institute of Fundamental Research, Homi Bhabha Road, Colaba, Mumbai 400005, IN}
\email{sujayvijayjadhav@gmail.com}

\author[0000-0002-6737-5267]{Mayra Osorio}
\affiliation{Instituto de Astrof{\'i}sica de Andaluc{\'i}a, CSIC, Glorieta de la Astronom{\'i}a s/n, E-18008 Granada, ES}
\email{osorio@iaa.es}

\author[0000-0002-2585-0111]{B. Shridharan}
\affiliation{Department of Astronomy and Astrophysics, Tata Institute of Fundamental Research, Homi Bhabha Road, Colaba, Mumbai 400005, IN}
\email{shridharan.1997@gmail.com}

\author[0000-0003-2300-8200]{Amelia M.\ Stutz}
\affiliation{Departamento de Astronom\'{i}a, Universidad de Concepci\'{o}n,Casilla 160-C, Concepci\'{o}n, Chile}
\email{amelia.stutz@gmail.com}

\author[0000-0002-5812-9232]{Thomas Stanke}
\affiliation{Max-Planck Institut f\"ur Extraterrestrische Physik, Garching bei München, DE}
\email{tstanke049@gmail.com}

\author[0000-0002-6195-0152]{John J. Tobin}
\affil{National Radio Astronomy Observatory, 520 Edgemont Rd., Charlottesville, VA 22903 USA} 
\email{jtobin@nrao.edu}

\author[0000-0002-9470-2358]{Lukasz Tychoniec}
\affiliation{Leiden Observatory, Universiteit Leiden, Leiden, Zuid-Holland, NL}
\email{tychoniec@strw.leidenuniv.nl}

\author[0000-0002-0826-9261]{Scott Wolk}
\affiliation{Center for Astrophysics Harvard \& Smithsonian, Cambridge, MA, US}
\email{swolk@cfa.harvard.edu}

\author[0009-0001-7523-8887]{Manya Arora}
\affiliation{Department of Astronomy and Astrophysics, Tata Institute of Fundamental Research, Homi Bhabha Road, Colaba, Mumbai 400005, IN}
\email{manya.arora@tifr.res.in}





\begin{abstract}

Molecular winds {may} play a key role in governing angular momentum transport and accretion during the early evolution of protostars. We present the morphology and kinematic properties of the \hydrogenmol\ emission in five young, envelope-dominated, protostars across a broad bolometric luminosity range, from 0.2 to $10^4$~\lsun, observed with the NIRSpec/IFU and MIRI/MRS onboard JWST as part of the Investigating Protostellar Accretion (IPA) program. A rich set of pure rotational lines of \hydrogenmol, up to $v=0-0$ S(18), and a few ro-vibrational lines are detected in the winds, revealing bipolar structures. The \hydrogenmol\ lines show a stratified/onion-like structure morphologically and kinematically, where the lines with higher \Eup\ show a higher degree of collimation and higher velocities. Additionally, the wind velocity scales with the \lbol\ of the host protostellar system. In 4 out of 5 protostars, \hydrogenmol\ emission fills the outflow cavity without showing pronounced limb brightening. We also report a tentative detection of \hydrogenmol\ wind rotation in IRAS 16253, which suggests a launch radius of $\sim4$ au and the magnetic lever arm parameter of $\sim5-10$. Taken together, these properties of the \hydrogenmol\ winds can be explained by the magnetohydrodynamic disk wind models. We detect a collimated, high-velocity \hydrogenmol\ jet toward HOPS 370, which is more evolved than the extremely young source HH 211, but is accreting at a high accretion rate. This suggests that the presence of collimated molecular jets in protostars is more closely connected to accretion rate than system age.

\end{abstract}


\section{Introduction}\label{Intro}
The birth of stars and planetary systems begins with the collapse of a rotating dense core under self-gravity \citep[][]{Shu1977ApJ...214..488S, Terebey1984ApJ...286..529T, StahlerBook}. However, the conservation of angular momentum prevents the direct infall of envelope material onto the protostar, leading to the formation of a circumstellar disk \citep{Ulrich1976ApJ...210..377U, Cassen1981Icar...48..353C, Terebey1984ApJ...286..529T, Shu1987ARA&A..25...23S, StahlerBook, HartmannBook}. The central star then accumulates material via disk-mediated accretion \citep[e.g.,][]{Hartmann2016ARA&A..54..135H}, and most of this happens in an early, envelope-dominated, phase typically associated with the Class 0 spectral energy distribution (SED) \citep[e.g,][]{Andre1993ApJ...406..122A, Andre2000prpl.conf...59A, Evans2009ApJS..181..321E, Offner2011ApJ...736...53O, Fischer2017ApJ...840...69F, Mayank2023JApA...44...92N, Federman2023ApJ...944...49F, Fiorellino2026FrASS..1319945F}.
The overall protostellar phase is characterized by three mass flows that regulate the process of star and planet formation in these initial stages. These flows are: 1. infall of the envelope onto the disk, 2. accretion from the disk to the star, and 3. outflows \citep[e.g.,][and references therein]{Tobin2024arXiv240315550T}. 

Outflows {may} play a key role in star formation as they remove excess angular momentum from the disk to facilitate the accretion and prevent star from rapid rotation \citep[][]{Pudritz1983ApJ...274..677P, Pudritz1986ApJ...301..571P, Pudritz2007prpl.conf..277P, Shang2007prpl.conf..261S, Hartmann2016ARA&A..54..135H, Pudritz2019FrASS...6...54P, Ray2021NewAR..9301615R}. 
Protostellar winds and jets also play a role in shaping IMF \citep[e.g.,][]{Hennebelle2024ARA&A..62...63H}. 
Some studies have shown the direct proportionality between mass accretion and ejection \citep[see e.g.,][]{1995Hartigan, Cabrit2007LNP...723...21C, Ellerbroek2013A&A...551A...5E, Watson2016ApJ...828...52W, Oliva2023A&A...669A..80O, Oliva2023A&A...669A..81O}.
Observations of outflows from protostars have identified three distinct flows:  
1. the jets, fast-moving, highly collimated flows of gas, 2. winds, slower flows of gas with wide opening angles,
and 3. the cold ambient material entrained by the jets and the winds. For details, the reader is referred to \citet{Bachiller1999, Richer2000prpl.conf..867R, Arce2007prpl.conf..245A, Frank2014prpl.conf..451F, Bally2016ARA&A..54..491B, Lee2020A&ARv..28....1L, Ray2021NewAR..9301615R, Pascucci2023ASPC..534..567P} and references therein. 

Despite the ubiquitous detection of jets and winds in various evolutionary stages of star formation, their launching and propagation mechanism is still debated. The magnetocentrifugal models are generally used to describe the launch of jets and winds. The most discussed models in the literature are the X-winds  \citep{Shu1988ApJ...328L..19S, Shu1994ApJ...429..781S, Shu2000prpl.conf..789S, Shang2007prpl.conf..261S, Cai2008ApJ...672..489C} and the magnetohydrodynamics (MHD) disk winds  \citep{Pudritz1983ApJ...274..677P, Pudritz1986ApJ...301..571P, Wardle1993ApJ...410..218W, Konigl2000prpl.conf..759K, Pudritz2007prpl.conf..277P}.
In the X-wind scenario, material is launched from a narrow inner disk region, close to the co-rotation radius ($\sim$0.05 au to 0.5 au), with the help of the stellar magnetic field lines, threading the inner disk. Disk winds, in contrast, are launched from a wider range of disk radii, spanning several tens of au, accelerated by open magnetic field lines inherited from the parent cloud. 

Despite their differences, both models predict the presence of a wide-angle slow-moving component enveloping the fast-moving collimated jet. In the X-wind models, the wide angle component emerges due to the entrainment of the ambient material and infalling envelope 
in shocks generated by a fast wind with a strong angular dependence in the flow rate, launched from the inner disk \citep[e.g.,][and references therein]{Masson1993ApJ...414..230M, Raga1993A&A...278..267R, Chernin1994ApJ...426..204C, Ostriker1997ApJ...486..291O, Ostriker2001ApJ...557..443O, Arce2007prpl.conf..245A, Lee2000ApJ...542..925L, Lee2015ApJ...805..186L, Lee2021ApJ...907L..41L, Shang2020ApJ...905..116S, Lopez2024ApJ...977..126L}.

In the {MHD disk wind} model, on the other hand, the slow-moving wide angle material is naturally launched from a large range of disk radii \citep[e.g.,][and references therein]{Pudritz1983ApJ...274..677P, Pudritz1986ApJ...301..571P, Machida2004MNRAS.348L...1M, Bjerkeli2016Natur.540..406B, Tabone2017A&A...607L...6T, Kolligan2018A&A...620A.182K, Launhardt2023A&A...678A.135L, Pascucci2023ASPC..534..567P, Tsukamoto2023ASPC..534..317T}. One of the predictions of the {MHD disk wind} models is their nested structure in both velocity and chemical excitation \citep[e.g.,][and references therein]{Pascucci2023ASPC..534..567P}.
Moreover, disk winds may also play a fundamental role in accretion. Recent studies have shown that the magneto-rotational instability (MRI) is suppressed in the disks, suggesting that the MHD disk winds efficiently remove disks' angular momentum to facilitate accretion, especially in the outer disk radii \citep[e.g.,][]{Bai2013ApJ...769...76B, Gressel2015ApJ...801...84G, Flaherty2015ApJ...813...99F, Teague2018ApJ...864..133T, Pudritz2019FrASS...6...54P, Tabone2017A&A...607L...6T, Lesur2023ASPC..534..465L, Pascucci2025NatAs...9...81P, Villenave2025A&A...697A..64V}. 

Recent unified models of X-winds that take into account interactions with the core and its magnetic field can also reproduce these nested structures in morphology and velocity as wind gas slowed down via reverse shocks \citep{Shang2020ApJ...905..116S, Shang2023ApJ...944..230S}. 
{Nested structure of the winds is also a key feature of the MHD disk wind models \citep[][]{Frank2014prpl.conf..451F, Pascucci2023ASPC..534..567P}.}
Such nested structure has been routinely observed at various stages of star formation, thanks to the improved sensitivity and angular resolutions of modern infrared and submillimeter telescopes (e.g., {\citealt{Beck2007AJ....133.1221B, GarciaLopez2010A&A...511A...5G, Agra2014A&A...564A..11A, Frank2014prpl.conf..451F, Tabone2017A&A...607L...6T, deValon2020A&A...634L..12D, Lee2021ApJ...907L..41L, Pascucci2023ASPC..534..567P, Federman2023arXiv, Arulanantham2024ApJ...965L..13A, Delabrosse2024A&A...688A.173D, Garatti2024A&A...691A.134C, Nazari2024A&A...686A.201N_COMOutflow, Omura2024ApJ...963...72O, Tychoniec2024A&A...687A..36T, Pascucci2025NatAs...9...81P, Tazaki2025ApJ...980...49T, Bacciotti2025A&A...704A.157B, Blazquez-Calero2025NatAs.tmp..254B} }and references therein. See also the MHD models where fast material is launched into intrinsically narrow cones e.g., \citealt{Ferreira2000MNRAS.312..387F, Konigl2011MNRAS.416..757K, Zanni2013A&A...550A..99Z, Rabenanahary2022A&A...664A.118R, Rivera-Ortiz2023A&A...672A.116R}).


The origin of both the jets and the wider angle winds, which are ubiquitously detected in protostars, remains an open question. A systematic study is essential to characterize the properties of protostellar winds. This need is now being met by the unprecedented sensitivity, spatial, and spectral resolution of the James Webb Space Telescope in near to mid-IR spectral coverage \citep[e.g.,][]{Ray2023Natur.622...48R, Garatti2024A&A...691A.134C, Narang2024ApJ...962L..16N, Nisini2024ApJ...967..168N}. 

While jets are traced in ionic, atomic, and molecular tracers, historically, \hydrogenmol\ has been used as a tracer of shocks in molecular outflows (e.g., \citealt{Neufeld1998ApJ...506L..75N, Stanke2000PhDT........12S, Stanke2002A&A...392..239S, Neufeld2006ApJ...649..816N, Cabrit2004Ap&SS.292..501C, Garatti2006A&A...449.1077C, Maret2009ApJ...698.1244M}; also see \citealt{Davis2010A&A...511A..24D} for a catalog of \hydrogenmol\ emission objects related to YSO jets and outflows). 
The simultaneous spectral coverage and integral field unit (IFU) capabilities of the Near Infrared Spectrograph \citep[NIRSpec/IFU;][]{Boker2022A&A...661A..82B, Jakobsen2022A&A...661A..80J} and the Mid-Infrared Instrument \citep[MIRI/MRS][]{Rieke2015PASP..127..584R, Wright2015PASP..127..595W, Argyriou2023A&A...675A.111A, Wright2023PASP..135d8003W} provide an opportunity for the first time to spatially resolve multiple pure rotational and ro-vibrational lines of molecular hydrogen (\hydrogenmol) at high angular resolutions ($\sim0.2$\arcsec - 0.67\arcsec) in the immediate vicinity of the driving source. 
{Due to these capabilities, recent JWST studies have actively investigated the launching and propagation of \hydrogenmol\ winds and outflows from young protostars. These studies, primarily focused on low-mass protostars, have frequently reported a nested morphology in \hydrogenmol\ outflows (e.g., \citealt{Delabrosse2024A&A...688A.173D, Tychoniec2024A&A...687A..36T, Vleugels2025A&A...695A.145V, Narang2026arXiv260209837N}; see also \citealt{Harsono2023ApJ...951L..32H, Nisini2024ApJ...967..168N}). Similar results have also been observed in more evolved protoplanetary disks \citep[e.g.,][]{Arulanantham2024ApJ...965L..13A, Pascucci2025NatAs...9...81P,Narang2026arXiv260507016N}. However, such nested structures were not identified in \hydrogenmol\ observations of Ced110 IRS 4 or HH 46 IRS \citep{Narang2025AJ....169..192N, Navarro2025ApJ...995..199N}. In the very young protostar HH 211, \citet{Garatti2024A&A...691A.134C} found that \hydrogenmol\ emission dominates the mass ejection. Furthermore, a recent large survey by \citet{Francis2026arXiv260413773F} underscored both the prevalence of nested morphologies and the overall dynamical importance of \hydrogenmol\ in low-mass protostellar winds.}

{Most JWST studies of \hydrogenmol\ winds to date have focused on low-mass protostars with bolometric luminosity (\lbol) $<100$~\lsun. This paper presents a systematic investigation of the morphology and kinematics of the molecular hydrogen winds as a function of the host protostar's bolometric luminosity. As part of the Investigating Protostellar Accretion (IPA) across the mass spectrum, a JWST Cycle 1 medium GO program (PID: 1802; P.I.: S Thomas Megeath; \citealt{ipa_pro, Federman2023arXiv, Narang2024ApJ...962L..16N, Rubinstein2024ApJ...974..112R, Neufeld2024arXiv240407299N, Brunken2024, Nazari2024A&A...686A..71N, Slavicinska2024arXiv240415399S, Tyagi2025ApJ...983..110T}), we observed a sample of five close to edge-on Class 0 protostars in a broad range of bolometric luminosity (\lbol$\sim0.16-10^4$ \lsun; Table \ref{Table:ObsLog}).}
{While \citet{Federman2023arXiv} presented an overview of the NIRSpec observations for the full sample, comparing the morphology of multiple gas tracers (including the \hhnu(11)\ line), and \citet{Neufeld2024arXiv240407299N} presented a detailed analysis of the prompt OH lines in a shocked knot within the HOPS 370 outflow, this work focuses specifically on the \hydrogenmol\ emission. Furthermore, although the \hydrogenmol\ and CO outflows of IRAS 16253-2429 were recently analyzed by \citet{Narang2026arXiv260209837N}, we include this source to investigate trends across the entire luminosity range using uniform analysis tools.}

\begin{deluxetable*}{lcccccchhc}
\tablecaption{Sample Properties \label{Table:ObsLog}}

\tablehead{
\colhead{Object} &
\colhead{$L_{\rm bol}$} &
\colhead{$T_{\rm bol}$} &
\colhead{Inclination} &
\colhead{Disk PA} &
\colhead{Distance} &
\colhead{$M_\ast$} &
\nocolhead{Pointing R.A.} &
\nocolhead{Pointing Dec.} &
\colhead{Refs} \\
& 
\colhead{(\lsun)} &
\colhead{(K)} &
\colhead{($^\circ$)} &
\colhead{($^\circ$)} &
\colhead{(pc)} &
\colhead{($\rm M_{\odot}$)} &
\nocolhead{(ICRS)} &
\nocolhead{(ICRS)} &
\colhead{}
}

\startdata
IRAS 16253$-$2429 &
0.16 &
42 &
65 &
113 &
140 &
0.12--0.17 &
16$^{\rm h}$28$^{\rm m}$21.620$^{\rm s}$ &
$-$24$^\circ$36$'$24.16$''$ &
(1,1,2,2,2) \\
B335 &
1.4 &
37 &
68 &
5 &
165 &
0.25 &
19$^{\rm h}$37$^{\rm m}$00.894$^{\rm s}$ &
+07$^\circ$34$'$09.59$''$ &
(3,1,4,5,13,3) \\
HOPS 153 &
3.8 &
39.4 &
74.5 &
$-$33 &
390 &
0.6 &
05$^{\rm h}$37$^{\rm m}$57.023$^{\rm s}$ &
$-$07$^\circ$06$'$56.27$''$ &
(6,8,6,6,6,14) \\
HOPS 370 &
315.7 &
71.5 &
72.2 &
109.7 &
390 &
2.5 &
05$^{\rm h}$35$^{\rm m}$27.634$^{\rm s}$ &
$-$05$^\circ$09$'$34.42$''$ &
(7,8,6,6,6,7) \\
IRAS 20126+4104 &
$10^{4}$ &
60 &
82 &
56 &
1550 &
12 &
20$^{\rm h}$14$^{\rm m}$26.030$^{\rm s}$ &
+41$^\circ$13$'$32.57$''$ &
(9,10,11,12,9) \\
\enddata

\tablerefs{
(1) \citet{Ohashi2023ApJ...951....8O};
(2) \citet{Aso2023ApJ...954..101A};
(3) \citet{Evans2023ApJ...943...90E};
(4) \citet{Hodapp2026arXiv260212060H};
(5) \citet{Bjerkeli2023AA...677A..62B};
(6) \citet{Tobin2020ApJ...890..130T};
(7) \citet{Tobin2020ApJ...905..162T};
(8) \citet{Furlan2016ApJS..224....5F};
(9) \citet{Chen2016ApJ...823..125C};
(10) \citet{Cesaroni1999AA...345..949C};
(11) \citet{Massi2023AA...672A.113M};
(12) \citet{Reid2019ApJ...885..131R};
(13) \citet{Watson2020}
(14) J.J. Tobin P. Com.
}
\tablecomments{An inclination of $0^\circ$ corresponds to a face-on disk.}

\end{deluxetable*}

The paper is structured as follows: In Section \ref{ObservationSection}, we discuss the sample, observations, and data reduction. Section \ref{Section:Linemaps} to \ref{Section:Kinematics} present the observational results. In Section \ref{Discussion}, we discuss our findings and their implications. We summarize our conclusions in Section \ref{ConclusionSection}. {Finally, the availability of the dataset used in this work is described in Section \ref{Section:Data-Availability}.}

\section{Sample, Observations, and Data Reduction}
\label{ObservationSection}


Our sample of five protostars was observed during the JWST Cycle 1 with NIRSpec/IFU and MIRI/MRS as part of the IPA program. The detailed source properties are given in Table \ref{Table:ObsLog} \citep[see also][]{Federman2023arXiv, Rubinstein2024ApJ...974..112R}. The observation setup and data reduction methods are discussed below:

     
    NIRSpec observations were carried out between July 2022 and October 2022. All of the observations used the F290LP/G395M filter/grating combination in IFU mode, covering a spectral range from $\rm \sim2.87~to~5.01$ \micron\ with spectral resolution ranging from $\rm \sim700~to~\sim1300$. 
    {The angular resolution of NIRSpec IFU is slightly larger than the theoretical diffraction limit, ranging from $\sim0.12$\arcsec–$0.2$\arcsec\ over 3-5~\micron, and the point spread function (PSF) is known to be elongated along the instrument slicers \citep[][]{DEugenio2024NatAs...8.1443D, Bentz2025RNAAS...9..128B}.}
    To cover a larger field of view (FOV), we employed a $2\times2$ mosaic with 10\% overlap in a 4-point dither pattern. With these settings, our FOVs cover a region of 6\arcsec$\times$6\arcsec. We used NRSIRS2RAPID readout mode during all the observations. The raw data\footnote{Obtained from MAST \url{https://mast.stsci.edu/portal/Mashup/Clients/Mast/Portal.html}} were reduced using JWST pipeline version 1.9.5.dev7+gbf7d3c9b and CRDS context files `jwst\_1069.pmap'. We have applied custom bad pixel masks for flagging and astrometry corrections to the data. For the data reduction process details, the reader is referred to \citealt{Federman2023arXiv}.

    All five of our sources were observed with MIRI/MRS between July 2022 and July 2023. There is a gap of $\sim$6 to 10 months between our NIRSpec and MIRI observations for each object except for {IRAS 16253-2429 (hereafter IRAS 16253)}, which was observed nearly simultaneously,{ with a time gap of $\sim13$ hours}. These observations utilize all four channels of MIRI/MRS, covering a spectral range from $\sim4.9$ \micron\ to 28 \micron\ at spectral resolution ranging from $\rm R\sim4000~to~1500 $ \citep{Jones2023MNRAS.523.2519J}. 
    {Based on the in-flight performance, angular resolution of MIRI/MRS is also slightly larger than the theoretical diffraction limit and ranges from $\sim0.31$\arcsec at 5 \micron\ to $\sim1$\arcsec\ at 28 \micron\ \citep[][]{Argyriou2023A&A...675A.111A}.}
    Similar to NIRSpec observations, we used a $2\times2$ mosaicing footprint with 10\% overlap and a 4-point-dither pattern, except IRAS 20126, where a mosaic of $3\times1$ was used along the outflow axis. The SLOWR1 readout pattern was used for IRAS 16253, B335, and HOPS 153, while the FASTR1 readout pattern was used for HOPS 370 and {IRAS 20126+4104 (hereafter IRAS 20126)}. 
    {Dedicated background observations with similar exposure parameters, but without mosaicking, were obtained for each target to enable background subtraction. However, our background frames are contaminated by diffuse \hydrogenmol\ emission in the \hhnu(1) and S(2) lines. Therefore, to avoid over-subtraction of line flux, we did not apply background subtraction.}
    One drawback of this approach is an incorrect continuum level. However, this does not affect the measurements of line flux, velocity, or the line maps. Our data reduction steps for MIRI/MRS are discussed in \citealt{Narang2024ApJ...962L..16N} and \citealt{Neufeld2024arXiv240407299N}.
We refer the reader to \citet{ipa_pro} for a detailed description of the observational setup and configurations.

\label{ResultsSection}

\section{Detected \texorpdfstring{\hydrogenmol~}\ lines and their Morphology \label{Section:Linemaps}}

To determine the full set of \hydrogenmol\ emission lines that are detected and their spatial morphology in our sample, we first generated the continuum-subtracted line maps.
The continuum was estimated by fitting a baseline to each spaxel, using line-free regions on either side of the target emission line as anchor points. Following continuum subtraction, {a Gaussian profile was used to fit} to the residual \hydrogenmol\ emission at each spaxel. The integrated intensity derived from these Gaussian fits across all spaxels was then used to construct the spatially resolved line maps.

To compile a comprehensive inventory of \hydrogenmol\ lines, we used the \hhnu(11) line map as our guide due to its high S/N and extracted a one-dimensional spectrum from a 0.3\arcsec\ radius aperture centered on the brightest emission region in the line map for each protostar. At these brightest \hydrogenmol\ emission regions, we have detected several pure rotational lines, S($J$), in the ground ($v=0$) and first excited ($v=1$) vibrational states.
Additionally, we detected ro-vibrational lines of $v=1\rightarrow0$ O($J$) and $v=2\rightarrow1$ O($J$) in all five protostars. A detailed list of the detected lines is provided in Table \ref{Table:DetectedLines}.

\begin{figure*}[htbp]
    \centering
    \includegraphics[width=\linewidth]{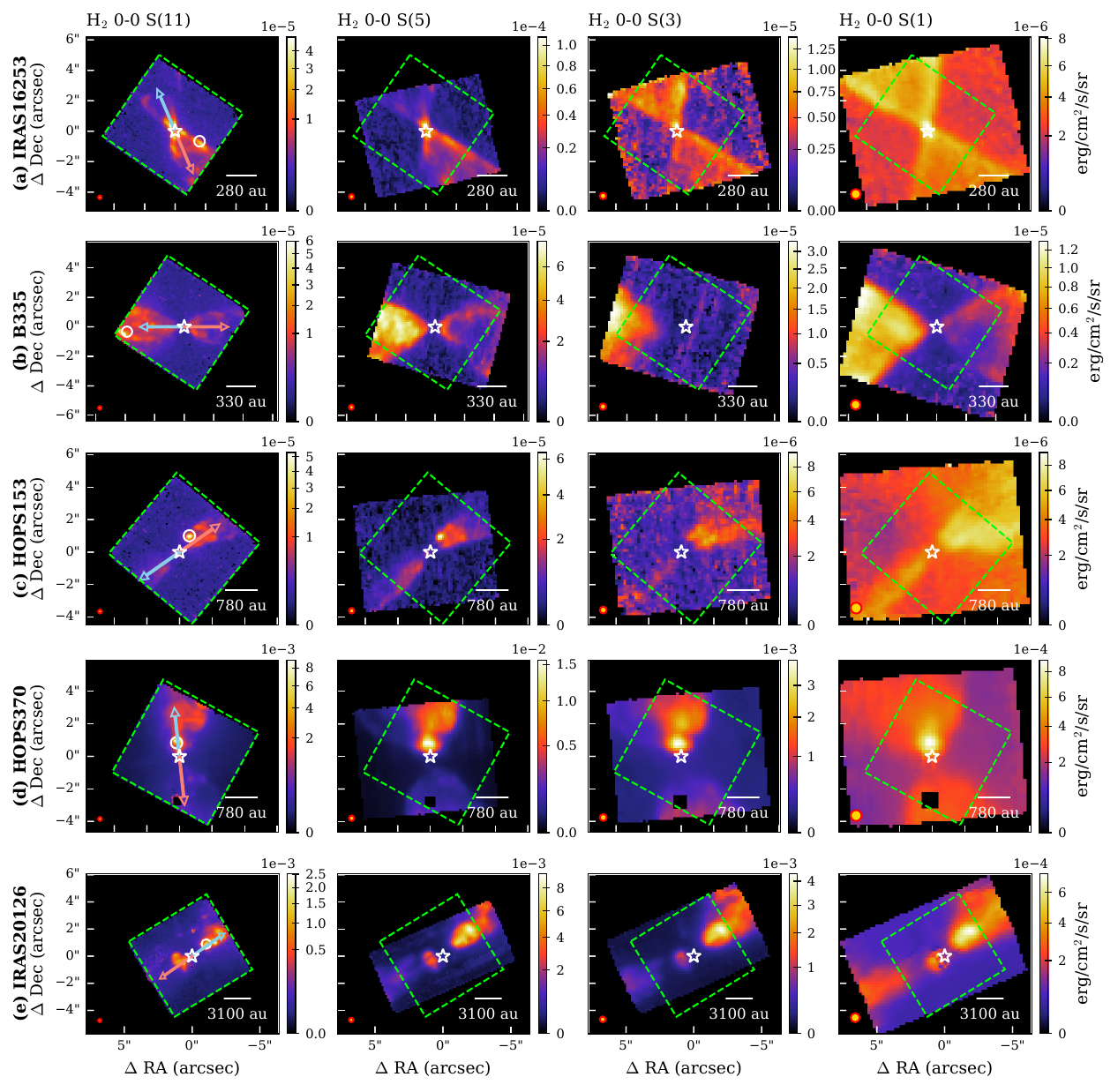}
    \caption{Line maps of the \hhnu(11), S(5), S(3), and S(1) transitions (labeled at the top of each panel) at 4.18, 6.91, 9.66, and 17.03~\micron, respectively, for IRAS~16253, B335, HOPS~153, HOPS~370, and IRAS~20126 (rows a–e). The white star marks the ALMA 870~\micron\ continuum position adopted from \citet{Tobin2020ApJ...890..130T}. The golden circles outlined in red at the lower left of each panel indicate the JWST beam size, and the dashed lime rectangle shows the NIRSpec field of view. In row (d), the bright foreground pre–main-sequence star [MGM2012]~2301 is masked to enhance the image contrast. In the \hhnu(11) maps, the positions of the collimated [Fe II] jets are indicated by blue and red arrows; their position angles are adopted from \citet{Narang2024ApJ...962L..16N} and \citet{Federman2026arXiv260109587F}. {White circle in the S(11) line maps mark apertures used to calculate the line fluxes reported in Table \ref{Table:DetectedLines}.}}
    \label{fig:LineMaps}
\end{figure*}

\begin{figure*}[htbp]
    \centering
    \includegraphics[width=\linewidth]{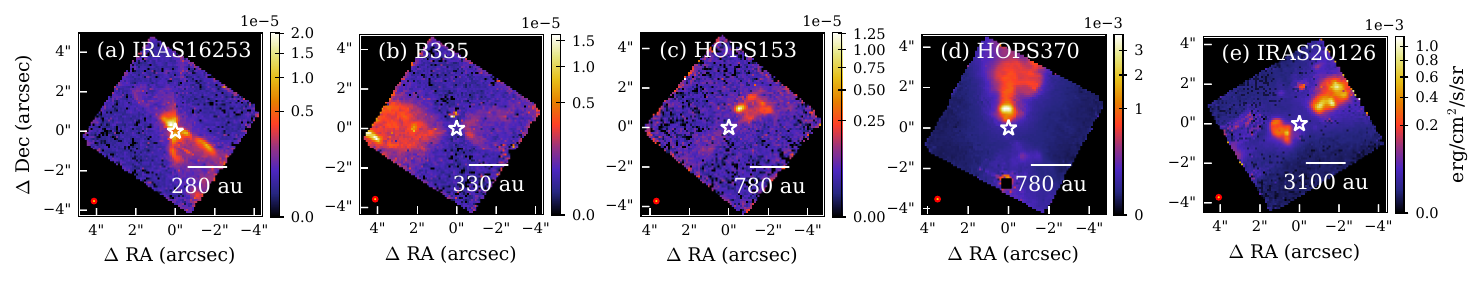}
    \caption{Line map of the H$_2$ $v=1$–0 O(7) transition at 3.81~\micron. The color scheme and annotation conventions are the same as in Figure~\ref{fig:LineMaps}.}
    \label{fig:H2-Ro-Vib}
\end{figure*}

All the maps of \hydrogenmol\ lines show bipolar morphology; however, the details of emission morphology vary depending on the upper state energy (\Eup) and the protostar (Figure \ref{fig:LineMaps} and Appendix \ref{LineMapAppendix}).
Based on visual inspection of the line maps, we determine that lines \hhnu(1), S(3), S(5), and S(11) (\Eup$\rm ~=~$1015, 2503, 4586,~and~13702 K, respectively) capture all the differences in morphology due to \Eup. They are shown in Figure~\ref{fig:LineMaps}, and below we discuss the details of the wind morphology of each protostar in our sample using these lines. 
Note that in Figure~\ref{fig:LineMaps}, we present only the ortho lines, as they are relatively brighter than their nearest para counterparts. Morphologically, however, the ortho and para lines closely trace each other. 
All the other line maps of pure-rotational \hydrogenmol\ transitions are presented in Appendix \ref{LineMapAppendix} in Figures \ref{fig:I16253Linemap}, \ref{fig:B335Linemap}, \ref{fig:HOPS153Linemap}, \ref{fig:HOPS370Linemap}, and \ref{fig:IRAS20126Linemap}. 
For completeness, we also include maps of the ro-vibrational H$_2$ $v=1$–0 O(7) transition at 3.81~\micron\ for all five sources in Figure \ref{fig:H2-Ro-Vib}, which show morphologies similar to those of the pure-rotational lines observed with NIRSpec.
For the further discussion, we classify pure rotational \hydrogenmol\ lines into three categories based on the rotational quantum number of the lower energy state ($J$) of each transition: (1) high-$J$ lines ($J>7$), (2) mid-$J$ lines ($7 \geq J \geq 4$), and (3) low-$J$ lines ($J\leq3$).

\subsection{IRAS 16253-2429}

{IRAS 16253 is the lowest luminosity (\lbol=0.16\lsun) and the lowest stellar mass ($M_{*} \sim 0.12-0.17~\msun$) source in the IPA sample, located in the Ophiuchus molecular cloud \citep[][]{Narang2024ApJ...962L..16N, Narang2026arXiv260209837N}.} The bipolar, hourglass-shaped morphology of the line emission is distinctly outlined by bright limbs traced in \hydrogenmol\ emission (Figure \ref{fig:LineMaps}(a) and \ref{fig:I16253Linemap}; {see also \citealt{Federman2023arXiv, Narang2026arXiv260209837N}}). 
{The bright limbs likely originate from oblique shocks driven by winds along the cavity walls \citep[][]{Narang2026arXiv260209837N}.} 
We further note that {low-$J$ ($J\leq3$)} lines show some morphological differences compared to {mid-$J$ ($7 \geq J \geq 4$) and high-$J$ ($J>7$)} \hydrogenmol\ lines. {As also shown by \citet{Narang2026arXiv260209837N}, the low-$J$  \hydrogenmol\ lines are brighter and spatially more extended in the northern cavity than the higher-$J$ lines, and their emission appears to fill the outflow cavity. }

{Due to the higher angular resolution of JWST at shorter wavelengths, the high-$J$ \hydrogenmol\ line maps reveal more collimated emission within the northern cavity and fragmented shocked emission along the cavity limbs (see \citealt{Federman2023arXiv, Narang2026arXiv260209837N})}. In the northern cavity, the high-$J$ \hydrogenmol\ lines appear to spatially connect to the [Fe II] shock knot \citep[][]{Federman2023arXiv, Narang2024ApJ...962L..16N}.

{The \hhnu(11) line map reveals weak emission outside the outflow limbs in the FOV. This emission is also detected in the \hhnu(1) line map}, as evident by the non-zero \hydrogenmol\ emission outside the cavity boundary (Figure \ref{fig:LineMaps}; also see Section \ref{Section:AmbientH2} and Figure \ref{fig:H2_outside}). All the pure rotational \hydrogenmol\ line maps are shown in Figure \ref{fig:I16253Linemap}.

\begin{longtable}{lcccccccccccc}
\caption{List of detected H$_2$ lines. Notation -- (!) blended/tentative detection.} \label{Table:DetectedLines} \\
\toprule
Transition & $\lambda$ & $E_u/k$ & $A$ & $g_u$ & IRAS 16253 & B335 & HOPS 153 & HOPS 370 & IRAS 20126 \\
\cline{6-10}
 & $\mu$m & (K) & (s$^{-1}$) &  & \multicolumn{5}{c}{($10^{-17}\ {\rm erg~s^{-1}cm^{-2}}$)} \\\midrule
\endfirsthead
\caption[]{List of detected H$_2$ lines. Notation -- (!) blended/tentative detection.} \\
\toprule
Transition & $\lambda$ & $E_u/k$ & $A$ & $g_u$ & IRAS 16253 & B335 & HOPS 153 & HOPS 370 & IRAS 20126 \\
\cline{6-10}
 & $\mu$m & (K) & (s$^{-1}$) &  & \multicolumn{5}{c}{($10^{-17}\ {\rm erg~s^{-1}cm^{-2}}$)} \\\midrule
\endhead
\midrule
\multicolumn{10}{r}{Continued on next page} \\
\midrule
\endfoot
\bottomrule
\endlastfoot
0 - 0 S(18) & 3.439 & 27643.26 & 3.65e-06 & 41 & 0.64$\pm$0.04 & 0.96$\pm$0.07 & 0.35$\pm$0.06 & 416.54$\pm$36.9 & 70.03$\pm$2.79 \\
0 - 0 S(17) & 3.486 & 25538.82 & 3.24e-06 & 117 & 0.58$\pm$0.02 & 1.22$\pm$0.08 & 0.25$\pm$0.02 & 163.02$\pm$23.77 & 45.4$\pm$2.39 \\
0 - 0 S(16) & 3.548 & 23459.03 & 2.82e-06 & 37 & 0.38$\pm$0.02 & 0.55$\pm$0.05 & 0.18$\pm$0.03 & 99.99$\pm$5.26 & 20.81$\pm$0.49 \\
0 - 0 S(15) & 3.626 & 21411.28 & 2.40e-06 & 105 & 2.13$\pm$0.06 & 3.09$\pm$0.07 & 1.43$\pm$0.06 & 575.11$\pm$18.25 & 124.8$\pm$3.77 \\
0 - 0 S(14) & 3.724 & 19403.37 & 1.99e-06 & 33 & 1.4$\pm$0.03 & 1.78$\pm$0.06 & 1.12$\pm$0.07 & 434.24$\pm$24.43 & 87.57$\pm$3.98 \\
0 - 0 S(13) & 3.846 & 17443.51 & 1.61e-06 & 93 & 5.46$\pm$0.09 & 8.23$\pm$0.07 & 4.79$\pm$0.17 & 1497.65$\pm$71.91 & 324.4$\pm$12.04 \\
0 - 0 S(12) & 3.996 & 15540.28 & 1.26e-06 & 29 & 2.8$\pm$0.03 & 3.53$\pm$0.11 & 2.79$\pm$0.13 & 762.71$\pm$28.63 & 169.83$\pm$4.82 \\
0 - 0 S(11) & 4.181 & 13702.64 & 9.60e-07 & 81 & 13.13$\pm$0.31 & 16.57$\pm$0.22 & 14.3$\pm$0.17 & 3538.97$\pm$69.45 & 829.45$\pm$26.87 \\
0 - 0 S(10) & 4.410 & 11939.86 & 7.01e-07 & 25 & ! & ! & ! & ! & ! \\
0 - 0 S(9) & 4.695 & 10261.47 & 4.89e-07 & 69 & ! & ! & ! & ! & ! \\
0 - 0 S(8) & 5.053 & 8677.17 & 3.23e-07 & 21 & 8.5$\pm$0.47 & 11.69$\pm$0.82 & 16.36$\pm$0.92 & 3708.2$\pm$138.74 & 892.4$\pm$25.39 \\
0 - 0 S(7) & 5.511 & 7196.72 & 2.00e-07 & 57 & 30.52$\pm$0.8 & 49.95$\pm$1.55 & 47.3$\pm$0.45 & 10175.79$\pm$362.25 & 3714.36$\pm$43.7 \\
0 - 0 S(6) & 6.109 & 5829.85 & 1.14e-07 & 17 & 7.06$\pm$0.27 & 11.8$\pm$0.56 & 6.01$\pm$0.32 & 3441.34$\pm$176.08 & 1234.53$\pm$42.6 \\
0 - 0 S(5) & 6.910 & 4586.07 & 5.87e-08 & 45 & 29.43$\pm$1.23 & 46.46$\pm$2.34 & 26.24$\pm$1.33 & 10843.1$\pm$638.07 & 4114.03$\pm$155.32 \\
0 - 0 S(4) & 8.025 & 3474.51 & 2.64e-08 & 13 & 12.38$\pm$0.2 & 26.43$\pm$0.54 & 12.4$\pm$0.19 & 4005.67$\pm$76.48 & 1784.04$\pm$18.83 \\
0 - 0 S(3) & 9.665 & 2503.75 & 9.83e-09 & 33 & 4.68$\pm$0.12 & 11.66$\pm$0.26 & 2.62$\pm$0.11 & 2999.27$\pm$98.22 & 1069.68$\pm$15.53 \\
0 - 0 S(2) & 12.279 & 1681.64 & 2.75e-09 & 9 & 4.92$\pm$0.07 & 10.03$\pm$0.15 & 3.94$\pm$0.05 & 1023.22$\pm$16.97 & 594.94$\pm$7.12 \\
0 - 0 S(1) & 17.035 & 1015.09 & 4.76e-10 & 21 & 3.73$\pm$0.09 & 6.87$\pm$0.17 & 3.99$\pm$0.08 & 571.72$\pm$45.8 & 385.37$\pm$9.51 \\
1 - 1 S(17) & 3.698 & 30152.96 & 2.61e-06 & 117 & 0.28$\pm$0.03 & 0.58$\pm$0.06 & 0.3$\pm$0.03 & 51.66$\pm$4.17 & 15.96$\pm$0.92 \\
1 - 1 S(16) & 3.760 & 28193.66 & 2.30e-06 & 37 & $<$0.13 & $<$0.16 & $<$0.14 & 30.41$\pm$5.68 & 6.32$\pm$0.74 \\
1 - 1 S(15) & 3.841 & 26262.65 & 1.99e-06 & 105 & $<$0.45  & $<$0.33 & $<$1.34 & $<$349.92 & $<$76.32 \\
1 - 1 S(14) & 3.942 & 24367.55 & 1.67e-06 & 33 & 0.38$\pm$0.07 & 0.54$\pm$0.04 & 0.5$\pm$0.08 & 39.78$\pm$4.21 & 17.73$\pm$2.25 \\
1 - 1 S(13) & 4.068 & 22516.32 & 1.37e-06 & 93 & 1.07$\pm$0.02 & 2.02$\pm$0.18 & 1.35$\pm$0.12 & 262.29$\pm$33.84 & 68.31$\pm$3.86 \\
1 - 1 S(12) & 4.224 & 20717.31 & 1.09e-06 & 29 & 0.23$\pm$0.02 & 0.77$\pm$0.08 & 0.79$\pm$0.09 & 94.05$\pm$12.29 & 22.14$\pm$1.65 \\
1 - 1 S(11) & 4.417 & 18979.17 & 8.36e-07 & 81 & ! & ! & ! & ! & ! \\
1 - 1 S(10) & 4.656 & 17310.84 & 6.17e-07 & 25 & ! & ! & ! & ! & ! \\
1 - 1 S(9) & 4.954 & 15721.51 & 4.35e-07 & 69 & ! & ! & ! & ! & ! \\
1 - 1 S(8) & 5.330 & 14220.52 & 2.90e-07 & 21 & $<$0.64 & $<$12.91 & $<$0.89 & 167.47$\pm$30.76 & 28.46$\pm$2.04 \\
1 - 1 S(7) & 5.811 & 12817.29 & 1.81e-07 & 57 & $<$0.78 & $<$1.81 & $<$1.62 & 321.95$\pm$11.22 & 82.44$\pm$4.8 \\
1 - 1 S(6) & 6.438 & 11521.15 & 1.04e-07 & 17 & $<$0.49 & $<$1.03 & $<$1.58 & 78.53$\pm$6.06 & 30.12$\pm$3.7 \\
1 - 1 S(5) & 7.280 & 10341.27 & 5.41e-08 & 45 & $<$0.69 & $<$0.64 & 1.4$\pm$0.25 & 246.5$\pm$15.7 & 59.45$\pm$2.83 \\
1 - 1 S(4) & 8.453 & 9286.45 & 2.45e-08 & 13 & $<$0.51 & $<$0.46 & $<$0.56 & 39.13$\pm$4.04 & 9.68$\pm$0.81 \\
1 - 1 S(3) & 10.178 & 8364.95 & 9.16e-09 & 33 & $<$0.23 & $<$0.26 & $<$0.18 & 23.92$\pm$3.58 & 5.91$\pm$0.92 \\
1 - 1 S(2) & 12.928 & 7584.36 & 2.58e-09 & 9 & $<$0.12 & $<$0.16 & $<$0.11 & $<$10.28 & $<$4.98 \\
1 - 1 S(1) & 17.932 & 6951.31 & 4.47e-10 & 21 & ! & ! & ! & ! & ! \\
1 - 0 O(4) & 3.004 & 6471.40 & 2.89e-07 & 5 & $<$0.18 & $<$0.31 & $<$0.37 & 340.78$\pm$5.69 & 162.19$\pm$2.26 \\
1 - 0 O(5) & 3.235 & 6951.31 & 2.08e-07 & 21 & 1.36$\pm$0.04 & 2.57$\pm$0.08 & 0.48$\pm$0.06 & 1936.92$\pm$32.48 & 540.34$\pm$8.46 \\
1 - 0 O(6) & 3.501 & 7584.36 & 1.50e-07 & 9 & 1.43$\pm$0.05 & 1.46$\pm$0.04 & 0.56$\pm$0.06 & 842.16$\pm$29.78 & 181.43$\pm$7.2 \\
1 - 0 O(7) & 3.807 & 8364.95 & 1.06e-07 & 33 & 4.02$\pm$0.12 & 4.92$\pm$0.11 & 2.98$\pm$0.08 & 1907.68$\pm$52.36 & 410.59$\pm$6.69 \\
1 - 0 O(8) & 4.162 & 9286.45 & 7.35e-08 & 13 & 0.7$\pm$0.03 & 0.86$\pm$0.08 & 0.99$\pm$0.08 & 289.98$\pm$5.69 & 69.62$\pm$1.77 \\
1 - 0 O(9) & 4.575 & 10341.27 & 4.96e-08 & 45 & ! & ! & ! & ! & ! \\
2 - 1 O(3) & 2.974 & 11789.07 & 6.36e-07 & 9 & $<$0.07 & $<$0.18 & $<$0.24 & 101.11$\pm$8.45 & 36.03$\pm$0.89 \\
2 - 1 O(4) & 3.190 & 12094.90 & 4.38e-07 & 5 & $<$0.1 & $<$0.21 & 0.21$\pm$0.03 & 40.05$\pm$3.2 & 12.9$\pm$0.49 \\
2 - 1 O(5) & 3.438 & 12550.05 & 3.15e-07 & 21 & ! & ! & ! & ! & ! \\
\end{longtable}

\subsection{B335}
\begin{figure}[htbp]
    \centering
    \includegraphics[width=\linewidth]{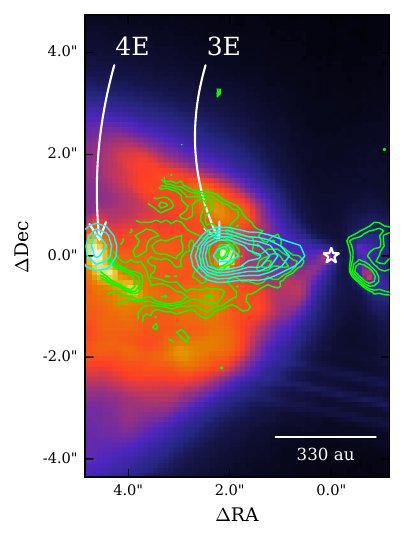}
    \caption{{\hhnu(11)} contours (lime) and [Fe II] 17.9 \micron\ contours (cyan) overplotted on the NIRCam  F444W image of B335. The NIRCam image is taken from GTO program 1187 (P.I.: Hodapp, K. W.). 
    The white arrows mark the location of shocks 4E and 3E as labeled by \citealt{Hodapp2024AJ....167..102H}. White star marks the ALMA 870 \micron\ continuum position in 2023 (Kim, C. et al. in prep.).}
    \label{fig:B335_NIRCAM}
\end{figure}
{B335 is the most isolated source in our sample as it is forming in a Bok globule \citep[][]{Keene1983ApJ...274L..43K}, while the others are forming in clusters in larger molecular clouds.
It has \lbol~$=1.4$~\lsun\ for a distance of 165 pc \citep{Watson2020} and $M_{*}=0.25~\msun$.} \hydrogenmol\ winds in B335 also show a bipolar, hourglass-shaped morphology (Figures \ref{fig:LineMaps}(b) and \ref{fig:B335Linemap}; {\citealt{Federman2023arXiv}}). The eastern cavity shows bright emission, with no distinct limb-brightening observed. However, the western lobe shows limb brightening. The eastern wind emission {is} brighter than the western emission in all the line maps because the eastern cavity is inclined toward the observer{, resulting in relatively lower extinction towards the eastern cavity than the western cavity }\citep[][]{Stutz2008, Yen2010ApJ...710.1786Y, Evans2023ApJ...943...90E}.  
The {low-$J$ ($J\leq3$)} lines, e.g., S(1), show much wider and conical morphology, while the {high-$J$ ($J>7$)} lines show narrower and curved shell-like morphology. 
{
The line emission of \hhnu(3) and the $v=0-0$ transitions with $J>14$ is faint within $\sim1.8$\arcsec\ east of the central protostar and in the western cavity due to relatively higher extinction toward the protostar and the western cavity, as explained below (see Figure \ref{fig:B335Linemap}).
The $0-0$ S(3) line falls on the trough of the silicate absorption and, therefore, suffers higher extinction and is harder to detect. $J>$14 lines are intrinsically weaker relative to their lower-$J$ counterparts and suffer higher extinction as they fall towards shorter wavelengths ($<3.7$ \micron). In addition, $J>14$ lines lie in the broad 3.0 \micron\ \water\ ice absorption band, providing additional extinction and making their detection difficult in the central region and in the western cavity.}

{High-$J$ lines reveal detailed structures and multiple shock fronts in the outflow cavity \citep[see also][]{Federman2023arXiv, Hodapp2024AJ....167..102H}.} 
{Among these features, a bright shell structure in the east is enhanced in the high-$J$ \hydrogenmol\ lines (e.g., see \hhnu(11) line map in Figure \ref{fig:LineMaps}(b); also see Figure \ref{fig:B335Linemap}). A shocked knot, traced by the [Fe II], is associated with the head of this feature. Such structures are explained by one-sided bow shocks (e.g., \citealt{Lee2000MNRAS.315...11L}; see also \citealt{Gustafsson2010A&A...513A...5G})}
The associated [Fe II] shocked knot is shown in Figure~\ref{fig:B335_NIRCAM}~(labeled as 4E in \citealt{Hodapp2024AJ....167..102H}; also see \citealt{Federman2026arXiv260109587F}).

Further, NIRSpec observations detect an extended faint emission, outside the hourglass bright emission, everywhere in the FOV (first panel from left in Figure \ref{fig:LineMaps}). This faint emission, however, is only weakly observed in MIRI data, due to the lower sensitivity of MIRI/MRS relative to NIRSpec/IFU \citep[see Figure 8 in][]{Rigby2023PASP..135d8001R}.

\begin{figure*}
    \centering
    \gridline{
              \fig{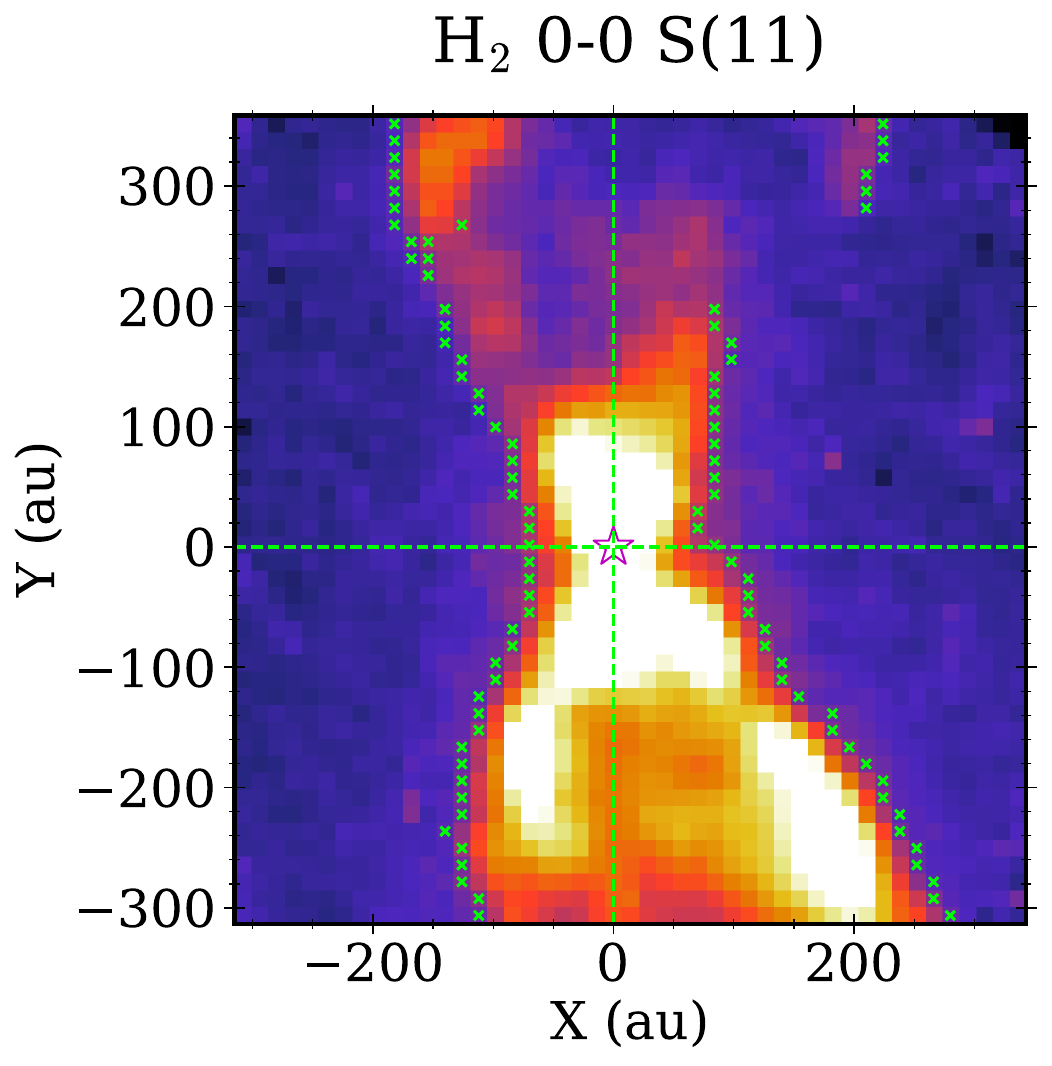}{0.3\textwidth}{}
              \fig{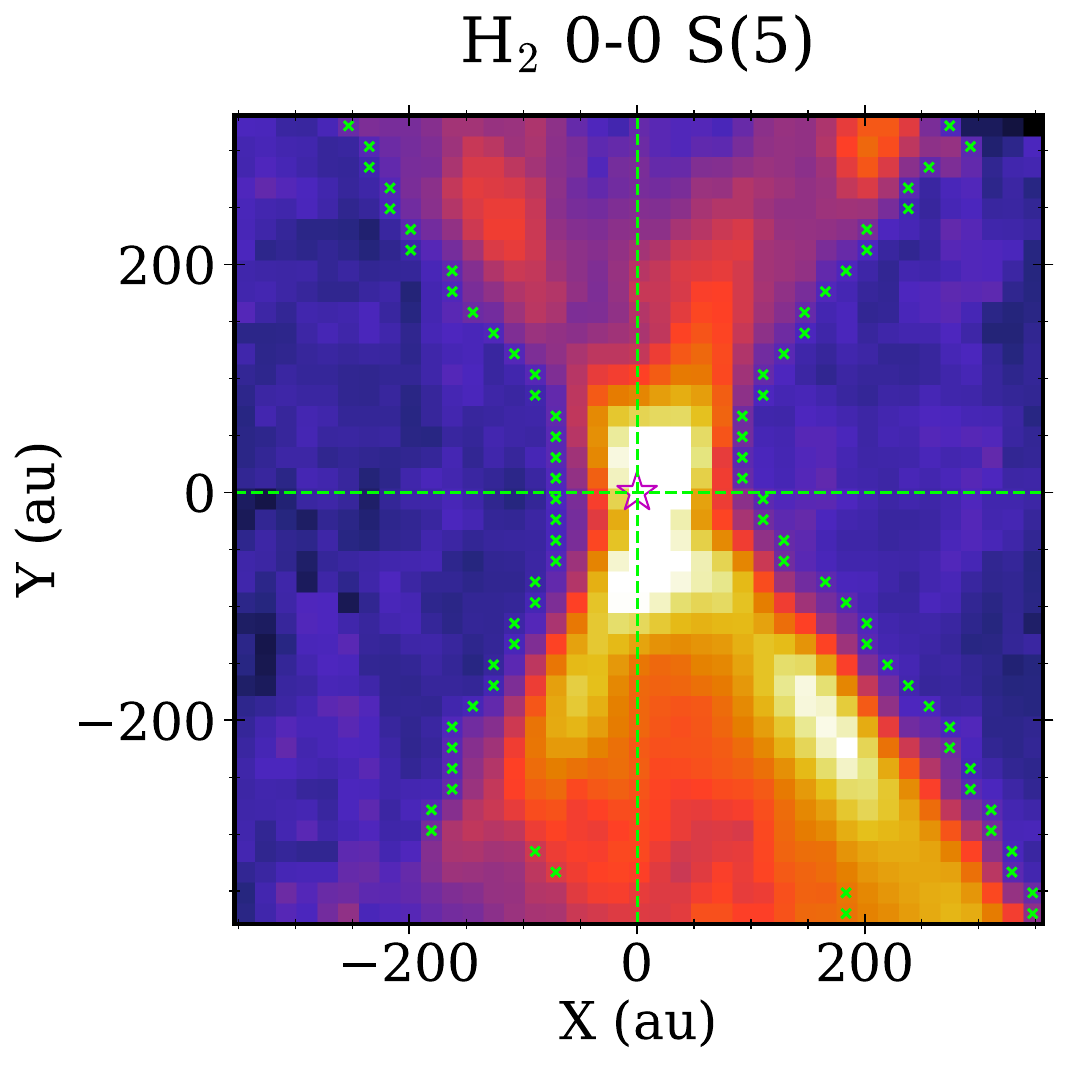}{0.3\textwidth}{}
              \fig{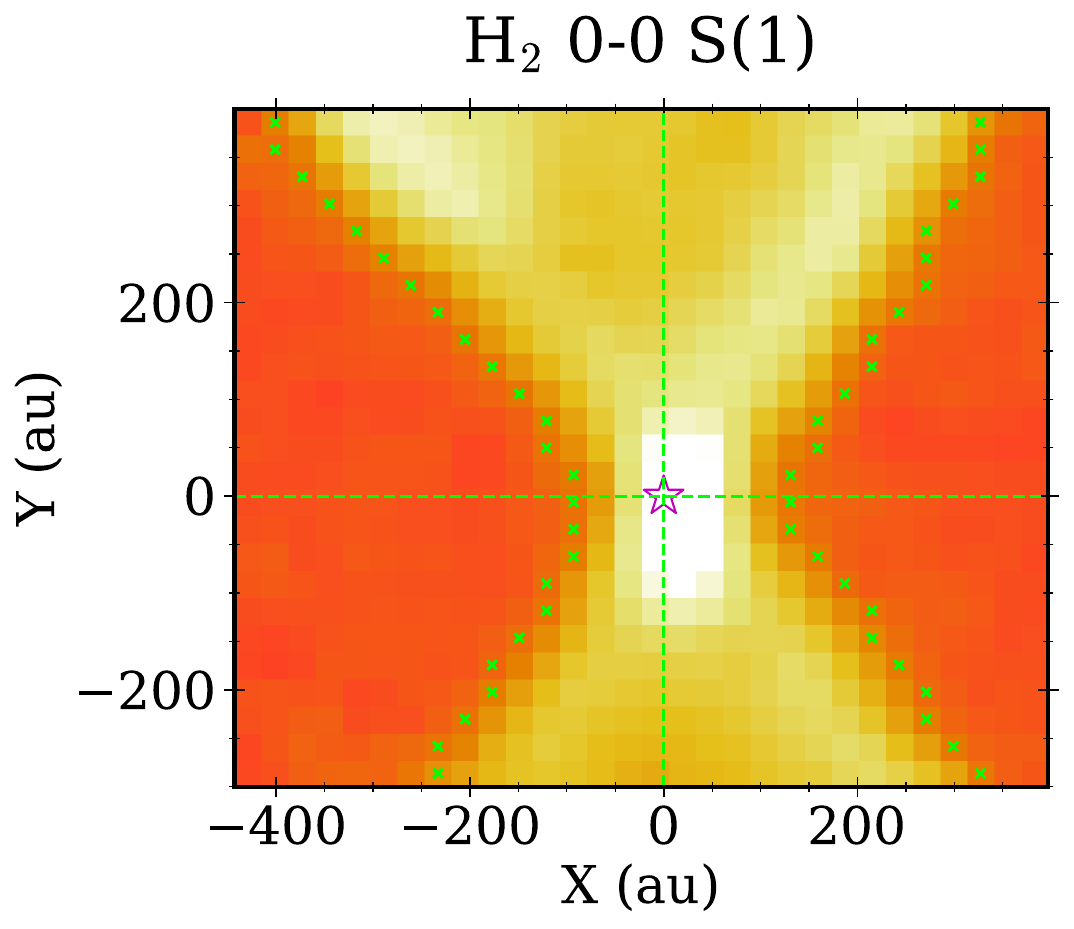}{0.3\textwidth}{}
              }
    \caption{Example of edge detection in IRAS 16253. The color scale depicts the rotated line maps of {\hhnu(11), S(5), and S(1)} lines (left to right). The lime crosses mark the edges of the \hydrogenmol\ emission. The magenta star marks the ALMA 870 \micron\ peak.}
    \label{fig:EdgeDetection}
\end{figure*}

\subsection{HOPS 153}
{HOPS 153 is the third most luminous source in our sample with \lbol=3.8\lsun\ and $M_*=0.6~\msun$.} It is located in the Orion A molecular cloud, at a distance of 390 pc \citep{Tobin2020ApJ...905..162T}.
\hydrogenmol\ line maps of HOPS 153 also reveal a bipolar hourglass morphology (Figures \ref{fig:LineMaps}(c) and \ref{fig:HOPS153Linemap}; see also \citealt{Federman2023arXiv}). 
{In this case, the southeastern cavity, which is inclined towards the observer, {is} fainter than the northwestern cavity.}
Notably, only the northwestern cavity is clearly visible in $J>13$ \hydrogenmol\ lines. The northwestern cavity is also broader than its southeastern counterpart (Figure \ref{fig:HOPS153Linemap}). 
{The southeastern lobe shows a lateral brightness asymmetry, with its lower half brighter than the upper half.}
Similar to IRAS 16253 and B335, {low-$J$ ($J\leq3$)} \hydrogenmol\ lines display a wider morphology than the {mid-$J$ $(7\geq J \geq4)$ and high-$J$ ($J>7$)} lines. The highly collimated [Fe II] jet also shows a similar orientation as the \hydrogenmol\ and bisects the \hydrogenmol\ lobes \citep[][]{Federman2026arXiv260109587F}.

{The northwestern cavity reveals complex emission morphology in mid- and high-$J$ \hydrogenmol\ lines and lacks the clear limb-brightened structure as observed in IRAS 16253 \citep[see also][for high-$J$ lines]{Federman2023arXiv}.} Further, the northwestern cavity reveals shell-like morphology with a bright knot located north of the collimated jet axis \citep[see also][]{Federman2023arXiv}. Surprisingly, this bright knot is not present in the S(3), S(2), and S(1) line maps (Figure \ref{fig:HOPS153Linemap}). 
{Furthermore, the \hhnu(1) and S(2) lines and other pure rotational \hydrogenmol\ lines within the NIRSpec spectral coverage, from S(8) to S(15),} exhibit faint, extended emission outside the outflow cavity (see Section \ref{Section:AmbientH2}).

\subsection{HOPS 370} \label{section: HOPS 370 Linemaps}
The intermediate-mass protostar, HOPS 370 (\lbol$=315.7$\lsun, $M_{\star}=2.5\msun$; {\citealt{Tobin2020ApJ...905..162T}}), also shows a bipolar-hourglass-shaped morphology in \hydrogenmol\ (Figure \ref{fig:LineMaps}(d) and \ref{fig:HOPS370Linemap}; {\citealt{Federman2023arXiv}}). 
The northern cavity, inclined towards the observer, is brighter than the southern cavity.
However, the details inside the hourglass cavity are very complex. The line maps of the highest resolution and {high-$J$ ($J>7$)} \hydrogenmol\ lines show a collimated, jet-like structure with a bright shocked knot in the middle at $\sim0.8$\arcsec\ north of the ALMA continuum peak \citep[see also][]{Federman2023arXiv, Neufeld2024arXiv240407299N, Tyagi2025ApJ...983..110T}. 
{A mushroom-like structure is observed close to the collimated, jet-like structure, at $\sim2$\arcsec\ north of the ALMA continuum \citep{Federman2023arXiv}.} 
This mushroom-like structure {appears to expand} predominantly towards the western side of the {[Fe II] jet axis (marked by the blue arrow in the first panel of Figure \ref{fig:LineMaps} from \citep[][]{Federman2026arXiv260109587F}). }
Even in {mid-$J$ $(7\geq J \geq4)$ and low-$J$ ($J\leq3$)} \hydrogenmol\ lines, the central shock knot and the mushroom-shaped structure remain visible despite their relatively lower {angular resolution}.
The southern cavity is fainter but also shows structured emissions in \hydrogenmol\ west of the jet axis. 

Surrounding these highly structured features is a faint, hourglass-shaped enveloping emission without limb brightening, which is more pronounced in the \hhnu (1) to S(7) lines (Figure \ref{fig:LineMaps}(d), \ref{fig:HOPS370Linemap}). This enveloping emission is asymmetric in the northern cavity, with the eastern side more extended relative to the western cavity. The low-$J$ \hydrogenmol\ lines show wider morphology than the high-$J$ lines. In HOPS 370 also, we detect faint \hydrogenmol\ emission outside the outflow cavity. This extended emission is detected in all \hhnu($J$)\ lines. 

\subsection{IRAS 20126+4104}\label{section: IRAS20126 Linemaps}
{The highest luminosity and stellar mass source in our sample, IRAS 20126 (\lbol~$=10^4$~\lsun and $M_*=12~\msun$),} also shows a bipolar outflow similar {to the low-mass (IRAS 16253, B335, and HOPS 153) and intermediate-mass (HOPS 370) protostars} (Figure \ref{fig:LineMaps}(e) and \ref{fig:IRAS20126Linemap}). However, the morphology is not strictly hourglass-shaped, perhaps due to its distance (1.5 kpc), as we are tracing the outflow out to much larger distances than the other sources.
{The high-$J$ \hydrogenmol\ line maps, observed with NIRSpec, reveal a highly fragmented structure in the emission due to JWST's higher angular resolution at shorter wavelengths (see e.g., \hhnu(11) line map in Figure \ref{fig:LineMaps}(e)).}
The northwestern outflow shows multiple shock fronts, while the southeastern outflow, fainter than the northwestern outflow, shows a bubble-like structure very close to the protostar. 

The MIRI observations, which cover a larger spatial region along the outflow, show the \hydrogenmol\ wind bending towards the north in the northwestern cavity. Further in the southeast, MIRI observations show a very faint but extended flow. Recently, a similar morphology was observed by \citet{Massi2023AA...672A.113M} in the \hydrogenmol\ 2.12 \micron\ line. 
The complexity of the observed \hydrogenmol\ morphology may additionally be influenced by the presence of multiple companions in the IRAS~20126 system, which might be driving multiple outflows \citep[][]{Federman2023arXiv}.
Similar to low-mass protostars, an extended fainter emission outside the wind is also detected in all the \hhnu($J$)\ lines.

\subsection{Measuring the Spatial Extent of the \texorpdfstring{\hydrogenmol}\~~Emission}
\label{Section:OA_section}


\begin{deluxetable*}{llcccccc}
\tablecaption{Half opening angles ($\Delta\theta$) and outer radii ($R_{\rm out}$) for different {H$_2$ 0-0} transitions.\label{Table:OpeningAngle}}
\tabletypesize{\small}
\tablehead{
\colhead{Source} &
\colhead{Lobe} &
\colhead{$\Delta\theta$ S(11)} &
\colhead{$\Delta\theta$ S(5)} &
\colhead{$\Delta\theta$ S(1)} &
\colhead{$R_{\rm out}$ S(11)} &
\colhead{$R_{\rm out}$ S(5)} &
\colhead{$R_{\rm out}$ S(1)} \\
\colhead{} &
\colhead{} &
\colhead{(deg)} &
\colhead{(deg)} &
\colhead{(deg)} &
\colhead{(au)} &
\colhead{(au)} &
\colhead{(au)}
}
\startdata
IRAS16253 & Blue & 28 $\pm$ 1 & 38 $\pm$ 1 & 41 $\pm$ 1 & 28 $\pm$ 8  & 18 $\pm$ 5  &  51 $\pm$ 8  \\ 
           & Red  & 25 $\pm$ 1 & 31 $\pm$ 4 & 38 $\pm$ 1 & 68 $\pm$ 11  & 66 $\pm$ 30  &  44 $\pm$ 7  \\ 
B335 & Blue & 38 $\pm$ 2 & 37 $\pm$ 1 & 44 $\pm$ 1 & 21 $\pm$ 12  & 50 $\pm$ 11  &  44 $\pm$ 15  \\ 
           & Red  & 18 $\pm$ 1 & 28 $\pm$ 1 & 39 $\pm$ 1 & 104 $\pm$ 17  & 106 $\pm$ 14  &  81 $\pm$ 13  \\ 
HOPS 153   & Blue & --         & --         & --         & --          & --          & --          \\
          & Red & 9 $\pm$ 1 & 10 $\pm$ 1 & 28 $\pm$ 2 & 201 $\pm$ 34  & 217 $\pm$ 42  &  142 $\pm$ 47  \\ 
HOPS370 & Blue & 32 $\pm$ 1 & 50 $\pm$ 1 & 46 $\pm$ 2 & 343 $\pm$ 31  & 336 $\pm$ 28  &  475 $\pm$ 53  \\ 
           & Red  & 33 $\pm$ 2 & 35 $\pm$ 2 & 41 $\pm$ 2 & 120 $\pm$ 36  & 181 $\pm$ 44  &  147 $\pm$ 46  \\ 
IRAS20126 & Blue & 25 $\pm$ 2 & 32 $\pm$ 3 & 41 $\pm$ 6 & 25 $\pm$ 77  & 17 $\pm$ 144  &  52 $\pm$ 389  \\ 
           & Red  & --         & --         & --         & --          & --          & --          \\
\enddata
\end{deluxetable*}

Across the morphologies discussed above, the \hydrogenmol\ line emission systematically becomes narrower with increasing \Eup\ (Figure \ref{fig:LineMaps}).
To quantify this nested morphology observed in the line maps, we measure the half-width and opening angle of the \hydrogenmol\ wind with respect to the axis perpendicular to the disk plane. We perform this analysis using the \hhnu(11), S(5), and S(1) line maps, which probe a broad range of \Eup\ 
{(from $\sim1\times10^3$ to $\sim1.3\times10^4$ K).}
First, the line maps are rotated such that the major axis of the ALMA disk aligns horizontally, placing the disk projected on the sky at a position angle of $90^{\circ}$ from the Y-axis in the rotated frame \citep[see e.g.,][]{Habel2021ApJ...911..153H}. Position angles of the disks are adopted from the literature (See Table \ref{Table:ObsLog}).
To determine the wind width, we use an edge-detection algorithm. The maps are initially smoothed with a median filter to suppress noise. A Sobel filter 
is then applied to identify intensity gradients and to extract the emission edges \citep{Sobelhttps://doi.org/10.13140/rg.2.1.1912.4965}. Finally, we use the Otsu thresholding method \citep{Otsu1979ITSMC...9...62O} to trace the outermost edges of the emission, from which the outflow width is measured. An example of the output of this edge detection procedure is shown in Figure \ref{fig:EdgeDetection} (see also Figure \ref{fig:All-edge-detection}).

\begin{figure}[htbp]
    \centering
    \includegraphics[width=\linewidth]{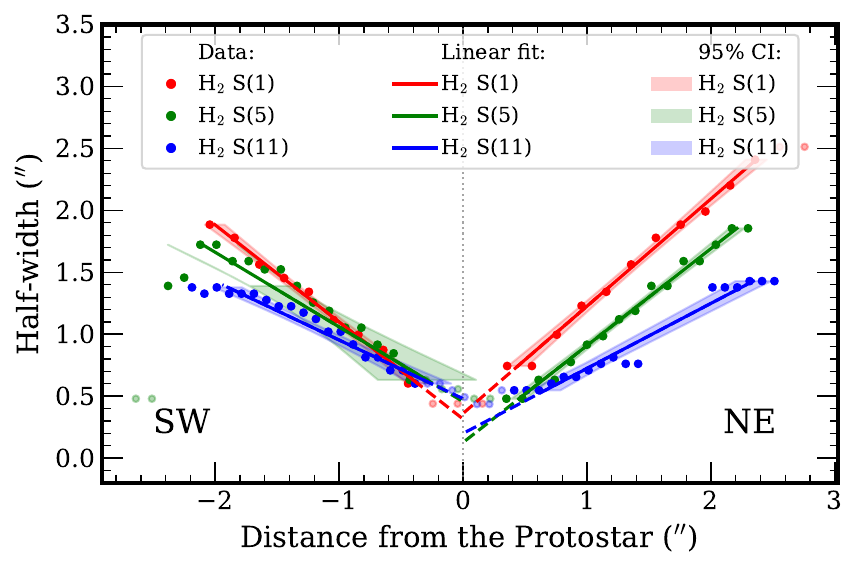}
    \caption{{Deconvolved half-width} of the outflow vs distance from the protostellar disk plane of IRAS 16253. The red, green, and blue dots show the measured half-widths as a function of distance for {\hhnu(1), S(5), and S(11)}, respectively. The gray dotted vertical line marks the disk plane. The best linear fits for the dotted data points of various \hydrogenmol\ lines are shown in solid lines, sharing the color of the data points. The colored shaded regions around the best-fit lines show the 95\% confidence interval. The best fits are extrapolated up to zero distance from the disk plane to determine the $R_{\rm out}$. Data marked in the fainter dots were not used in the linear fitting (discussed in Section \ref{Section:OA_section}).}
    \label{fig:OA_fitting_example_IRAS16253}
\end{figure}
We determine the half-width of the wind as a function of distance along the outflow axis from the ALMA disk plane using the previously identified \hydrogenmol\ emission edges. 
{The measured half-widths were subsequently corrected for varying angular resolution of the JWST. Using the empirical relation from \citet{Law2023AJ....166...45L}, the beam FWHM values at 4.18, 6.91, and 17~\micron\ are $\sim0.24''$, $0.33''$, and $0.67''$, respectively. To remove the effect of varying resolution from our measurements, we assume the JWST PSF as a 2D Gaussian and subtract the beam FWHM in quadrature from the observed widths as follows:
\begin{equation}
    \theta_{\rm Deconvolved} = \sqrt{\theta_{\rm Observed}^2-\theta_{\rm PSF}^2}
\end{equation} 
This deconvolution method successfully recovers the intrinsic wind widths to within 10\% of their true values. Full details demonstrating the robustness of this approach are provided in Appendix \ref{Appendix:Deconvolution}.}

To estimate the wind’s semi-opening angle, we fit a linear function to the deconvolved half-widths. For this fit, we exclude points within one PSF radius of the central protostar ({also see Appendix \ref{Appendix:Deconvolution}}), as well as regions where the edge detection does not work properly due to lack of signal, FOV coverage, or complicated outflow structure (see e.g., the bottom pixels in the middle panel of Figure \ref{fig:EdgeDetection}).
The semi-opening angle ($\theta_{open}$) is then derived from the slope ($m$) of the best-fit line, such that $\theta_{open}=\arctan(m)$. 
We note that the opening angle measurement adopted here differs from other commonly used definitions, which are defined by an angle from the central protostar to the cavity wall at a given height above the plane \citep[e.g.,][]{Habel2021ApJ...911..153H}. 
The best-fit linear function, on extrapolation, intersects the disk plane, providing an upper limit on the outermost launch footprint ($R_{\rm out}$) of the \hydrogenmol\ winds on the disk. To demonstrate an example of such fitting, we show the plot of the measured half-width as a function of distance from the protostar in Figure \ref{fig:OA_fitting_example_IRAS16253}. Measured points using the edge detection shown in Figure \ref{fig:EdgeDetection} are shown in red, blue, and green dots for the S(1), S(5), and S(11) lines, respectively. The solid lines in the figure represent the best linear fit to the half-width as a function of the distance from the protostellar disk plane. These best-fit lines are extrapolated up to the distance of zero to determine $R_{\rm out}$.

The results of this exercise for all the sources are presented in Table \ref{Table:OpeningAngle}. 
{The semi-opening angle increases as a function of decreasing \Eup, demonstrating the nested morphology of the \hydrogenmol\ wind. 
However, a robust trend of the nested morphology with $R_{\rm out}$ is not observed. This may be because the winds are not strictly conical and with uncertainties in the opening angle measurements, simple radial extrapolation might not accurately constrain $R_{\rm out}$ (see, e.g., Figures \ref{fig:LineMaps} and \ref{fig:B335_NIRCAM}).}
Because of faint \hydrogenmol\ emission in HOPS 153’s blue-shifted flow and the complex, faint red-shifted flow in IRAS 20126, the edges could not be determined robustly. 
Due to the inclination of the sources, the true opening angles should be smaller than the observed opening angles. Following \citet{Habel2021ApJ...911..153H}, the distance along the y-axis is reduced to $y_{\rm obs} = y \sin(i)$.  In other words, the true angle is given by $\theta = \arctan(\sin(i) dx/dy)$, where $dx/dy =m$ is the observed slope.

\begin{figure*}
    \centering
    \includegraphics[width=\textwidth]{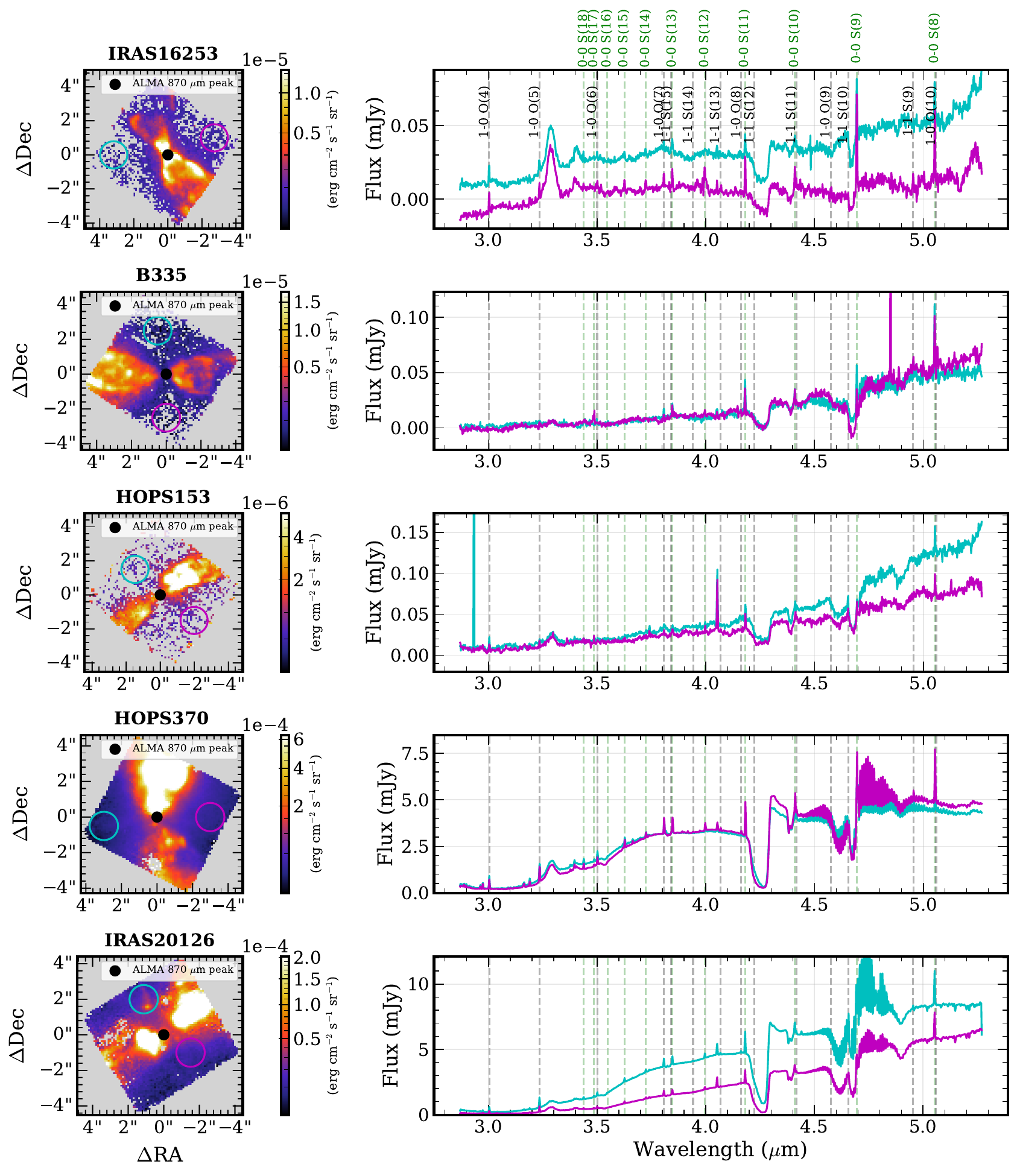}
    
    \caption{Spectra from the apertures outside the outflow cavities. \textit{left}: \hhnu(11) line map depicting the \hydrogenmol\ emission. All spaxels with S/N $ \leq$ 3 are masked. The cyan and magenta circles mark the apertures used to extract the spectrum. The black dot marks the ALMA 870 \micron\ continuum peak. \textit{right}: Spectra extracted from the magenta and cyan apertures in the same colors as their respective apertures. The vertical green and black lines mark the lab wavelengths of various \hydrogenmol\ transitions, labeled in the top panel.}
    \label{fig:H2_outside}
\end{figure*}
\section{\texorpdfstring{\hydrogenmol}\~ Emission Outside the Outflow Cavity}
\label{Section:AmbientH2}
As discussed in Section \ref{Section:Linemaps}, we detect \hydrogenmol\ emission outside the outflow cavity in all five protostars. This ambient emission is particularly detected in the NIRSpec observations, due to NIRSpec's higher sensitivity.
To demonstrate the detection of these \hydrogenmol\ lines and other spectral features, we present the spectrum extracted from apertures outside the outflow cavity, showing the presence of many pure rotational and ro-vibrational \hydrogenmol\ emission lines in all five protostars in Figure \ref{fig:H2_outside}. {We have detected these ambient \hydrogenmol\ lines with MIRI observations also, particularly the \hhnu(1) and S(2) transitions are detected in all five protostars.}
Further, to show the distribution of the ambient \hydrogenmol, we present \hhnu(11) line maps in Figure~\ref{fig:H2_outside}, where all spaxels with ${\rm S/N}\leq3$ are masked. In the resulting maps, the ambient \hydrogenmol\ emission appears roughly uniform in intensity across the field of view. This ambient emission is also detected at the disk plane of B335, which is heavily extincted, suggesting that the ambient \hydrogenmol\ emission surrounds the outflow cavity and forms a thin foreground sheet in front of the outflowing gas from the protostars. As a result, even limb-brightened cavities might appear to be filled with faint \hydrogenmol\ emission (see Section \ref{sec:H2_limb_bright} for details). 
The origin of this ambient emission will be further discussed in a future paper.

These off-cavity apertures also show faint continuum emission with the prominent ice absorption features of  \coo, CO, \ocn, and OCS imprinted onto it. 
Additionally, the off-cavity apertures in all five sources show emission from Polycyclic Aromatic Hydrocarbons (PAH) features at 3.3 \micron\ and the Br $\alpha$ line at $4.05$ \micron\ (Figure \ref{fig:H2_outside}). The absorption feature of \water\ ice is also visible in HOPS 370 and IRAS 20126; however, it does not appear in other sources due to their fainter and relatively low S/N continuum.

\section{Kinematics of \texorpdfstring{\hydrogenmol~}\ Winds}
\label{Section:Kinematics}
To investigate the kinematics of \hydrogenmol\ winds, we generated velocity maps and position-velocity (PV) diagrams. 
We primarily used the MIRI/MRS observations of the brightest \hydrogenmol\ lines for the kinematic study, as MRS has a higher spectral resolution relative to the NIRSpec/IFU G395M grating used in our observations (see Section \ref{ObservationSection}).

\subsection{Velocity Maps}

\begin{figure*}[htbp]
    \centering
    
    
    
    
    \includegraphics[width=\linewidth]{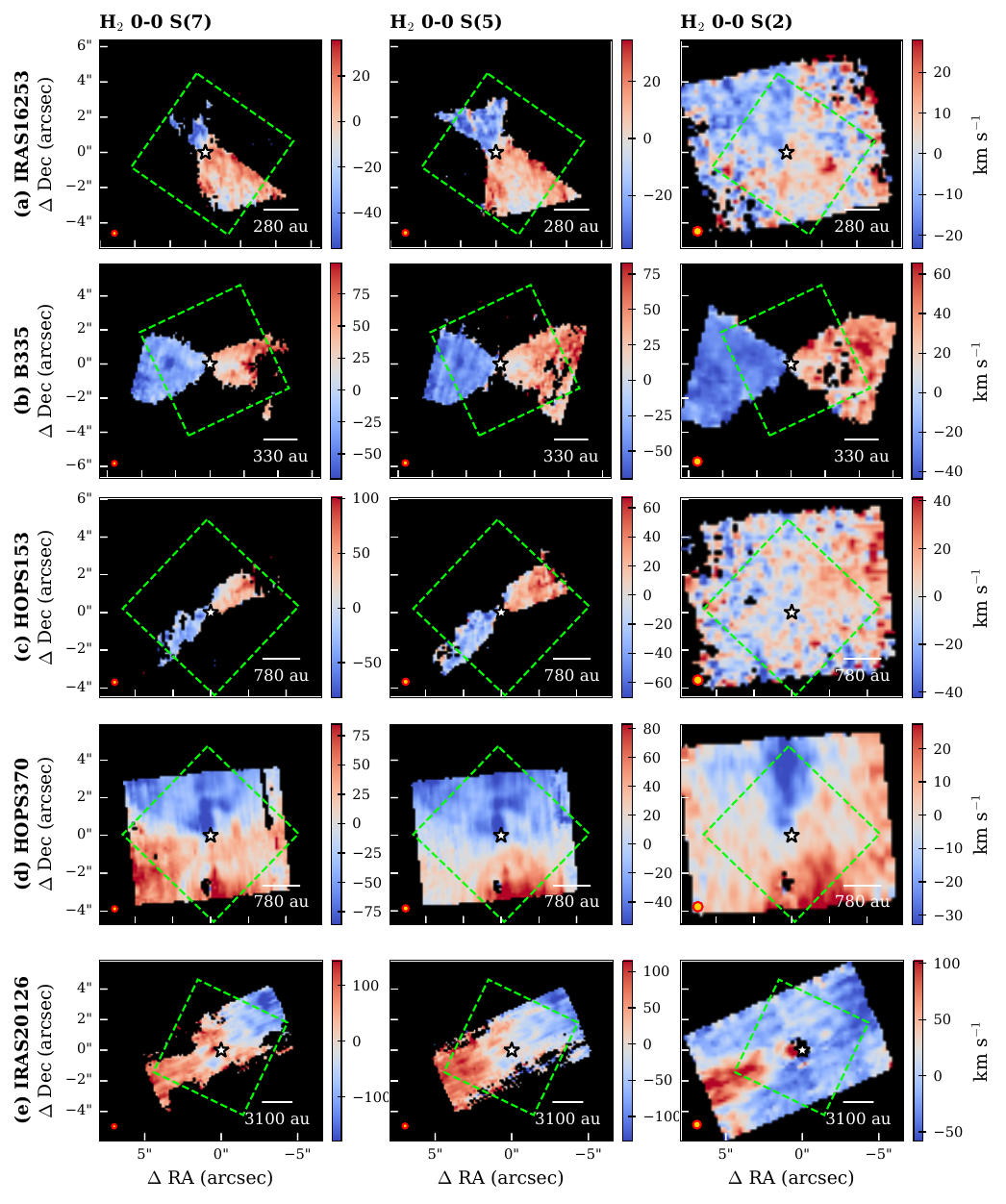}

    \caption{Inclination-corrected velocity maps of {\hhnu(7), S(5), and S(2)} lines for IRAS 16253, B335, HOPS 153, HOPS 370, and IRAS 20126. 
    The white star marks the ALMA continuum peak. Golden circles, outlined in red, at the bottom left edge mark the beam size of JWST. The dashed rectangle in lime shows the NIRSpec FOV.}
    \label{fig:mom1}
\end{figure*}

We explore the spatial distribution of wind velocities with velocity maps of \hhnu(7), S(5), and S(2) lines at 5.51, 6.909, and 12.28 \micron, respectively. These lines were selected for their high S/N relative to other \hydrogenmol\ lines. The velocity maps are generated as part of the line-mapping procedure described in Section \ref{Section:Linemaps}. For each spaxel, the centroid of a Gaussian fit to the line profile is determined and converted to a radial velocity, thereby constructing a spatially resolved radial velocity map. 
{This centroid-based velocity measurement is highly robust. It is well established that for well-sampled profiles with high S/N, the line center can be determined with a precision significantly better than the instrumental spectral resolution \citep[e.g.,][]{Landman1982ApJ...261..732L, Bouchy2001A&A...374..733B}. 
We used the in-flight wavelength calibration of the JWST MIRI/MRS (implemented in CRDS version 1179 and later), which is refined to a level of a few \kms\ across Channels 1-3 \citep{Argyriou2023A&A...675A.111A, Harkett2024JGRE..12908415H, Pontoppidan2024ApJ...963..158P}.  
Restricting our kinematic analysis to these strong, pure rotational transitions ensures that our sub-resolution velocity measurements remain robust.
Furthermore, several published studies have used the same approach to measure velocities in the JWST observations \citep[e.g.,][]{Garatti2024A&A...691A.134C, Delabrosse2024A&A...688A.173D, Navarro2025ApJ...995..199N, vanDishoeck2025A&A...699A.361V, Vleugels2025A&A...695A.145V, Narang2026arXiv260209837N, Federman2026arXiv260109587F, Francis2026arXiv260413773F}.}

We note that the \hydrogenmol\ line velocities are offset by $\sim10-20$ \kms\ between MIRI sub-bands due to systematic uncertainties in the absolute wavelength calibration \citep[also reported by][]{Narang2024ApJ...962L..16N}, making it difficult to subtract the protostellar systemic velocity from the velocity maps.
To remove the protostellar systemic velocity, we assume that the blue- and red-shifted winds have similar kinematics and subtract the mean velocity of all spaxels from each spaxel of the velocity maps \citep[similar to][]{Narang2024ApJ...962L..16N}. This procedure to remove the systematic velocity was adopted due to the issue of absolute wavelength calibration. Here, we assume that the relative wavelength calibration is accurate enough to measure the velocities.
Subsequently, the velocities are corrected for inclination by dividing by the cosine of the inclination angle ($i$). The analysis is restricted to spaxels with $\rm S/N>7$. 
The resulting velocity maps are presented in Figure \ref{fig:mom1}, and a detailed discussion of each source's velocity structure follows. 

Figure \ref{fig:mom1}(a) shows the velocity maps of IRAS 16253, revealing a clear bipolar velocity structure. The northern emission is predominantly blue-shifted, while the southern emission is distinctly red-shifted. This is similar to the collimated jet traced with ionic lines \citep[][]{Narang2024ApJ...962L..16N}. Among the \hydrogenmol\ lines presented in velocity maps, the S(7) line shows the highest velocities, followed by the S(5) and S(2) lines. The red-shifted outflow edges display higher velocities in the S(7) and S(5) velocity maps, as also discussed by \citet{Narang2026arXiv260209837N}.  Additionally, extended emission detected outside the outflow cavity, as seen in the S(2) velocity map, exhibits a range of velocities, likely originating from ambient gas surrounding the protostar.

Measurements of the flow velocities are sensitive to the assumed inclination angle. Published estimates of the inclination of B335 span a wide range. Based on radiative transfer modeling, \citet{Stutz2008} derived a cavity inclination of 87\arcdeg. Using proper motions and radial velocities of [Fe II] shock knots, \citet{Federman2026arXiv260109587F} estimated an inclination of 57\arcdeg, suggesting jet wiggling and precession. \citet{Federman2026arXiv260109587F} further noted that using an inclination of 87\arcdeg\ results in unrealistically high velocities than those expected from their 3D velocity measurements. Kim et al. (in prep.), applying a similar method to ALMA CO observations, report an inclination angle of $\sim$73\arcdeg. Most recently, \citet{Hodapp2026arXiv260212060H}, using JWST observations, determined a lower limit of 68\arcdeg\ for the inclination of the molecular shock from 3E (see Figure~\ref{fig:B335_NIRCAM}). 
Given that the value reported by \citet{Hodapp2026arXiv260212060H} is based on the JWST observations of the molecular shock knot, relevant to our analysis, we adopt an inclination angle of 68\arcdeg\ throughout this paper.

A clear bipolar velocity structure is observed in B335, also with the eastern emission being blue-shifted and the western being red-shifted (Figure \ref{fig:mom1}(b)). As expected, this matches the collimated jet traced with ionic lines \citep[][]{Federman2026arXiv260109587F}, and molecular outflows observed with \submm\ telescopes (\citealt{Yen2010ApJ...710.1786Y, Bjerkeli2019}; see also \citealt{Galfalk&Olofsson2007}).  The bright knot (identified as 3E by \citealt{Hodapp2024AJ....167..102H}; also see Figure \ref{fig:LineMaps} and \ref{fig:B335_NIRCAM}) in the blue-shifted cavity shows a higher velocity than the rest of the blue-shifted emission.
This molecular knot was also identified by \citet{Federman2023arXiv} and appears to be part of the jet.
The \hydrogenmol\ shell connected to a jet knot identified in \citet{Federman2026arXiv260109587F}, and labeled 4E by \citet{Hodapp2024AJ....167..102H} (Figure \ref{fig:B335_NIRCAM}), also show higher velocity.
The outer edges along the cavity walls of the red-shifted cavity also show higher velocity than the central part.
Interestingly, unlike IRAS 16253, the velocity maps do not show a significant difference in velocity between S(7), S(5), and S(2) lines.

Figure \ref{fig:mom1}(c) depicts the HOPS 153 velocity maps, revealing a bipolar velocity structure with blue- and red-shifted components clearly traced in the S(7) and S(5) lines. In contrast, the S(2) line does not distinctly separate the blue- and red-shifted flows, likely due to the lower velocities. The wind directions align with those observed in the collimated jets (see \citealt{Federman2026arXiv260109587F}). Notably, the red-shifted emission is brighter in the line maps, a feature that distinguishes HOPS 153 from other protostars in the sample. The S(7) velocity map shows higher velocities than the S(5) velocity map, similar to those observed in IRAS 16253. 
The red-shifted flow within the velocity maps of S(5) and S(7) lines reveal a higher velocity component at the farthest region away from the source than in the immediate vicinity.   

 The velocity maps of HOPS 370 (Figure \ref{fig:mom1}(d)) reveal a fast-moving bipolar \hydrogenmol\ component ($\sim50$ \kms) that is enveloped by an extended, slower ($\sim10-20$ \kms) bipolar \hydrogenmol\ gas. However, the low-$J$ lines show less pronounced red/blue shifts due to extended emission.
 The collimated emission structure observed in the line map (Figure \ref{fig:LineMaps}(d)) shows the higher velocities in both S(7) and S(5) velocity maps, suggesting the presence of a collimated fast molecular \hydrogenmol\ jet.
 The shocked knot and the fish-hook structure in HOPS 370 also have relatively higher velocities \citep[see for more details][]{Neufeld2024arXiv240407299N, Tyagi2025ApJ...983..110T}. Furthermore, the S(7) and S(5) velocity maps reveal elevated blue-shifted velocities within the mushroom-shaped structure (see Section \ref{section: HOPS 370 Linemaps}), suggesting gas expansion due to molecular jet envelope/wind interaction.

The velocity map of IRAS 20126, shown in Figure \ref{fig:mom1}(e), highlights a complex structure of the outflow, with distinct blue- and red-shifted components traced in the S(7) and S(5) lines. Blue-shifted emission is primarily in the northwestern part, while the red-shifted emission extends in the southeast. The bubble-like structure directly southeast of the protostar (Figure \ref{fig:LineMaps}; Section \ref{section: IRAS20126 Linemaps}) shows blue-shifted edges, perhaps due to gas expansion driven by shocks or outflow driven by a companion (see Higgins, M. et al. in prep.). The S(2) map reveals a distinct high-velocity red-shifted flow in the southeastern direction. Additionally, the bent segment of the \hydrogenmol\ winds along the northwestern edge of the field of view displays higher velocities relative to other regions within the blue-shifted outflow. All the discussed velocity structures together underscore the kinetic complexity of the highest mass source in our sample. Furthermore, IRAS 20126 is located at $\sim1.5$ kpc, 4-5 times farther away than the other sources, resulting in significantly poorer angular resolution and making interpretation more difficult. Furthermore, source's multiplicity and the presence of multiple outflows introduce additional complexity (\citealt{Federman2023arXiv}; Higgins, M. et al. in prep.).

\subsection{PV Diagrams Along the Outflow Axis \label{section:PV-along}}

\begin{figure*}[htbp]
    \centering

    \begin{subfigure}{0.19\textwidth}
        \centering
        \includegraphics[width=\linewidth]{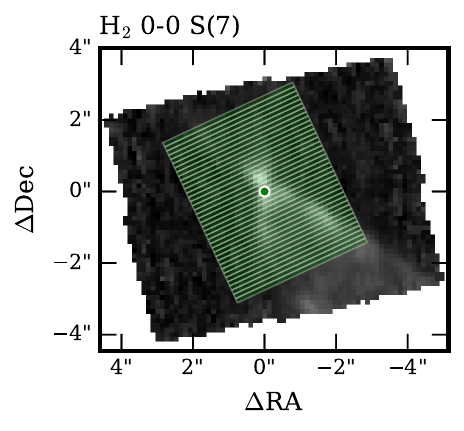}
    \end{subfigure}
    \begin{subfigure}{0.19\textwidth}
        \centering
        \includegraphics[width=\linewidth]{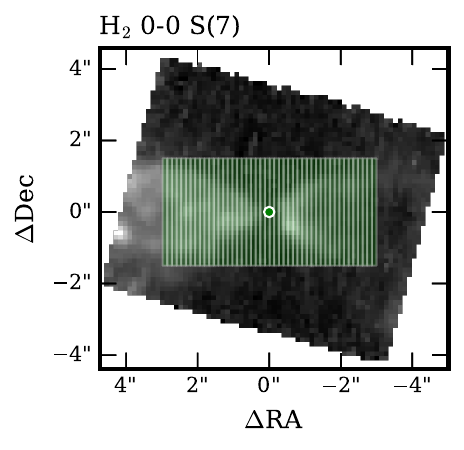}
    \end{subfigure}
    \begin{subfigure}{0.19\textwidth}
        \centering
        \includegraphics[width=\linewidth]{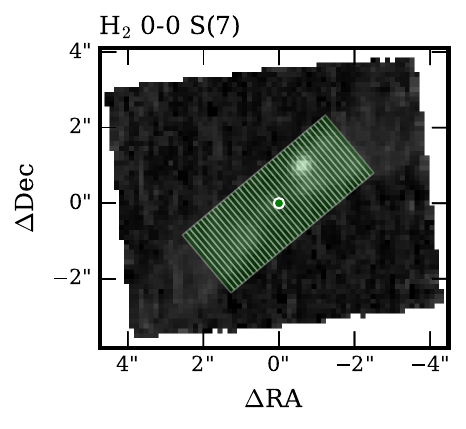}
    \end{subfigure}
    \begin{subfigure}{0.19\textwidth}
        \centering
        \includegraphics[width=\linewidth]{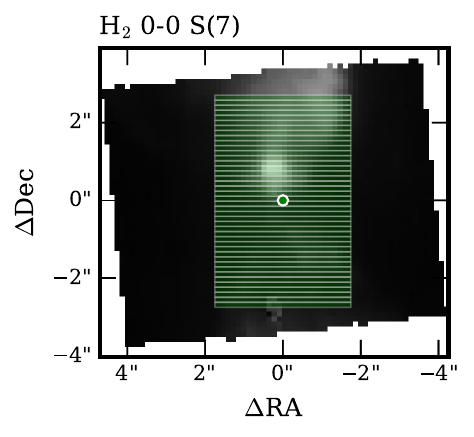}
    \end{subfigure}
    \begin{subfigure}{0.19\textwidth}
        \centering
        \includegraphics[width=\linewidth]{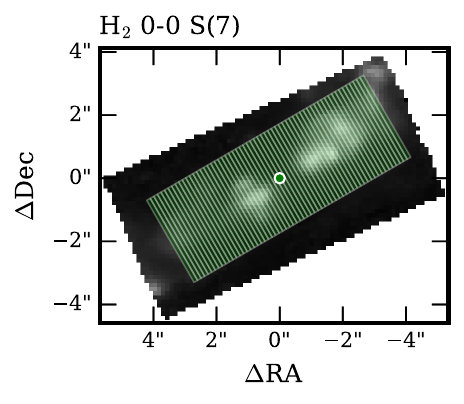}
    \end{subfigure}


    \begin{subfigure}{0.19\textwidth}
        \centering
        \includegraphics[width=\linewidth]{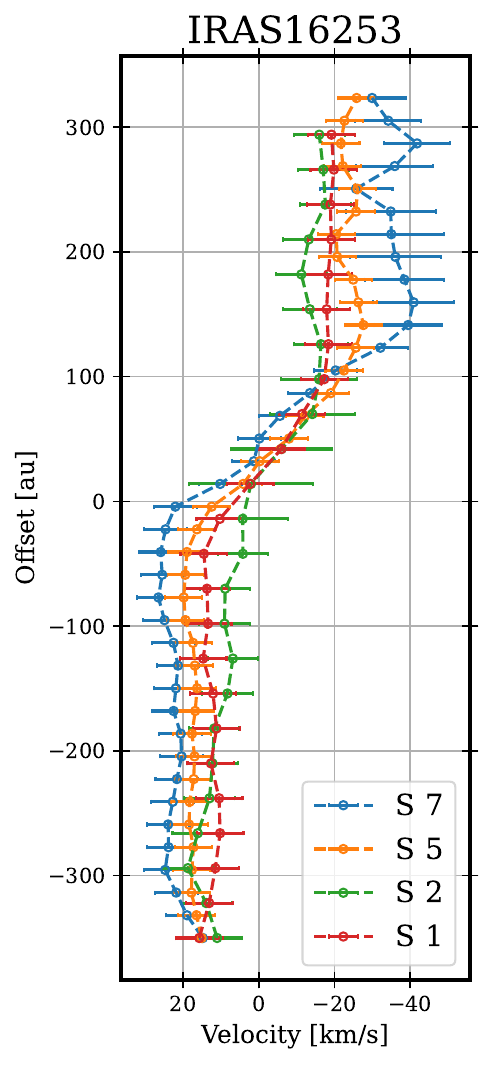}
        \caption{IRAS 16253}
        \label{fig:pv_iras16253}
    \end{subfigure}
    \begin{subfigure}{0.19\textwidth}
        \centering
        \includegraphics[width=\linewidth]{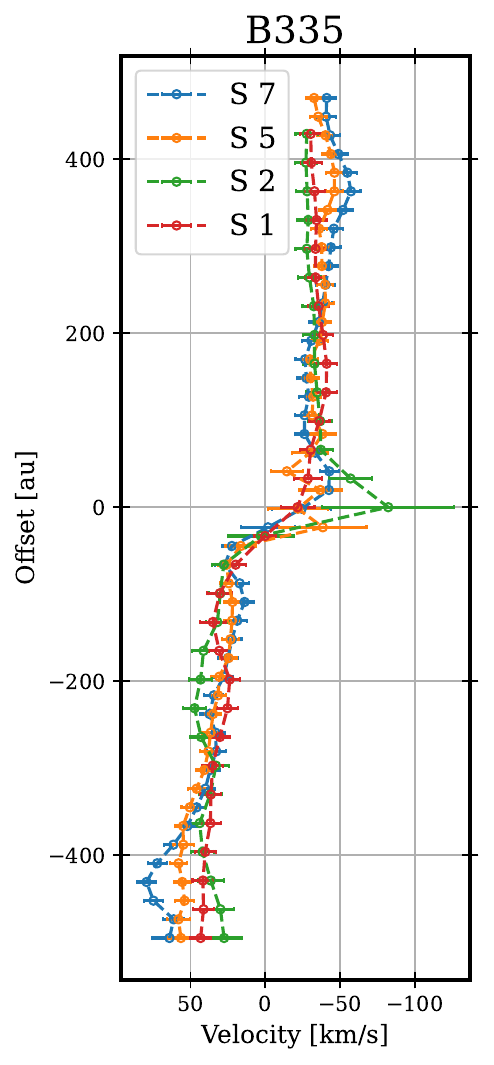}
        \caption{B 335}
        \label{fig:pv_b335}
    \end{subfigure}
    \begin{subfigure}{0.19\textwidth}
        \centering
        \includegraphics[width=\linewidth]{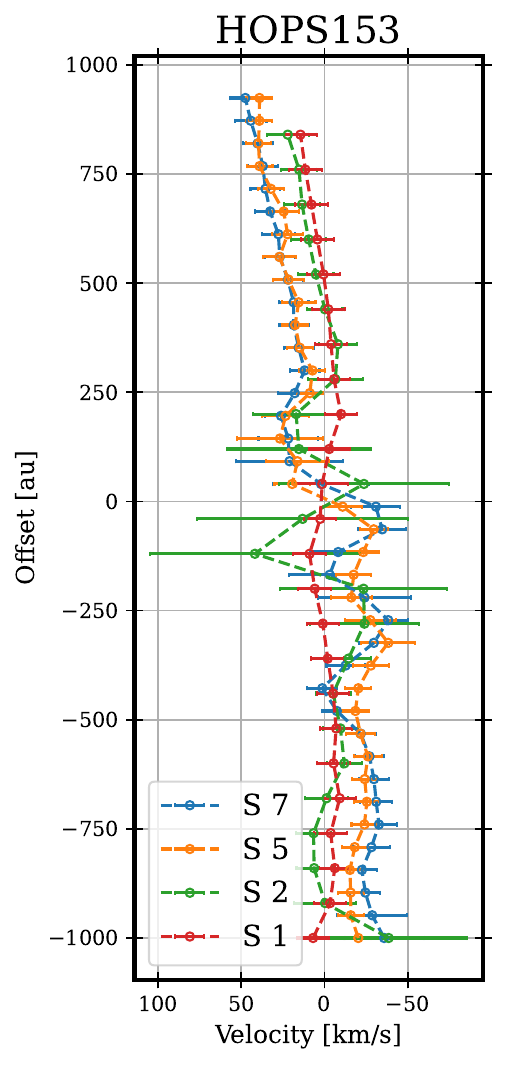}
        \caption{HOPS 153}
        \label{fig:pv_hops153}
    \end{subfigure}
    \begin{subfigure}{0.19\textwidth}
        \centering
        \includegraphics[width=\linewidth]{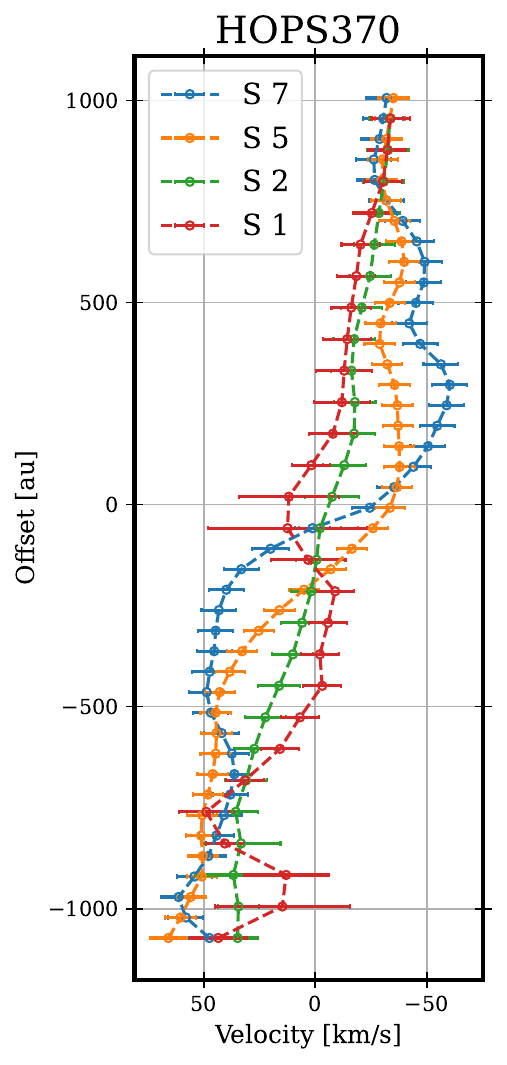}
        \caption{HOPS 370}
        \label{fig:pv_hops370}
    \end{subfigure}
    \begin{subfigure}{0.19\textwidth}
        \centering
        \includegraphics[width=\linewidth]{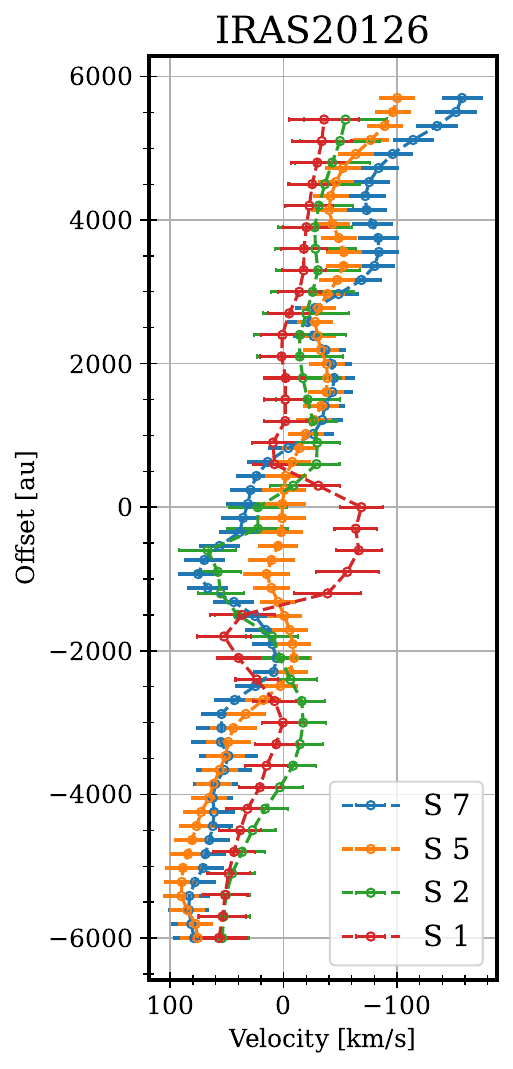}
        \caption{IRAS 20126}
        \label{fig:pv_iras20126}
    \end{subfigure}

    \caption{
    \textit{Top row:} Sky paths of the PV diagrams overlaid in green on {\hhnu(7)} line maps. White lines show 1-pixel-wide slices used to find the velocity centroids.
    \textit{Bottom row:} Inclination-corrected centroids of PV diagrams of {\hhnu(7), S(5), S(2), and S(1)} lines along the outflow axis. 
    The S(1) line in IRAS 20126 shows a sudden offset near zero due to bad pixels. 
    }
    \label{fig:PV_figure}
\end{figure*}

To understand the velocity structure more quantitatively, we construct PV diagrams of \hhnu(1), S(2), S(5), and S(7) lines along and perpendicular to the outflow axis. 
The faint emission outside the outflow cavity walls, discussed in Section \ref{Section:Linemaps}, forms a layer of foreground emission to the outflow cavity. Before generating the PV diagrams, we subtract out the foreground \hydrogenmol\ gas emission from the \hydrogenmol\ wind flow, except for IRAS 20126 (See Section \ref{Section:AmbientH2}). 
To subtract the ambient gas contribution, we take two apertures outside of the protostellar \hydrogenmol\ outflow. These apertures are shown in Figure \ref{fig:H2_outside}. The median of the extracted intensities is subsequently subtracted from each pixel of the science data cubes.
This step removes the contribution of the slow-moving ambient gas's velocity to the flow emission.

For the PV diagrams along the outflow direction, we averaged the velocities over the perpendicular extent of the molecular hydrogen wind, as indicated by the white slices in the green shaded regions (sky path) in the first row of  Figure \ref{fig:PV_figure}. 
{In addition to the PV diagram, we also create a PV diagram error map by summing up the errors of the pipeline cube in quadrature.}
We then fit a Gaussian profile, {weighted by the intensity uncertainty of the PV diagram error map}, to determine the velocity centroid of the PV diagram in all the spatial slices, marked by the white lines overplotted on the sky path (see Figure \ref{fig:PV_figure} top row). 
Similar to the previous section, we remove the systematic offset between MIRI sub-bands by subtracting the mean velocity of each line from the individual velocity measurements within the PV diagrams. This step assumes that the mean of the red- and blue-shifted velocities should be zero \citep[see also][]{Narang2024ApJ...962L..16N}. Finally, to facilitate the comparison of velocities across different sources, we further account for the inclination of each source ($i$) by dividing the derived velocities by $\cos{i}$.


\begin{deluxetable*}{llccccc}
\tablecaption{Velocity from PV diagrams along the outflow axis.\label{Table:PV}}
\tabletypesize{\small}
\tablehead{
\colhead{H$_2$ Line} & \colhead{} &
\colhead{IRAS 16253} &
\colhead{B 335} &
\colhead{HOPS 153} &
\colhead{HOPS 370} &
\colhead{IRAS 20126} \\
\colhead{$v = 0-0$} & \colhead{} &
\colhead{(km s$^{-1}$)} &
\colhead{(km s$^{-1}$)} &
\colhead{(km s$^{-1}$)} &
\colhead{(km s$^{-1}$)} &
\colhead{(km s$^{-1}$)}
}
\startdata
$v_{S(7)}$ & Mean of Peaks & 34 $\pm$ 5 & 68 $\pm$ 5 & 43 $\pm$ 7 & 61 $\pm$ 6 & 121 $\pm$ 12 \\
 & Mean & 26 $\pm$ 9 & 41 $\pm$ 16 & 25 $\pm$ 11 & 44 $\pm$ 9 & 59 $\pm$ 33 \\
[2pt]
$v_{S(5)}$ & Mean of Peaks & 24 $\pm$ 3 & 52 $\pm$ 5 & 39 $\pm$ 9 & 53 $\pm$ 5 & 95 $\pm$ 11 \\
 & Mean & 20 $\pm$ 4 & 39 $\pm$ 10 & 23 $\pm$ 8 & 38 $\pm$ 12 & 45 $\pm$ 29 \\
[2pt]
$v_{S(2)}$ & Mean of Peaks & 18 $\pm$ 5 & 65 $\pm$ 22 & 40 $\pm$ 39 & 35 $\pm$ 10 & 61 $\pm$ 22 \\
 & Mean & 13 $\pm$ 3 & 34 $\pm$ 6 & 11 $\pm$ 9 & 24 $\pm$ 10 & 29 $\pm$ 16 \\
[2pt]
$v_{S(1)}$ & Mean of Peaks & 18 $\pm$ 4 & 42 $\pm$ 5 & 12 $\pm$ 7 & 41 $\pm$ 7 & 63 $\pm$ 15 \\
 & Mean & 15 $\pm$ 3 & 35 $\pm$ 5 & 6 $\pm$ 3 & 21 $\pm$ 13 & 25 $\pm$ 18 \\
\enddata
\tablecomments{Error reported in the mean velocity is the standard deviation of the absolute velocities.}
\end{deluxetable*}

\begin{figure}[htbp]
    \centering
    \includegraphics[width=\linewidth]{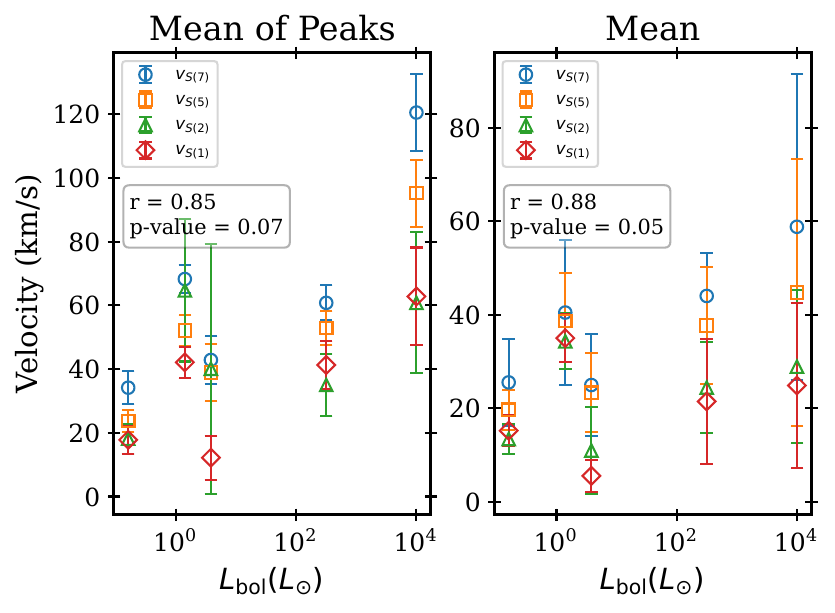}
    \caption{Velocity measured from PV diagrams along the outflow axis as a function of \lbol. \textit{Left:} Peak-to-peak velocities of the {\hhnu(7), S(5), S(2), and S(1)} lines are indicated by blue circles, orange squares, green triangles, and red diamonds, respectively. \textit{Right:} Same as the left panel, but showing the mean velocities. In both panels, the Pearson correlation coefficient and corresponding p-value are reported for the \hhnu(7) transition. }
    \label{fig:velocity_vs_lbol}
\end{figure}

The PV centroid diagrams are shown in Figure \ref{fig:PV_figure}, where the centroid velocities of the spatial slices along the outflow axis are shown as a function of the distance offset of those slices from the central source (ALMA position).
{The precision of these velocity measurements is limited by several instrumental systematics inherent to the MIRI/MRS. Specifically, spaxel-to-spaxel wavelength calibration, residual fringing in the data, and slit illumination effects across slicing mirrors might introduce artificial kinematic shifts. The wide-angled wind morphology is extended and relatively smooth and relatively less susceptible to the slit illumination effects than the compact structures or the point sources. We, however, note that the kinematics measurements might be affected by these instrumental systematics. To account for these effects, we adopt a minimum uncertainty floor of 1/20th of a spectral pixel for all the velocity measurements if their statistical uncertainty is smaller than 1/20th of a spectral pixel. We plot these uncertainties in Figure \ref{fig:PV_figure}.}

The blue- and red-shifted velocities are distinctly visible in all the sources.
All sources except B335 show progressively higher velocities for higher excitation energy lines (S(7) and S(5)) compared to lower excitation energy lines (S(1) and S(2)). In contrast, B335 displays similar velocities for all lines ($\sim20-40$ \kms)
(Figure \ref{fig:PV_figure}(b)). The red-shifted wind in HOPS 153 shows hints of velocity increment with distance from the central protostar. For all the other sources, velocities remain almost constant along the outflow axis. {High-frequency fluctuations in velocity, observed on smaller spatial scales, might be either due to the turbulent nature of the winds or due to instrumental systematics.}

To quantify the outflow velocities, we measure the peak-to-peak mean (mean of the peak velocities) of the velocity centroids, which is just the mean of the max blue-shifted and the max red-shifted velocities. We also measure the mean of the absolute values and the standard deviation of the velocity centroids, which are spatially located $>2\times{\rm FWHM_{PSF}}$ for a given line, away from the central protostar. These measurements for all the protostars are presented in Table \ref{Table:PV}.
To visually present the trend observed between the \lbol\ and \hydrogenmol\ flow velocity, we plot both the measured velocities (peak-to-peak and mean) as a function of \lbol\ in Figure \ref{fig:velocity_vs_lbol}.
Both the peak-to-peak mean velocity and the mean velocity along the outflow axis of the \hydrogenmol\ lines seem to increase with $E_{up}$ within a source.
We further find that the flow velocities seem to increase with the \lbol\ of the source.

\begin{figure*}[htbp]
    \centering

    \begin{subfigure}{\textwidth}
        \centering
        \includegraphics[width=\linewidth]{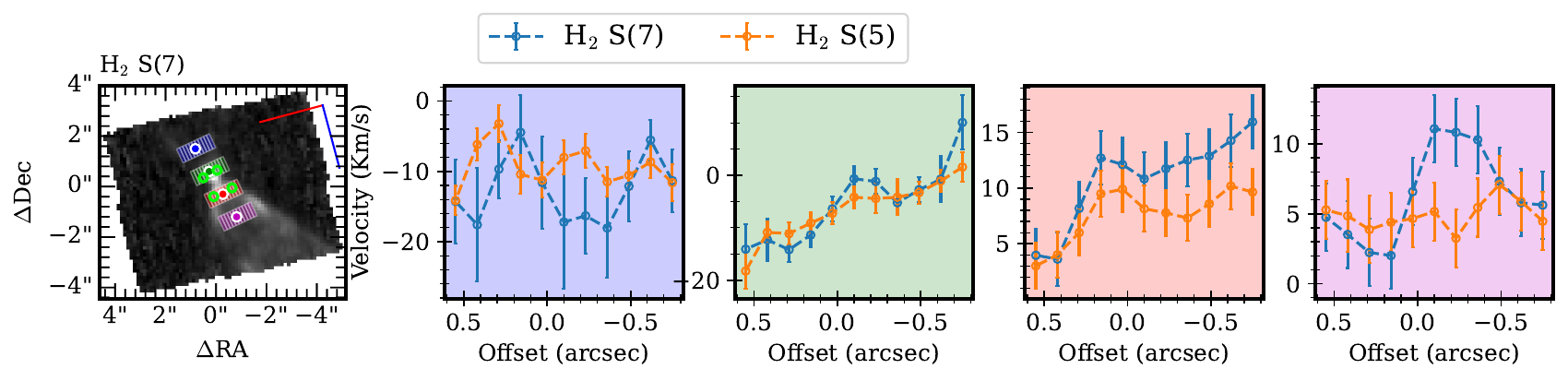}
        \caption{IRAS 16253}
        \label{fig:ortho_pv_iras16253}
    \end{subfigure}

    \vspace{0.5cm}

    \begin{subfigure}{\textwidth}
        \centering
        \includegraphics[width=\linewidth]{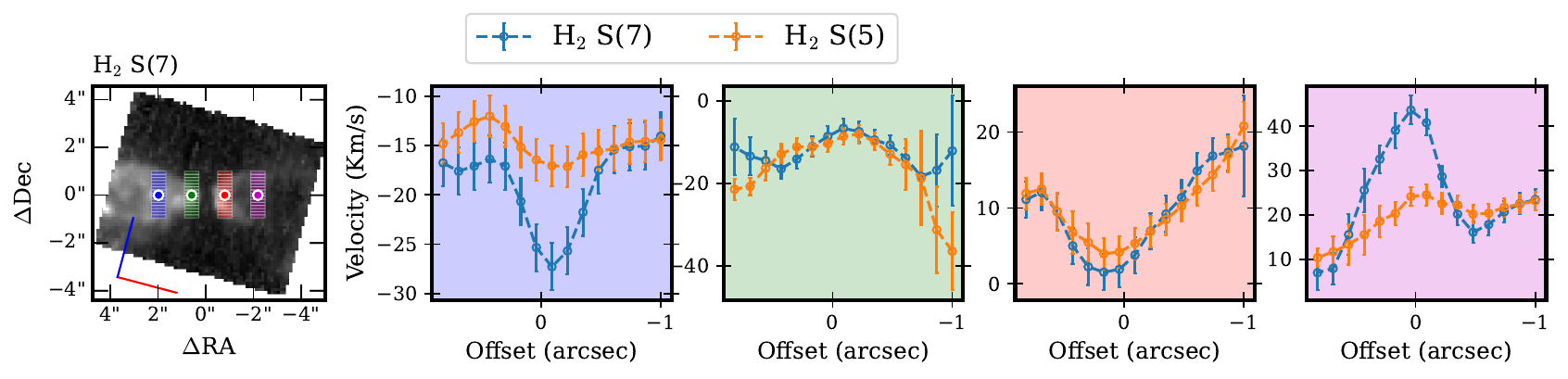}
        \caption{B 335}
        \label{fig:ortho_pv_b335}
    \end{subfigure}



    \caption{Velocity centroids of PV diagrams of the {\hhnu(7) and S(5)} lines (in blue and orange, respectively) for the sky paths orthogonal to the outflow axes. Each row corresponds to a specific protostar, as labeled in the sub-caption below each row. The selected sky paths used to generate the PV diagrams are overlaid as shaded regions on the {\hhnu(7)} line maps displayed in the first panel of each row. {The red and blue lines marked on the corners of the line maps depict the directions across and along the image slicer of the MIRI/MRS during the observations, respectively.} 
    The corresponding centroids of PV diagrams for these sky paths are shown in the subsequent panels, with the background color of each panel matching the shaded region of its corresponding sky path. 
    Lime circles overplotted on the IRAS 16253 mark the apertures used for specific momentum calculation (Section \ref{sec:wind-rotation}). (Continued)}
    \label{fig:Ortho-PV}
\end{figure*}

\begin{figure*}[htbp]\ContinuedFloat
    \centering

    \begin{subfigure}{\textwidth}
        \centering
        \includegraphics[width=\linewidth]{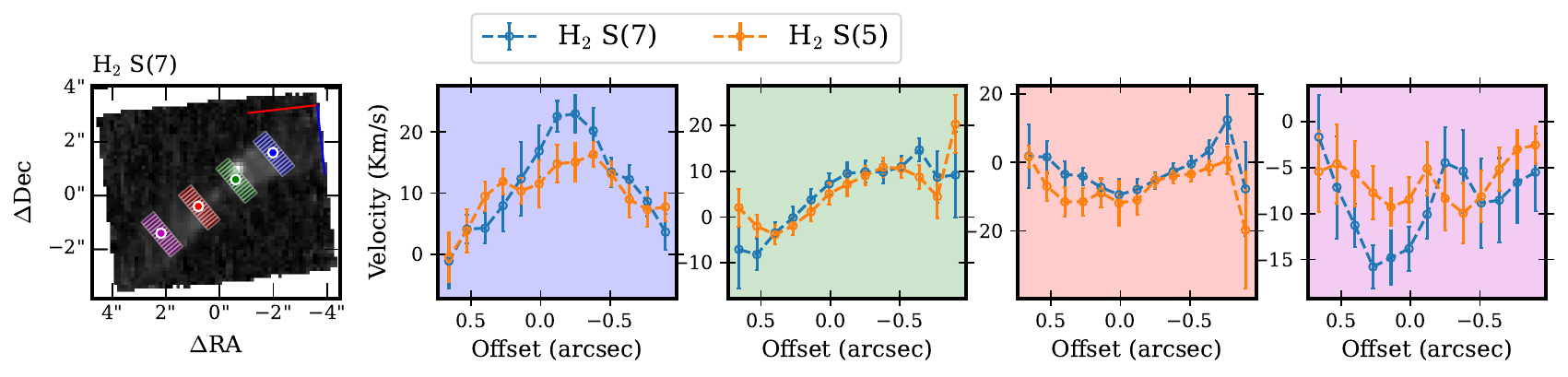}
        \caption{HOPS 153}
        \label{fig:ortho_pv_hops153}
    \end{subfigure}

    \vspace{0.5cm}

    \begin{subfigure}{\textwidth}
        \centering
        \includegraphics[width=\linewidth]{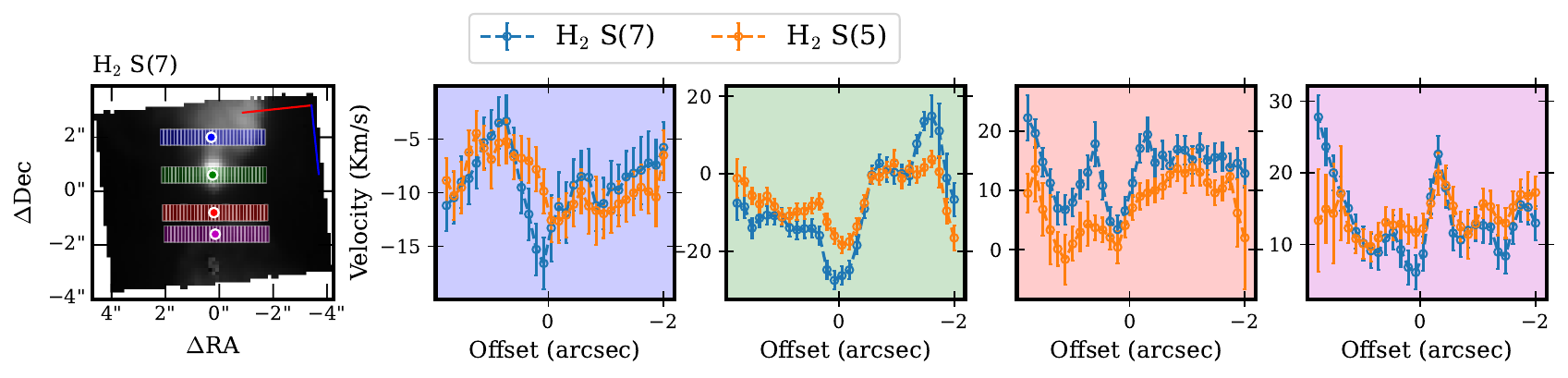}
        \caption{HOPS 370}
        \label{fig:ortho_pv_hops370}
    \end{subfigure}

    \vspace{0.5cm}

    \begin{subfigure}{\textwidth}
        \centering
        \includegraphics[width=\linewidth]{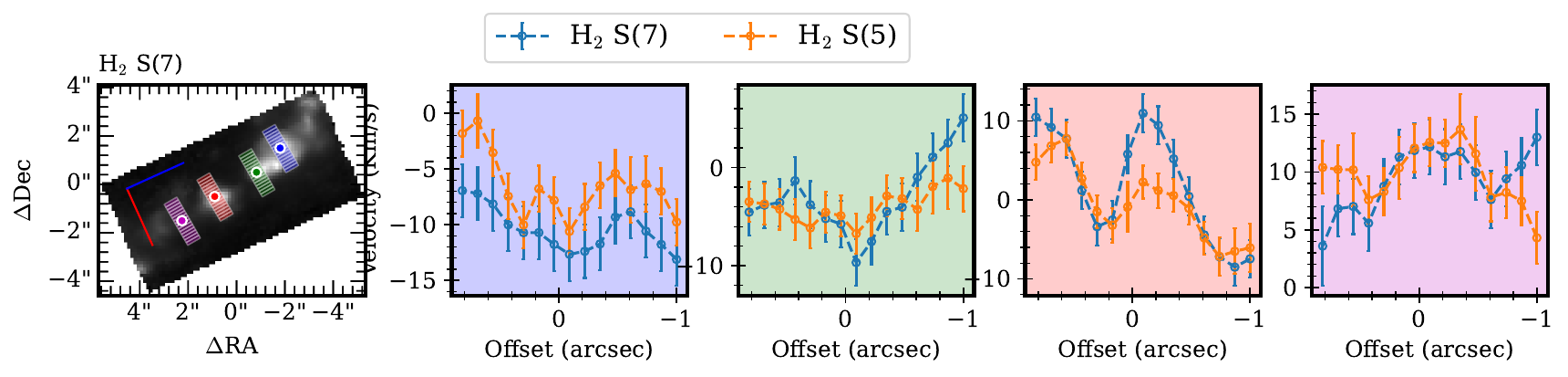}
        \caption{IRAS 20126}
        \label{fig:ortho_pv_iras20126}
    \end{subfigure}

    \caption{Figure~\ref{fig:Ortho-PV} continued.}
    \label{fig:Ortho-PV-2}
\end{figure*}

\subsection{PV Diagrams Orthogonal to the Outflow Axis}
To search for rotation and measure the lateral velocity structure across the \hydrogenmol\ flow, we generate PV diagrams orthogonal to the outflow axis using the prescription described above. 
These orthogonal PV diagrams are presented in Figure~\ref{fig:Ortho-PV}. In each diagram, the centroid velocity is measured for spatial slices one pixel wide (extracted perpendicular to the outflow axis) and plotted as a function of offset from the central axis.
{The displayed error bars show the statistical uncertainty in the centroid fitting and incorporate an uncertainty floor of 1/20th of a spectral pixel to account for MRS systematics (similar to as discussed in Section \ref{section:PV-along}).}
Four sampling locations along each outflow were selected, marked by colored points in the line maps of Figure~\ref{fig:Ortho-PV}; the corresponding regions of width 0.6\arcsec\ ($2\times$FWHM$_{\rm PSF}$ at 6.91 \micron) over which the velocities were averaged are shown by shaded areas of the same colors.
For this analysis, we used the \hhnu(7) and S(5) lines, as they are both bright and offer higher spectral resolution than the other \hydrogenmol\ transitions available in MIRI and NIRSpec.

PV diagrams of IRAS 16253, orthogonal to the outflow axis, reveal a complex velocity configuration (Figure \ref{fig:Ortho-PV}(a)). The green and red sky paths, which are closest to the central protostar, reveal roughly steady velocity gradients in both lines, with local velocity peaks observed at the center of each sky path. This steady gradient is observed over an angular scale of $\sim1''$ across the flow on both sides of the central protostar, located $>50$ au from the disk plane. The eastern part of the flow in these two sky paths is blue-shifted relative to its western counterpart. Very high angular resolution observations using ALMA of the disk in IRAS 16253 also show a similar velocity configuration, with the eastern part of the disk showing blue-shifted velocity relative to its western part due to disk rotation \citep{Aso2023ApJ...954..101A}. This suggests that we might be detecting rotation of the \hydrogenmol\ flow.

\begin{figure}
    \centering
    \includegraphics[width=\linewidth]{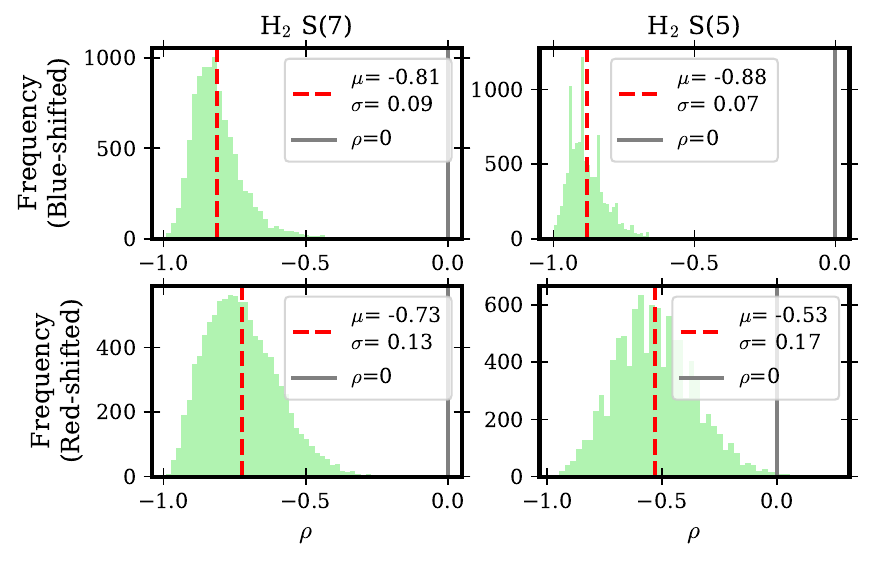}
    \caption{Distributions of the Spearman's rank correlation coefficients ($\rho$) to test the steady gradient in IRAS 16253. 
    The top two panels show the $\rho$ distributions for the \hhnu(7) and S(5) lines in the blue-shifted lobe (marked by the green region in Figure \ref{fig:ortho_pv_iras16253}. The bottom two panels show the $\rho$ distributions for the \hhnu(7) and S(5) lines in the red-shifted lobe(red region in Figure \ref{fig:ortho_pv_iras16253}. The vertical gray line in each panel marks $\rho=0$.
    }
    \label{fig:spearman}
\end{figure}
{
To test the statistical significance, given large uncertainties, of the steady gradients in the blue- and red-shifted lobes (indicated by the green and red paths in Figure \ref{fig:ortho_pv_iras16253}), we implemented a Monte Carlo error propagation technique. 
For each spatial position, we drew 10,000 random velocity values from a normal distribution centered on the measured velocity centroid, with a standard deviation equal to the uncertainty at that point. This process generated 10,000 synthetic PV diagrams that sample the full parameter space allowed by the error bars.
Simultaneously, we performed a Spearman's rank correlation test in each iteration, which provided us with a distribution of $\rho$. Although the steady gradients are not necessarily linear, we also performed a linear fit in each iteration. We carried out this analysis for both the S(7) and S(5) lines.}

{The resulting $\rho$ distributions are shown in Figure \ref{fig:spearman}. For the S(7) and S(5) lines in the blue-shifted lobe, the means of the $\rho$ distributions are $-0.81\pm0.09$ and $-0.88\pm0.07$, respectively. In the red-shifted lobe, the respective means are $-0.73\pm0.13$ and $-0.53\pm0.17$.
In all cases, the mean $\rho$ is separated from the null hypothesis of no monotonic correlation ($\rho=0$) by at least $5\sigma$ (and $>3\sigma$ for the red-shifted S(5) line), confirming that the observed steady gradients in IRAS 16253 are statistically significant. Furthermore, the linear fits of these Monte Carlo distributions yield mean velocity gradients of $-14\pm3$ and $-11\pm2$ \kms/arcsec for the S(7) and S(5) lines in the blue-shifted lobe, and $-8\pm2$ and $-4\pm2$ \kms/arcsec in the red-shifted lobe.
}

In contrast, the outermost sky paths (blue and magenta) do not show any noticeable velocity gradient monotonically continued across the flow. 
At the center of both the green and red sky paths, the S(7) line shows a velocity peak relative to the nearby region. 

Figure \ref{fig:Ortho-PV}(b) presents the PV diagrams orthogonal to the outflow axis for B335. None of the PV diagrams for B335 show evidence of wind rotation, as indicated by the absence of steady velocity gradients. The green and red sky paths, located closer to the protostar on the eastern and western sides, respectively, display higher velocities along the edges of the wind (see also Figure \ref{fig:mom1}(b)). In contrast, the magenta sky path, situated farther west, shows velocity peaking close to the center of the sky path. The blue sky path, outermost in the east, contains the shocked knot 3E and reveals the velocity peak coinciding with the location of the shock knot, and it is more pronounced in the S(7) line.

In HOPS 153, the green sky path, located closer to the protostar in the northwest direction, reveals a roughly steady velocity gradient across the green shaded region in both the S(7) and S(5) lines (Figure \ref{fig:Ortho-PV}(c)). However, this gradient is not observed in any other sky path. The outermost sky path in the northwest (in blue) shows a velocity peak at its center, which roughly aligns with the [Fe II] jet (also see Figure \ref{fig:mom1}(c)).

In HOPS 370, velocity complexity revealed by the velocity maps in Figure \ref{fig:mom1}(d) is further highlighted by the orthogonal PV diagrams, presented in Figure \ref{fig:Ortho-PV}(d). The northern cavity reveals a fast molecular jet, characterized by distinct velocity peaks at the centers of both the blue and green sky paths. The blue sky path PV diagram captures the high-velocity nature of the mushroom-like structure through its pronounced blue-shifted velocity wing. Additionally, a higher velocity structure is observed at the cavity wall in the green sky path, as evident by the peak in velocity wing around $\sim1$\arcsec. However, the steady velocity gradient, a supporting piece of evidence for rotation, is not observed in any of the orthogonal sky paths' PV diagrams.

The \hydrogenmol\ emission from IRAS 20126 also lacks evidence of velocity gradient, and hence wind rotation, in all four sky paths, orthogonal to the outflow (Figure \ref{fig:Ortho-PV}(e)). The orthogonal PV diagrams capture all the complex details of the velocity structure as discussed in the previous section (also see Figure \ref{fig:mom1}(e)).

\section{Discussion} 
\label{Discussion}



The unprecedented details of the protostellar winds and outflows revealed by our JWST observations raise many questions about the origin of \hydrogenmol\ emission and the nature of the flow. This section explores the broader implications of our observational results of the \hydrogenmol\ morphology and kinematics of five young envelope-dominated protostars.

\subsection{Origin of the \texorpdfstring{\hydrogenmol}~ emission} \label{sec:H2_limb_bright}

\begin{figure*}[htbp]
    \centering

    \begin{subfigure}{0.48\textwidth}
        \centering
        \includegraphics[width=\linewidth]{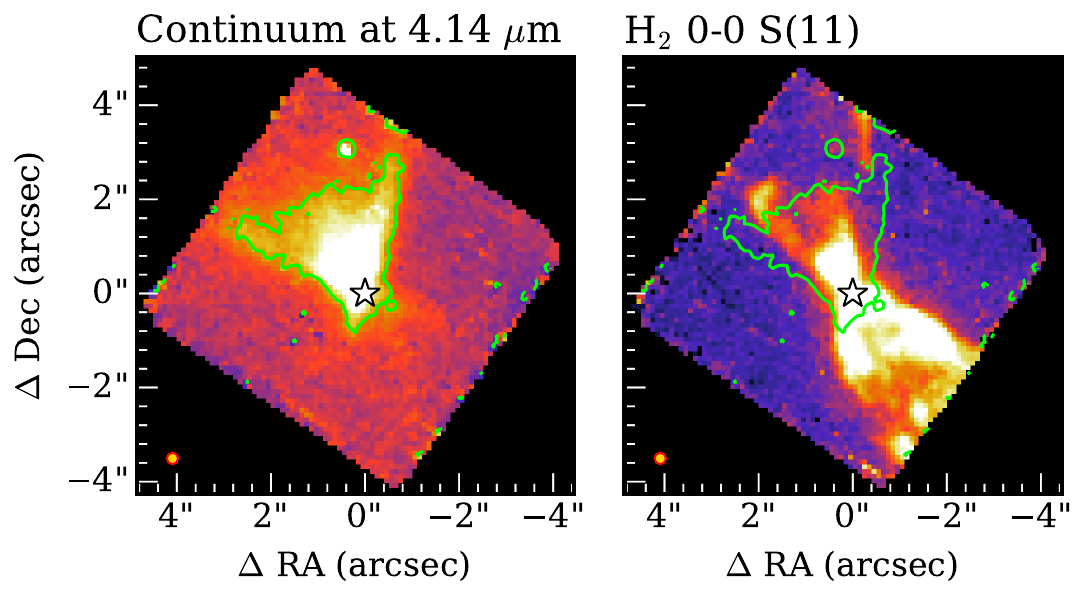}
        \caption{IRAS 16253}
        \label{fig:scattered_iras16253}
    \end{subfigure}
    \hfill
    \begin{subfigure}{0.48\textwidth}
        \centering
        \includegraphics[width=\linewidth]{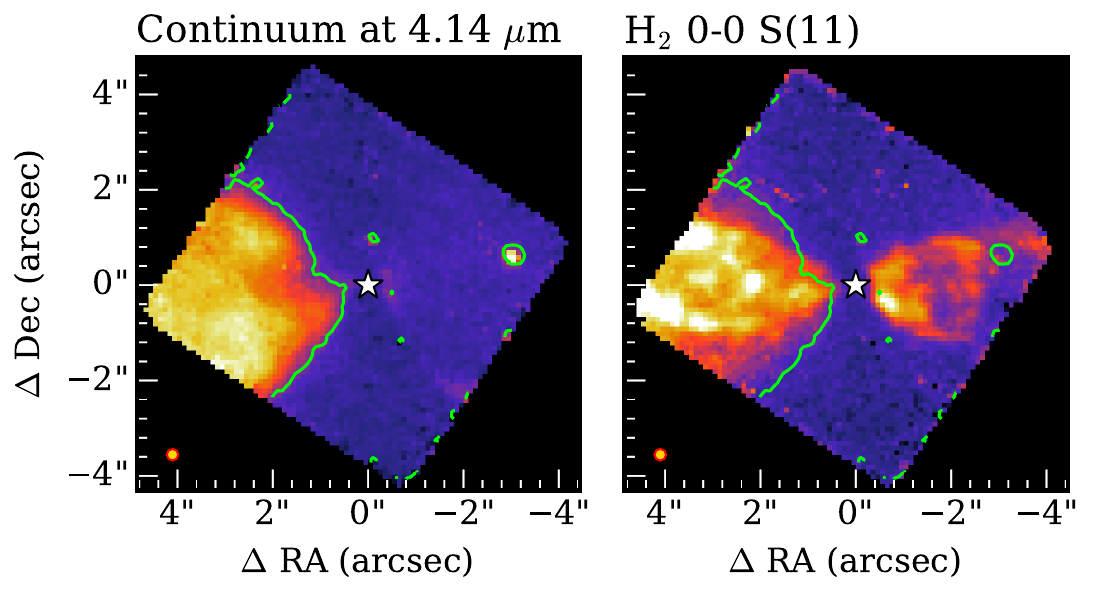}
        \caption{B 335}
        \label{fig:scattered_b335}
    \end{subfigure}

    \vspace{0.5cm}

    \begin{subfigure}{0.48\textwidth}
        \centering
        \includegraphics[width=\linewidth]{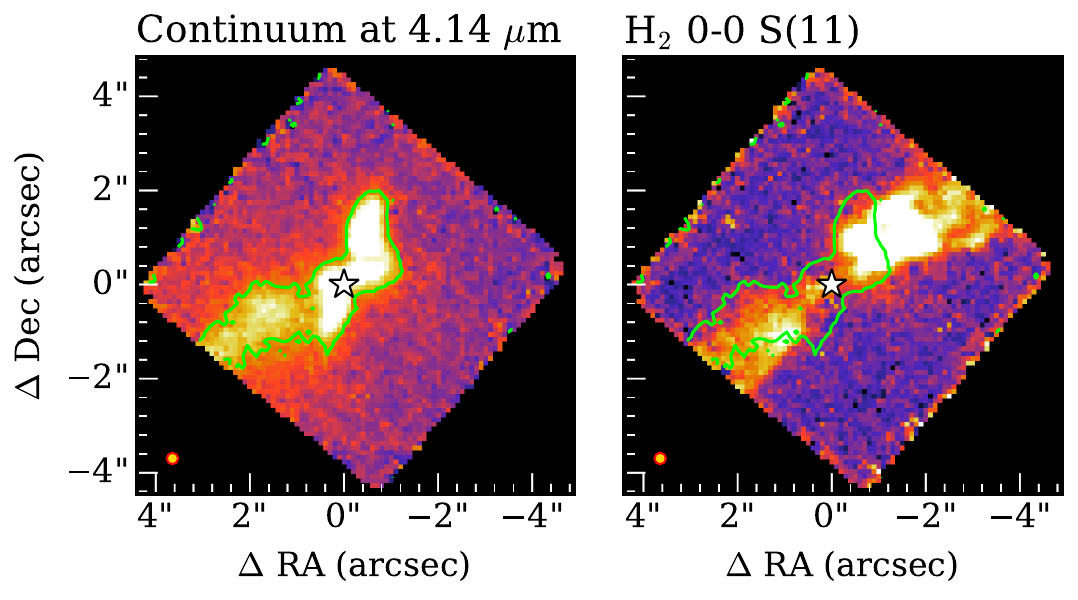}
        \caption{HOPS 153}
        \label{fig:scattered_hops153}
    \end{subfigure}
    \hfill
    \begin{subfigure}{0.48\textwidth}
        \centering
        \includegraphics[width=\linewidth]{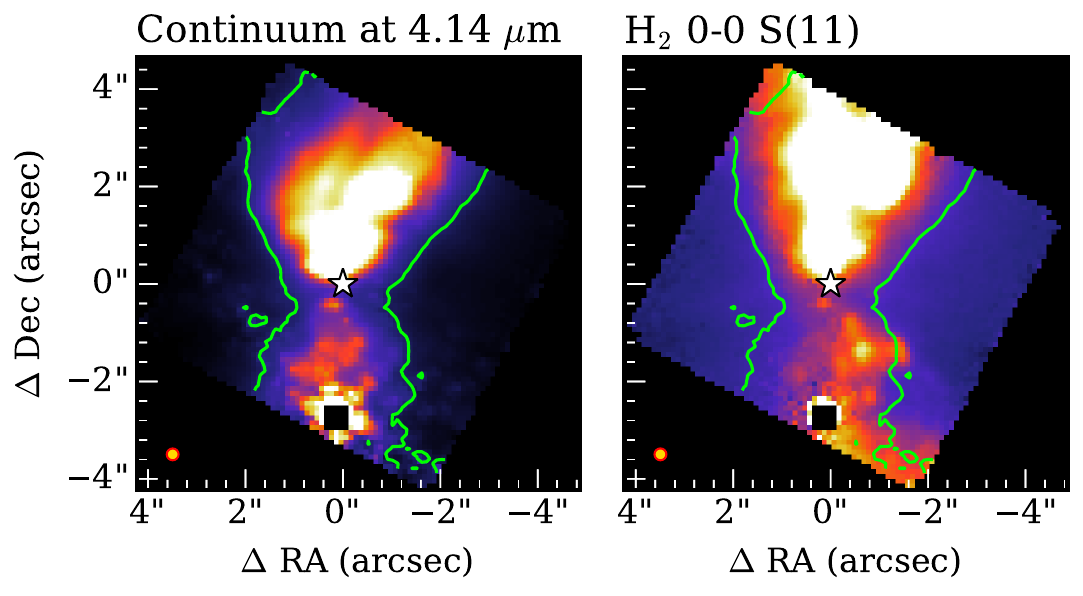}
        \caption{HOPS 370}
        \label{fig:scattered_hops370}
    \end{subfigure}

    \caption{Comparison of the extent of the scattered light at 4.14~\micron\ and the {\hhnu(11) line at 4.18\micron}. 
    The lime-colored contour ($9\times$ local rms of the scattered-light image) traces the outer edge of the scattered-light emission, 
    overplotted on the respective line images. A star in each panel indicates the position of the 870 \micron\ ALMA continuum source.}
    \label{fig:ScatteredvsH2}
\end{figure*}

Molecular hydrogen emission in protostellar outflows has been primarily attributed to shock-excited gas located along outflow cavity walls, 
internal shocks within collimated jets or bow shocks, where jets encounter the surrounding cloud
(e.g., see Figure 3 in \citealt{Bally2016ARA&A..54..491B}; see also \citealt{Lee2020A&ARv..28....1L}). 
However, the unprecedented angular resolution and sensitivity provided by JWST now reveal that \hydrogenmol\ emission fills the inner part of the outflow cavity, where the cavity walls are delineated in scattered light continuum from the central protostar \citep[e.g.,][]{Habel2021ApJ...911..153H, Federman2023arXiv, Narang2024ApJ...962L..16N}. 
To compare the spatial extent of \hydrogenmol\ emission with respect to the outflow cavity, we present the morphology of scattered light at 4.14 \micron\ continuum and the \hhnu(11) line emission in the left and right panels of Figure~\ref{fig:ScatteredvsH2}, respectively \citep[see also][]{Federman2023arXiv}. Here, we show only four sources, as the most luminous source, IRAS 20126, is a very complex region with multiple sources in the FOV (\citealt{Federman2023arXiv}; Higgins, M. et al. in prep.). The lime contours, tracing the edges of the scattered light cavity, are wider than the observed \hydrogenmol\ emission in all four protostars, indicating that the \hydrogenmol\ flow is narrower and well within the cavity.

\begin{figure*}[htbp]
    \centering
    \includegraphics[width=\linewidth]{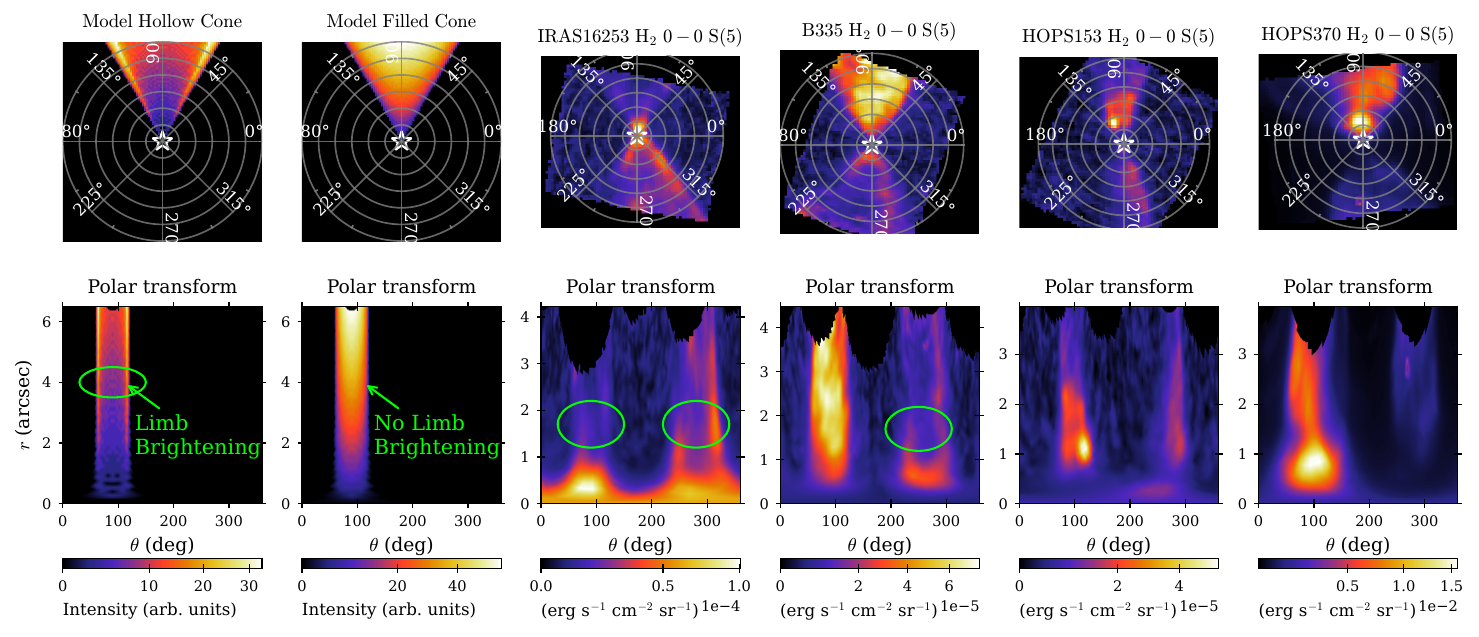}
    \caption{Comparison between the simulated toy outflow cavity and the observed \hydrogenmol\ emission.
    \textit{Top row}: Morphological comparison of the outflow. The first two panels show the projected density from the simulated cavity model, while the panel(s) to the right show the observed emission from the \hhnu(5) line. A white star in each panel indicates the position of the 870 \micron\ ALMA continuum source, which defines the origin for the overplotted polar grid.
    \textit{Bottom row}: The intensity distribution from each corresponding panel above, transformed into polar coordinates $(r, \theta)$. Regions showing prominent limb-brightening are highlighted with green ellipses.
    }
    \label{fig:Cavity Simulation}
\end{figure*}

Furthermore, \hydrogenmol\ emission does not strictly show limb brightening at the cavity edges (except for IRAS 16253), which would be expected if it originated in outflow shocks along the cavity edges. 
To illustrate this point, we construct a simple toy model for two scenarios: (1) emission arising from shocks along the cavity walls, represented by a hollow cone with thick walls (with a wall thickness of 5\arcdeg\ from the cone edge) and uniform \hydrogenmol\ density, and (2) emission uniformly filling the cavity, modeled as a solid cone. Assuming that the emission is optically thin so that we see each molecule emitting, we generate two-dimensional projections (Figure \ref{fig:Cavity Simulation}), which demonstrate that the observed \hydrogenmol\ emission is visually more consistent with a filled cavity, except for IRAS 16253 and the red lobe of B335.
The limb brightening becomes clearer and shows up like two bright parallel lanes in the intensity distribution map transformed into polar coordinates (marked by lime ellipses in Figure \ref{fig:Cavity Simulation}).
These bright lanes are clearly observed only in IRAS 16253 and in the red-shifted cavity of B335.

Altogether, our observations suggest that the winds traced by \hydrogenmol\ are narrower than the outflow cavity and generally fill the cavity. 
Similar results have also been observed in contemporary JWST studies of protostars (e.g., \citealt{Federman2023arXiv, Vleugels2025A&A...695A.145V, Narang2026arXiv260209837N}; Pathak, V. et al. in prep.). 
The broader extent of the continuum emission relative to the \hydrogenmol\ line emission may also arise from the fact that the outflow cavity wall is not a sharp physical boundary, but instead a stratified dust distribution whose optical depth increases with depth into the envelope. In this scenario, the continuum traces optically thick dust over a wider region, whereas the \hydrogenmol\ emission arises from more localized shocked or excited gas.

The observed morphology of \hydrogenmol\ emission can be indicative of three scenarios: (1) internal shocks in the wide-angled winds launched from the disk, (2) supersonic lateral splash from the collimated jet or the wide angle wind shocking the low-density cavity material, or (3) shocks in the outflow cavity resulting from the interaction of the collimated jet and the wide-angled winds launched from the disk. We further discuss these possibilities in the sections below using morphological and kinematical results observed.

\subsection{Nature of the \texorpdfstring{\hydrogenmol~}\ Winds}

The observed morphology and kinematics of the \hydrogenmol\ line emission reveal a clear stratified/nested structure. 
Morphologically, the emission becomes progressively more collimated with their increasing excitation energy (\Eup) (Figure \ref{fig:LineMaps} and Table \ref{Table:OpeningAngle}).
Kinematically, \hydrogenmol\ lines with higher \Eup, e.g., {\hhnu(7) and S(5),} show higher radial velocities relative to the lines with lower \Eup, e.g., {\hhnu(1) and S(2)} lines (Figure \ref{fig:PV_figure} and Table \ref{Table:PV}). 
This nested structure of \hydrogenmol\ winds in morphology and velocity is consistent with the MHD disk wind simulations \citep[see e.g.,][and references therein]{Ray2021NewAR..9301615R, Pascucci2023ASPC..534..567P}.
{However, models with pure-jet, X-winds, or combined X and disk winds may also reproduce the nested morphology and kinematic structure observed} \citep[see e.g.,][]{Rabenanahary2022A&A...664A.118R, Shang2023ApJ...944..230S, Lopez2024ApJ...977..126L}. 

The stratified structure of \hydrogenmol\ emission in morphology and kinematics is consistently seen in all five protostars in our sample, which spans a wide range in \lbol. 
The contemporary studies of more evolved sources in Class I and Class II phases and even in more embedded and (possibly) younger Class 0 protostars such as HH 211, have also found similar results  \citep[e.g.,][]{Tychoniec2024A&A...687A..36T, Pascucci2025NatAs...9...81P, Vleugels2025A&A...695A.145V, Ray2023Natur.622...48R, Garatti2024A&A...691A.134C}.
The common observation of nested structure across mass and age spectrum suggests that the launch and propagation of the \hydrogenmol\ winds are governed by a similar mechanism. Based on the similar launching and propagation mechanism argument and growing evidence of disk winds in more evolved sources, we attribute the origin of the \hydrogenmol\ flow in protostars to {MHD disk winds.} However, we also note that there are protostars, e.g., Ced 110 IRS4 and L1527, that lack the evidence of the nested structure in the winds (\citealt{Narang2025AJ....169..192N, Devaraj2026arXiv260117820D}; Drechsler et al. sub.).

Additionally, one noticeable difference in the morphology of the similar excitation \hydrogenmol\ lines across different sources is the lack of common detailed structures (Figure \ref{fig:LineMaps}).
Apart from the overall bipolar nested geometry of the \hydrogenmol\ winds, the details of features appear consistently different across the maps, suggesting complex interactions between the \hydrogenmol\ wind with the central jet or wind itself.
This is true not only for our sample but for other protostars observed with JWST \citep[e.g.,][]{Garatti2024A&A...691A.134C, Tychoniec2024A&A...687A..36T}. 
We further note that the inclination-corrected observed velocities of the \hydrogenmol\ winds appear to scale with the \lbol\ of its host protostar (Table \ref{Table:PV}; see Table 1 in \citealt{Federman2023arXiv} for mass). 
Our results seem to suggest that more massive protostars likely drive faster outflows. 
This result is not totally unexpected as the outflow escape velocity increases with the mass of the protostar and it is consistent with the previous studies \citep[e.g.,][]{Bontemps1996A&A...311..858B, Beuther2002A&A...383..892B, Beuther2025ARA&A..63....1B, Wu2004A&A...426..503W, Lopez2009A&A...499..811L, Maud2015MNRAS.453..645M}

\subsection{Search for Wind Rotation}
\label{sec:wind-rotation}

One of the key observational diagnostics for distinguishing between jet-launching models is the detection of rotation within the outflow \citep[e.g.,][]{Ray2021NewAR..9301615R, Lee2020A&ARv..28....1L}. In our sample, we find no clear evidence of rotational signatures in any of the sources, except for IRAS 16253, where a clear steady velocity gradient is observed close to the protostar (Figures \ref{fig:mom1} and \ref{fig:Ortho-PV}). In Figure \ref{fig:Ortho-PV}(a), the green and red sky paths perpendicular to the outflow axis reveal a distinct and steady velocity gradient across the outflow on both sides of the protostar. In contrast, sky paths (marked in blue and magenta) located farther out along the jet do not show such velocity gradient. Notably, the direction of the \hydrogenmol\ velocity gradient observed with JWST is consistent with the disk rotation observed in {high angular resolution} ALMA observations \citep[][]{Aso2023ApJ...954..101A}. 

The apparent gradient in velocity across the wind could be a signature of disk wind rotation. 
To investigate if the apparent \hydrogenmol\ velocity gradients in green and red sky paths of Figure \ref{fig:Ortho-PV} are indeed caused by the wind rotation, we quantify their specific angular momentum. 
We define $j_{\rm rot}=r \times v_{\rm rot}$, where $v_{\rm rot} = |v_{\rm left}-v_{\rm right}|/(2\sin{i})$ and $r$ is the half-distance between the points where velocities are measured.
Here, $v_{\rm left}$ and $v_{\rm right}$ are the line-of-sight velocities on the left and right of the jet axis. To improve the velocity estimation, we used spectra extracted from the apertures on the left and right sides of the jet axis on both sides of the disk (marked as lime circles in Figure \ref{fig:Ortho-PV}). Using apertures increases the S/N of the \hydrogenmol\ lines, improving line-center determination. The distance between the left and the right apertures provides the $2r$.
In the northern flow, values of the specific angular momentum traced by the {\hhnu(7) and S(5)} lines are $\rm 210\pm30~km~s^{-1}au$ and $\rm 165\pm44~km~s^{-1}au$, respectively. Similarly, the southern flow yields $\rm 181\pm20~km~s^{-1}au$ and $\rm 126\pm31~km~s^{-1}~au$ for the S(7) and S(5) lines, respectively. 
These values exceed the envelope's specific angular momentum \citep[$j_{\rm env} \sim \rm 45~km~s^{-1}~au$;][]{Hsieh2019ApJ...871..100H} by a factor of $\sim3-5$, suggesting that the observed gradient is not due to envelope rotation.

\begin{figure}[htbp]
    \centering
    \includegraphics[width=\linewidth]{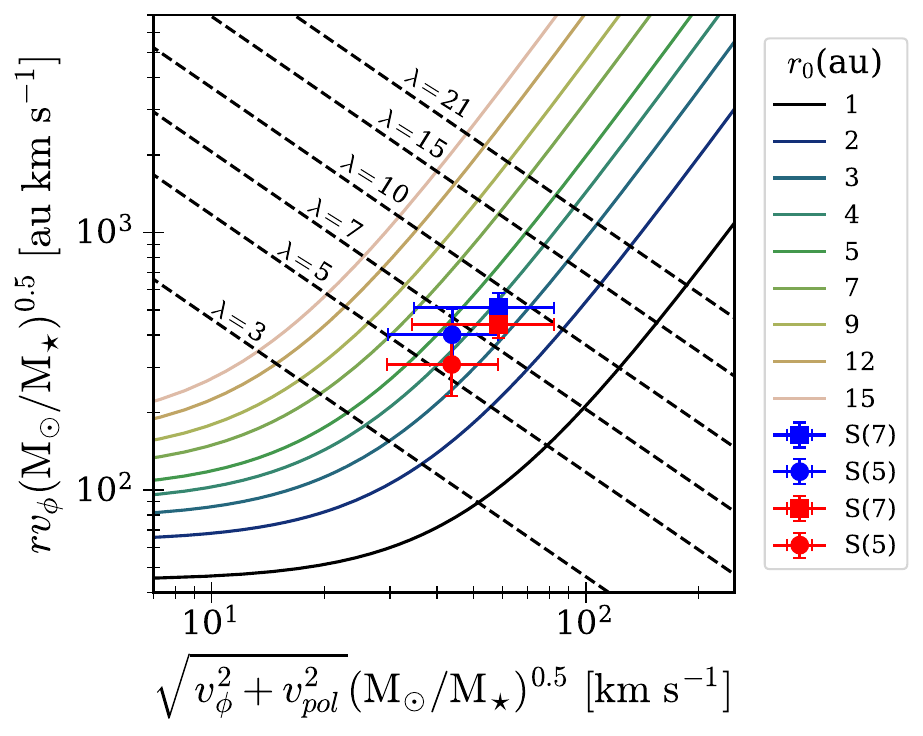}
    \caption{Relationship between the specific angular momentum of rotation vs the total velocity from the MHD wind models. The relation for constant magnetic lever arm parameters ($\lambda_{\phi}$) and for constant launch radii ($r_0$) are shown as black dashed lines and colored solid lines, respectively. The observed points {for transitions \hhnu(7) and S(5)} are marked in blue and red, corresponding to the blue- and red-shifted flows.}
    \label{fig:iras16253_launchradius}
\end{figure}

Assuming that the gradient detected in IRAS 16253 is indeed the wind rotation, we can constrain the launching radius ($r_0$) and the magnetic lever arm parameter ($\lambda_{\phi}$) of the \hydrogenmol\ winds.
To do this, we compare our observational measurements to predictions from MHD disk wind models, which provide a theoretical relationship between the specific angular momentum ($r v_{\phi}$) and the total wind velocity ($\sqrt{v_{\phi}^2+v_{pol}^2}$) for a given $r_0$ and $\lambda_{\phi}$ \citep{Anderson2003ApJ...590L.107A, Ferreira2006A&A...453..785F, Nazari2024A&A...686A.201N_COMOutflow}.
In Figure \ref{fig:iras16253_launchradius}, we plot these theoretical relations for constant values of $\lambda_{\phi}$ and $r_0$, calculated for IRAS 16253's stellar mass of 0.17 $M_{\odot}$ \citep{Aso2023ApJ...954..101A}.
We then overplot our measured values of the specific angular momentum and total velocity for {\hhnu(7) and S(5)} transitions. Here, $v_{pol}$ is taken as the mean velocity of the S(7) and S(5) lines, listed in Table \ref{Table:PV}. 

Figure \ref{fig:iras16253_launchradius} shows that the observed velocity gradients are consistent with a \hydrogenmol\ wind launching footprint at radii of $\sim4$~au, with inferred magnetic lever-arm parameters $\lambda_{\phi}$ in the range $\sim5$–10. 
The $\lambda_{\phi}$ value of 5 to 10, although moderate values for a few MHD models \citep[e.g.,][]{Ferreira2006A&A...453..785F, Kadam2025A&A...695A.167K}, is on the higher side than those typically observed with submm observations (\citealt{Tabone2017A&A...607L...6T, Tabone2020A&A...640A..82T, Lee2021ApJ...907L..41L}; See also \citealt{Zhang2018ApJ...864...76Z}). Moreover, most {MHD disk wind} models do not generally predict such large lever-arm parameters \citep[see, e.g., Section~3 and references therein of][]{Pascucci2023ASPC..534..567P}. Altogether, the launch radius and the magnetic lever arm parameter values estimated for IRAS 16253 suggest that observed kinematics of \hydrogenmol\ can be explained with the MHD disk wind model.

Despite the consistency of the inferred launching radii and lever-arm parameter with a disk wind interpretation, alternative explanations for the observed velocity gradients in IRAS~16253 remain plausible. In particular, IRAS~16253 is the only source (other than the redshifted cavity of B335) in our sample that displays limb brightening in \hydrogenmol\ emission, raising the possibility that the observed transverse velocity gradients arise from interactions between the wind and the inner cavity walls. 
In this scenario, angular momentum from the rotating envelope could be imparted to shocked molecular gas, producing apparent velocity gradients without requiring intrinsic wind rotation.
Similarly, asymmetric shocks against non-axisymmetric cavity walls may generate velocity structures that mimic rotation signatures. This scenario can naturally account for the large specific angular momentum inferred for the \hydrogenmol\ emission.

Such interpretations have been proposed for other protostellar outflows, including HH~46/47, where transverse velocity gradients are detected consistently along limb-brightened regions out to $\sim1000$~au with very large specific angular momentum, $\rm >3000~km~s^{-1}au$ \citep[][]{Birney2024A&A...692A.143B}.
In contrast, in IRAS~16253, the steady velocity gradient is detected only within $\lesssim200$~au of the disk plane on both sides of the disk, and is not observed along the extended limb-brightened regions. Also, the specific angular momentum in IRAS 16253 is similar to those reported for the rotating outflows in the ALMA observations \citep[e.g.,][]{Bjerkeli2016Natur.540..406B, Tabone2017A&A...607L...6T, Zhang2018ApJ...864...76Z}. 
This spatial confinement and comparable specific angular momentum favor an interpretation in which the observed gradients trace wind rotation close to the launching region, and the rotation signature vanishes due to wind interactions with the cavity/flow material. Nevertheless, we emphasize that wind rotation interpretation remains tentative, given the limited velocity resolution of the MIRI observations ($\sim80$~\kms).

The non-detection of rotation in the outer sky paths in IRAS 16253 suggests that the winds significantly interact with the material within the flow and the cavity wall. These complex interactions between winds and the cavity material/internal shocks, as well as their greater distance, may be a common issue contributing to the non-detection of rotation in the other sources in our sample \citep[see also][Section 3]{Lee2020A&ARv..28....1L}. The non-detection could also be because of a combination of other reasons, such as modest spectral resolution (best velocity resolution is $\sim80$ \kms)\footnote{https://jwst-docs.stsci.edu/jwst-calibration-status/miri-calibration-status/miri-mrs-calibration-status} of MIRI/MRS \citep[][]{Jones2023MNRAS.523.2519J, Argyriou2023A&A...675A.111A, Pontoppidan2024ApJ...963..158P}.

\subsection{Detection of a Molecular Jet in HOPS 370}

\begin{figure}
    \centering
    \includegraphics[width=\linewidth]{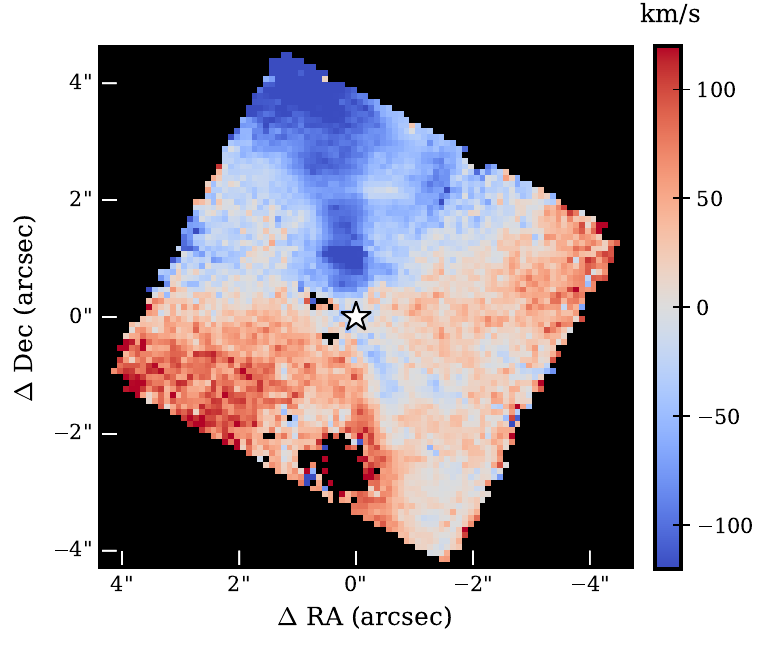}
    \caption{Inclination corrected velocity map of \hhnu(11) line toward HOPS 370. Spaxels with S/N below $5\sigma$ are masked. \textit{Caution}: The velocities of the ambient \hydrogenmol\ emission appear artificially large because the observed line-of-sight velocities are corrected by a factor of $1/\cos(72^\circ)$.}
    \label{fig:HOPS370-S11}
\end{figure}

The velocity maps of all five protostars in our sample clearly show the blue- and red-shifted winds (Figure \ref{fig:mom1}). Despite the wavelength calibration accuracy of MIRI/MRS being $\sim10$ \kms, we resolve coherent high-velocity structures within the outflow. For example, the molecular shocked knots in B335 and HOPS 370 show higher velocities relative to the rest of the flow in their respective outflow cavity (Figure \ref{fig:mom1}). 
A collimated, fast molecular hydrogen jet component is detected in HOPS 370 (Figure \ref{fig:mom1} (d); see also \citealt{Federman2023arXiv}).
In its northern (blue-shifted) cavity, the data reveal both a fast collimated \hydrogenmol\ jet and slow wide-angle \hydrogenmol\ winds enveloping the jet (Figure \ref{fig:mom1}(d)). The collimated jet is more pronounced in the velocity maps of higher excitation lines, as demonstrated by the \hhnu(11) velocity map, which confirms the fast component despite the lower velocity resolution of NIRSpec (Figure \ref{fig:HOPS370-S11}). The presence of the fast velocity molecular jet is further confirmed by the higher velocities in the middle of the PV diagrams orthogonal to the outflow axis (Figure \ref{fig:Ortho-PV}(d)). 

\begin{figure*}
    \centering
    \includegraphics[width=\linewidth]{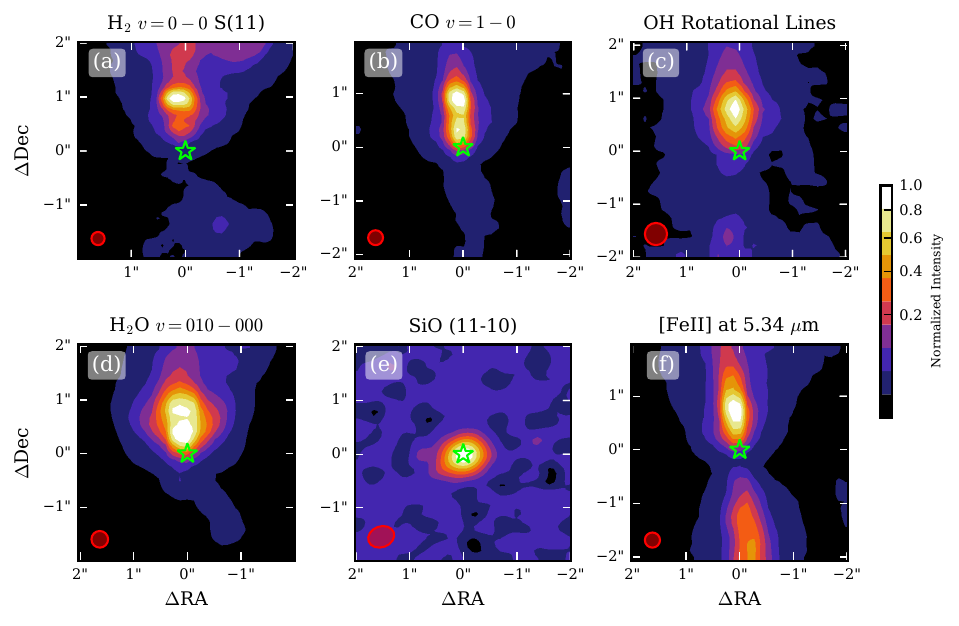}
    \caption{Molecular jet in various species detected in HOPS 370. (a) \hhnu(11) line map.
    Mean line images of CO 1-0 lines from P(38) to P(51),  OH pure rotational lines from 9.2 \micron\ to 10.07 \micron, and \water\ ro-vibrational lines from 6.11 \micron\ to 6.43 \micron\ are shown in panels (b), (c), and (d), respectively. (e) SiO ($J=11-10$) moment 0 map from -50 \kms\ to 50 \kms. For reference, a line map of [Fe II] jet at 5.34 \micron\ is shown in panel (f). A lime star marks ALMA 870 \micron\ continuum peak. The beam size is indicated on the bottom left in red.  }
    \label{fig:HOPS370-jets}
\end{figure*}
The detection of this prominent molecular jet in HOPS 370 is particularly intriguing because it has been suggested that the molecular component of the outflow dominates in the youngest protostars more than in their more evolved counterparts \citep[][]{Lee2020A&ARv..28....1L, Ray2023Natur.622...48R, Garatti2024A&A...691A.134C}.
HOPS 370, however, like the other protostars in our sample, appears more evolved than extremely young protostars like HH 211 \citep[][]{Ray2023Natur.622...48R}.
HOPS 370 is classified as a Class 0/I object based on its SED and does not appear less evolved than the other objects in the sample (\citealt{Furlan2016ApJS..224....5F}, see also \citealt{Federman2023ApJ...944...49F}), yet it is the only one in which we detect a collimated molecular jet. This finding, combined with a similar detection of a molecular jet in another evolved Class I source, HOPS 315 \citep[][]{Vleugels2025A&A...695A.145V}, suggests that protostellar age may not be the primary factor governing the presence of a molecular jet. 
We note that HOPS 315 may be classified as a Class I source due to its relatively low inclination. \citet{Federman2023ApJ...944...49F} report a disk-to-envelope mass ratio of 0.34 for HOPS 315, comparable to that of protostars in our sample (e.g., HOPS 153, with a ratio of 0.53); however, it appears to be more evolved than HH 211.

An alternative explanation may lie in the fact that the HOPS 370 is known to accrete at a very high accretion rate, $(1.7-3.2)\times10^{-5}$ \msunyr \citep[][]{Tobin2020ApJ...905..162T}, which would result in a higher outflow rate. 
High mass-loss rates can provide enough shielding to the ionized material so that molecules can reform at the base of the jet, thus rendering a molecular jet \citep[][]{Tabone2020A&A...636A..60T, Lee2020A&ARv..28....1L}. \citet{Tabone2020A&A...636A..60T} showed that \hydrogenmol\ formation efficiently initiates the formation of molecules such as CO, \water, SiO, and OH above kinetic temperature ($T_{\rm K}$) of $\sim$800 K. In such a scenario, one would expect all these species to be detectable simultaneously. Indeed, CO, \water, and OH have all been detected in the jet of HOPS 370 \citep[][Manoj, P. et al., in prep.]{Federman2023arXiv, Neufeld2024arXiv240407299N, Rubinstein2024ApJ...974..112R}. Line maps of these species showing molecular jets together with the SiO $J=11-10$ and [Fe II] line maps are shown in Figure \ref{fig:HOPS370-jets}. This would suggest that the appearance of the molecular jet correlates with the accretion/outflow mass-loss rate. A large statistical study of protostars is required to solidify this result.
Interestingly, SiO, whose reformation in molecular jets is well established \citep[e.g.,][]{Cabrit2012A&A...548L...2C, Podio2021A&A...648A..45P}, is not detected in the jet of HOPS 370 (Figure \ref{fig:HOPS370-jets}(e); obtained from ALMA Proposal ID: 2024.1.01623.S - P.I.: Lukasz Tychoniec).
Previous ALMA observations targeting the SiO ($J=5-4$) line similarly resulted in a non-detection of the collimated jet \citep{Sato2023ApJ...944...92S}.
Future studies will investigate the absence of the SiO jet.

We observed a key difference between the molecular and ionic jet in HOPS 370: while molecular jet emission is detected exclusively in the blue-shifted lobe, the [Fe II] jet is clearly visible in both the blue- and red-shifted components. This discrepancy raises the question of why molecular formation (or survival) is suppressed in the red-shifted lobe. 
Furthermore, \citet{Federman2023arXiv} noted that \hydrogenmol\ jet is wider than the [Fe II] jet.
We hypothesize that asymmetric conditions in the jet between the blue- and red-shifted lobes could account for this disparity. 
Such unipolar molecular jets have been observed with ALMA observations \citep[see e.g.,][]{Hsieh2023ApJ...947...25H}.

\section{Conclusions}
\label{ConclusionSection}
In this paper, we present the morphological and kinematical properties of the molecular hydrogen emission around five protostars in their envelope-dominated phase, spanning a broad range in \lbol\ (0.16 to 10$^4$\lsun), as observed with the IFU mode of NIRSpec and MIRI onboard the JWST as part of the Investigating Protostellar Accretion across the Mass Spectrum (IPA) GO program. Our primary findings are summarized as following:

\begin{enumerate}
    \item We detect a rich set of \hydrogenmol\ lines, including pure-rotational transitions ($v=0-0$) up to S(18) at the brightest locations and several ro-vibrational lines from the $v=1-1$, $v=1-0$, and $v=2-1$ transitions.
    \item The \hydrogenmol\ emission fills the outflow cavities traced in scattered light.  They show limb brightening along the cavity walls in only three of the eight cavities. 
    \item We measure the velocity of the gas and find it organized into blue- and red-shifted flows, which we interpret as winds from the central disk.
    
    \item The \hydrogenmol\ winds become more collimated and faster with increasing excitation energy (\Eup), demonstrating a nested, stratified structure in both morphology and kinematics. The \hydrogenmol\ emission is best explained by shocked material in a molecular wind driven by the disk, rather than shocked cavity walls.
    \item The half-opening angles of the \hydrogenmol\ winds range from $\sim8^{\circ}$ to $\sim40^{\circ}$ and show no correlation with the host protostar's bolometric luminosity.
    \item The inclination-corrected \hydrogenmol\ wind velocities scale with the host protostar's bolometric luminosity.
    \item 
    In IRAS~16253, we detected a systematic velocity gradient perpendicular to the outflow axis on both the blue- and red-shifted sides, confined to regions very close to the disk plane. This gradient may be interpreted as a signature of wind rotation. Under this assumption, the inferred kinematics correspond to a wind-launching radius of $\sim4$~au and a magnetic lever-arm parameter $\lambda_{\phi}\sim5-10$, broadly consistent with model predictions for {MHD disk winds}. No comparable, steady velocity gradients are detected in the remaining sources of our sample.
    
    \item We report the detection of a fast, collimated, blue-shifted molecular (\hydrogenmol, CO, and \water) jet from the Class 0/I intermediate-mass protostar HOPS 370. 
    SiO, however, is not detected in the jet by ALMA observations.
\end{enumerate}

JWST's unprecedented sensitivity and angular resolution have enabled, for the first time, spatially resolved observations of \hydrogenmol\ emission in the immediate vicinity of protostars.
Our results provide a detailed view of the \hydrogenmol\ morphology and kinematics in protostellar outflows.
With high angular and spectral resolution of NIRSpec/IFU and MIRI/MRS, we detect a nested wind structure in both morphology and velocity.
We also find that the outflow cavities are filled with the \hydrogenmol\ winds.  
Taken together, these properties offer observational support for an MHD disk wind origin.

\section{Data Availability} \label{Section:Data-Availability}
All of the data presented in this article were obtained from the Mikulski Archive for Space Telescopes (MAST) at the Space Telescope Science Institute. The specific observations analyzed can be accessed via \href{DOI: 10.17909/3kky-t040}{https://doi.org/10.17909/3kky-t040}.
\section{Acknowledgment} 
{The authors thank the anonymous reviewer for their careful reading and constructive comments, which helped improve the clarity of the paper.}
This work is based on observations made with the NASA/ESA/CSA James Webb Space Telescope. The data were obtained from the Mikulski Archive for Space Telescopes at the Space Telescope Science Institute, which is operated by the Association of Universities for Research in Astronomy, Inc., under NASA contract NAS 5-03127 for JWST. These observations are associated with program \#1802. 
H.T. and P.M. acknowledge the support of the Department of Atomic Energy, Government of India, under Project Identification No. RTI 4002. 
H.T. also acknowledges the support of the Sarojini Damodaran Fellowship and Infosys-TIFR Leading Edge Travel Grant.
Part of this  research  by MN was carried out at the Jet Propulsion
Laboratory, California Institute of Technology, under a contract with the National Aeronautics and Space Administration (80NM0018D0004).
Support for STM, RG, WF, JG, JJT, and DW in program \#1802 was provided by NASA through a grant from the Space Telescope Science Institute, which is operated by the Association of Universities for Research in Astronomy, Inc., under NASA contract NAS 5-03127. 
D.A.N. was supported by grant SOF08-0038 from USRA. 
A.C.G. and SF have been supported by PRIN-MUR 2022 20228JPA3A “The path to star and planet formation in the JWST era (PATH)”, by INAF-GoG 2022 “NIR-dark Accretion Outbursts in Massive Young stellar objects (NAOMY)”, and by Large Gran INAF-2024 “Spectral Key features of Young stellar objects: Wind-Accretion LinKs Explored in the infraRed (SKYWALKER)”. 
G.A. and M.O. acknowledge financial support from grants PID2023-146295NB-I00 and CEX2021-001131-S, funded by MCIN/AEI/10.13039/501100011033. 
Y.-L.Y. acknowledges support from Grant-in-Aid from the Ministry of Education, Culture, Sports, Science, and Technology of Japan (20H05845, 20H05844, 22K20389), and a pioneering project in RIKEN (Evolution of Matter in the Universe). 
Leiden astrochemistry thanks support from the European Research Council (ERC) under the European Union’s Horizon 2020 research and innovation programme (grant agreement No. 101019751 MOLDISK). 
AS gratefully acknowledges support by the Fondecyt Regular (project code 1220610), and ANID BASAL project FB210003.
The National Radio Astronomy Observatory and Green Bank Observatory are facilities of the U.S. National Science Foundation operated under cooperative agreement by Associated Universities, Inc.
RK acknowledges financial support via the Heisenberg Research Grant funded by the Deutsche Forschungsgemeinschaft (DFG, German Research Foundation) under grant no.~KU 2849/9, project no.~445783058.
P.N. acknowledges support from the ESO Fellowship and the IAU Gruber Foundation Fellowship programs.

\software{Astropy \citep{astropy:2013, astropy:2018, astropy:2022}; CARTA \citep{angus_comrie_2021_4905459}; Matplotlib \citep{matplotlibHunter:2007}; NumPy \citep{numpyharris2020array}}; SciPy \citep{scipy2020SciPy-NMeth}; scikit-image \citep{scikit-image}.

\bibliography{your_bib_file}{}
\bibliographystyle{aa_url}

\appendix

\restartappendixnumbering

\section{Line Maps}
\label{LineMapAppendix}
\begin{figure}[ht]
     \centering
    \includegraphics[width=0.9\linewidth]{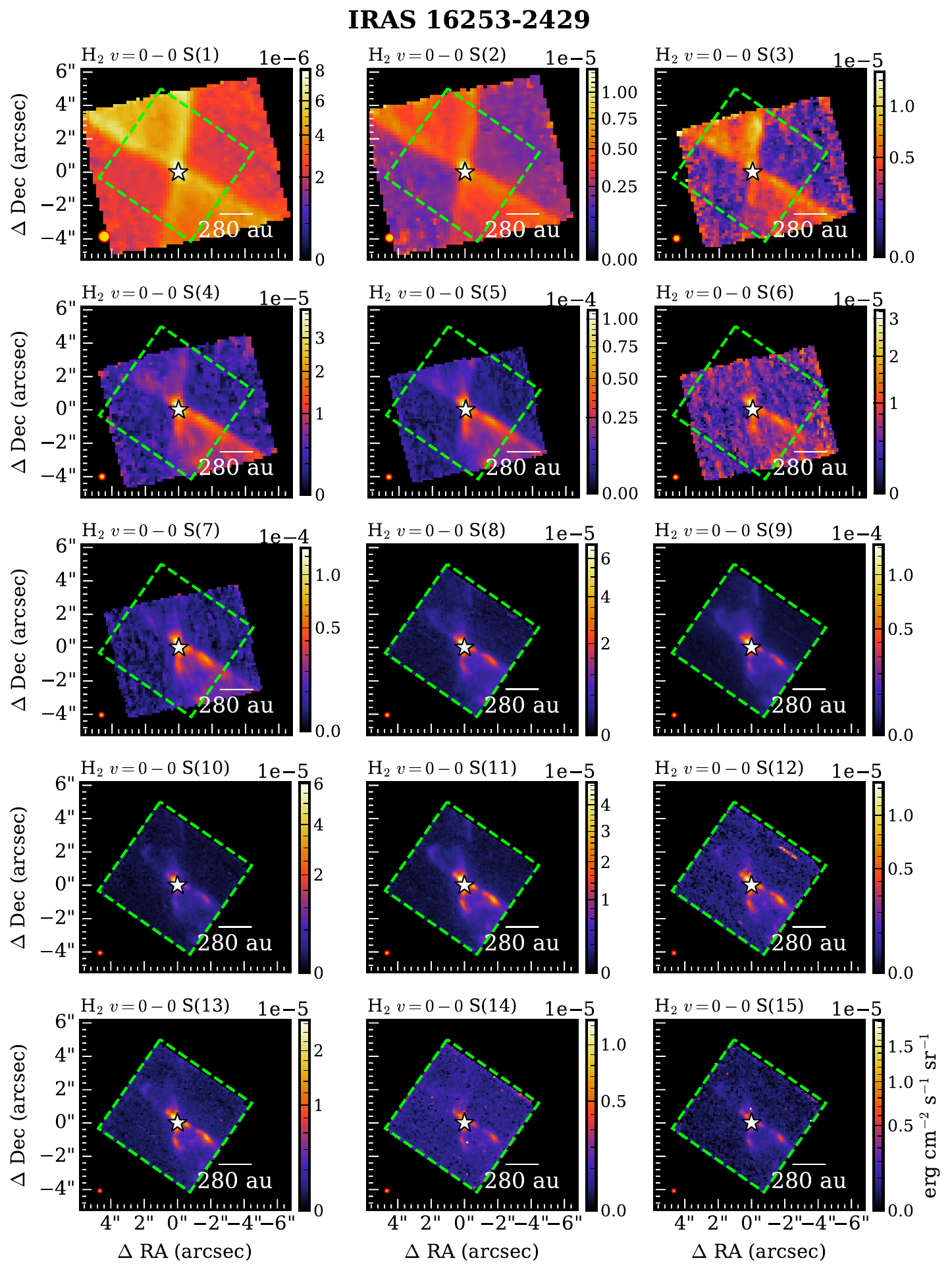}
    \caption{Line Maps of detected \hydrogenmol\ $\nu=0-0$ S($J$) lines (labeled on the top of images) in IRAS 16253. The white star marks the ALMA continuum position in all the panels. Golden circles at the bottom left edge mark the beam size. All the line maps are on the same spatial scale, and the NIRspec FOV is marked with a dashed lime rectangle.}
    \label{fig:I16253Linemap}
\end{figure}

\begin{figure}
     \centering
    \includegraphics[width=0.9\linewidth]{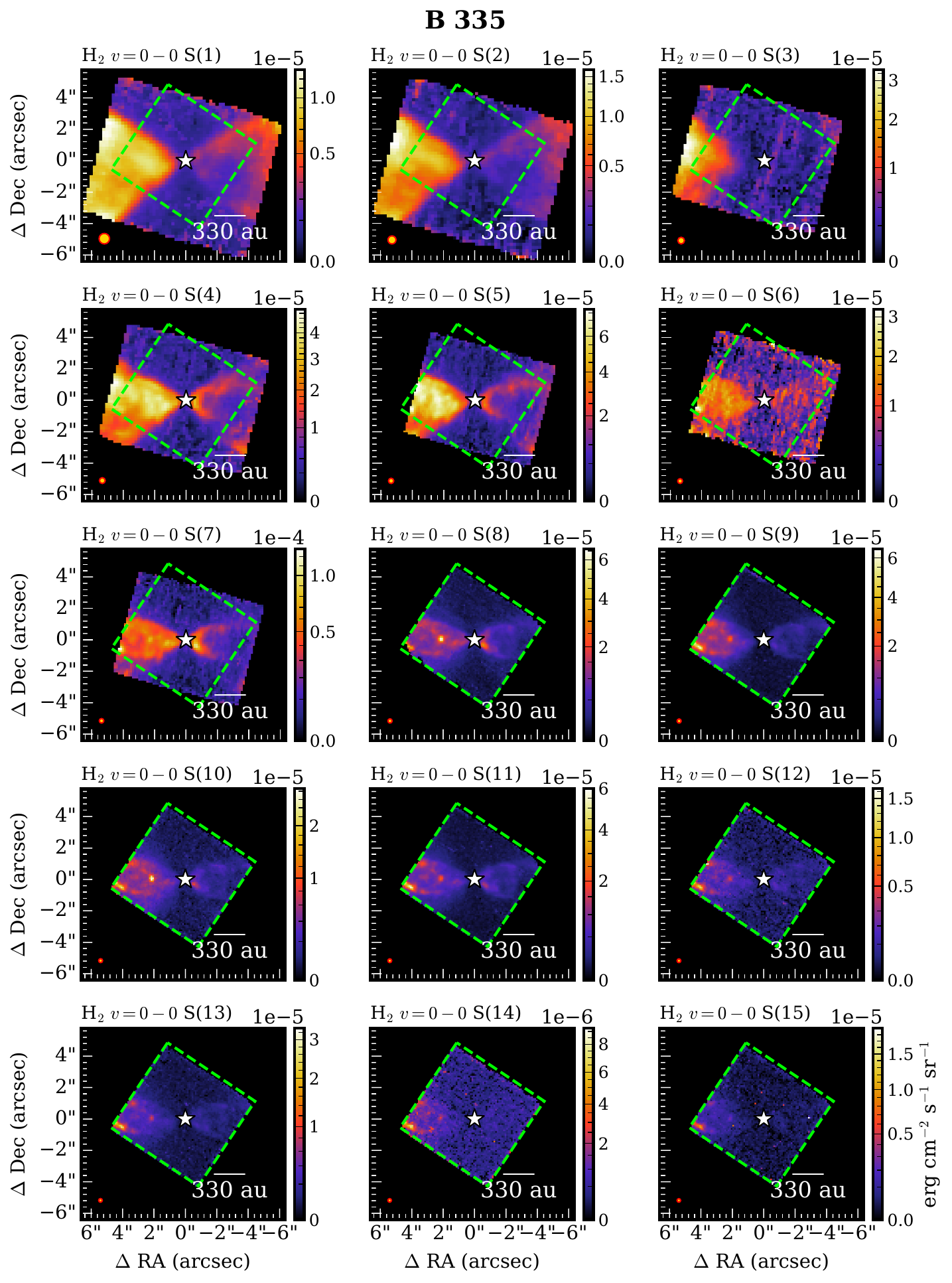}
    \caption{Line Maps of detected \hydrogenmol\ $\nu=0-0$ S($J$) lines (labeled on the top of images) in B 335. Other labels are same as in Figure \ref{fig:I16253Linemap}.}
    \label{fig:B335Linemap}
\end{figure}

\begin{figure}
     \centering
    \includegraphics[width=0.9\linewidth]{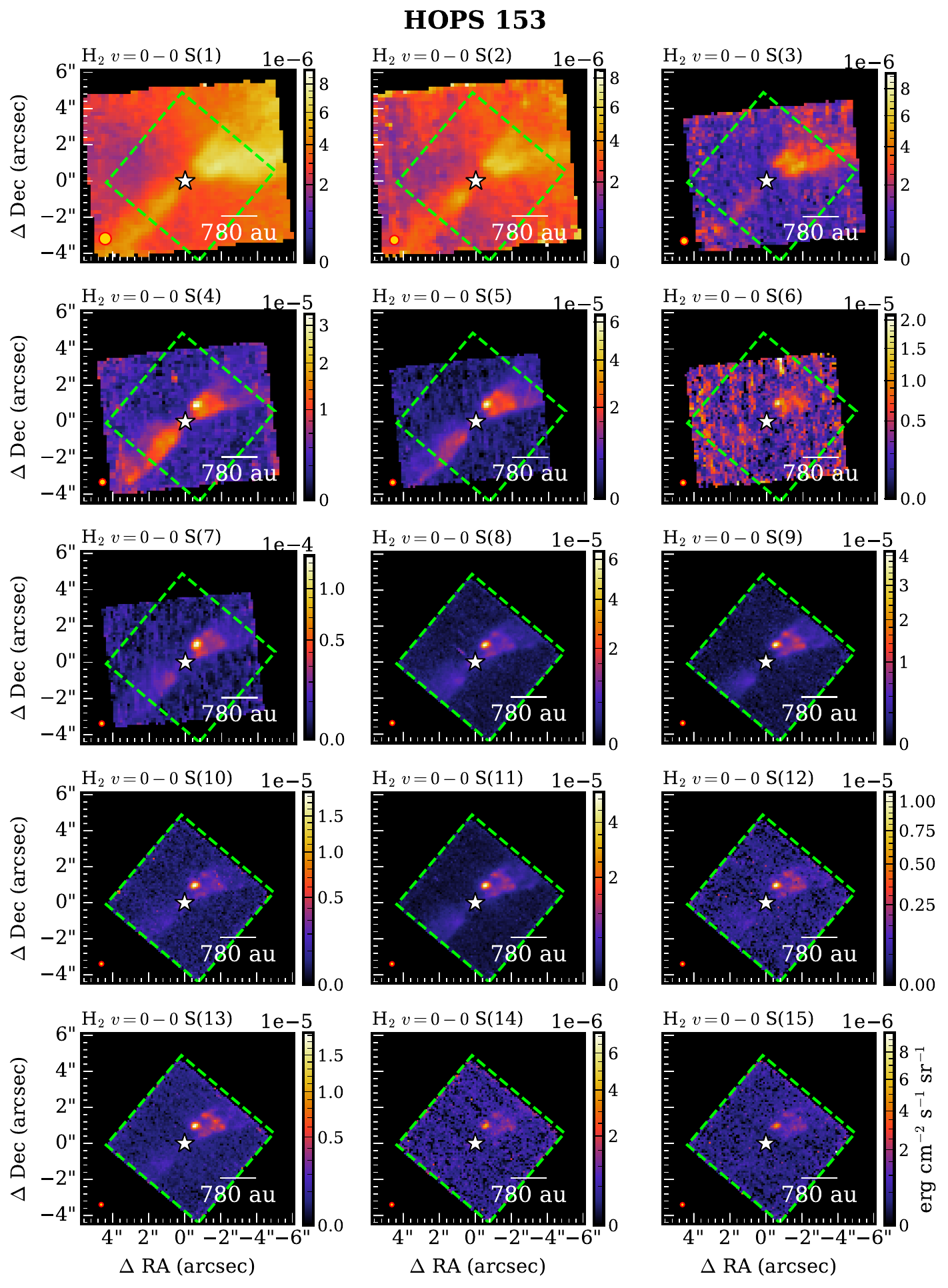}
    \caption{Line Maps of detected \hydrogenmol\ $\nu=0-0$ S($J$) lines (labeled on the top of images) in HOPS 153. Other labels are same as in Figure \ref{fig:I16253Linemap}.}
    \label{fig:HOPS153Linemap}
\end{figure}

\begin{figure}
     \centering
    \includegraphics[width=0.9\linewidth]{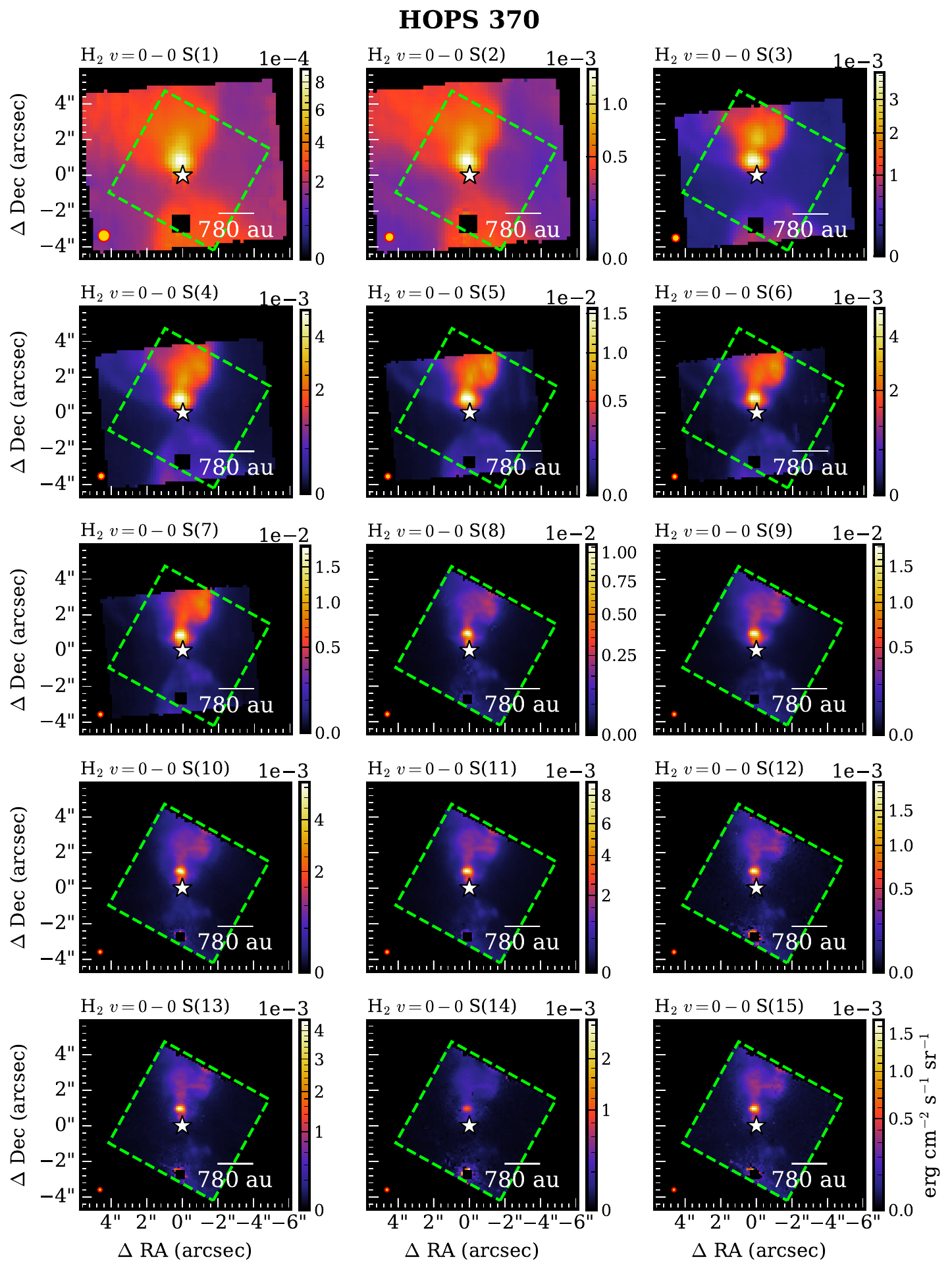}
    \caption{Line Maps of detected \hydrogenmol\ $\nu=0-0$ S($J$) lines (labeled on the top of images) in HOPS 370. Other labels are same as in Figure \ref{fig:I16253Linemap}.}
    \label{fig:HOPS370Linemap}
\end{figure}

\begin{figure}
     \centering
    \includegraphics[width=0.9\linewidth]{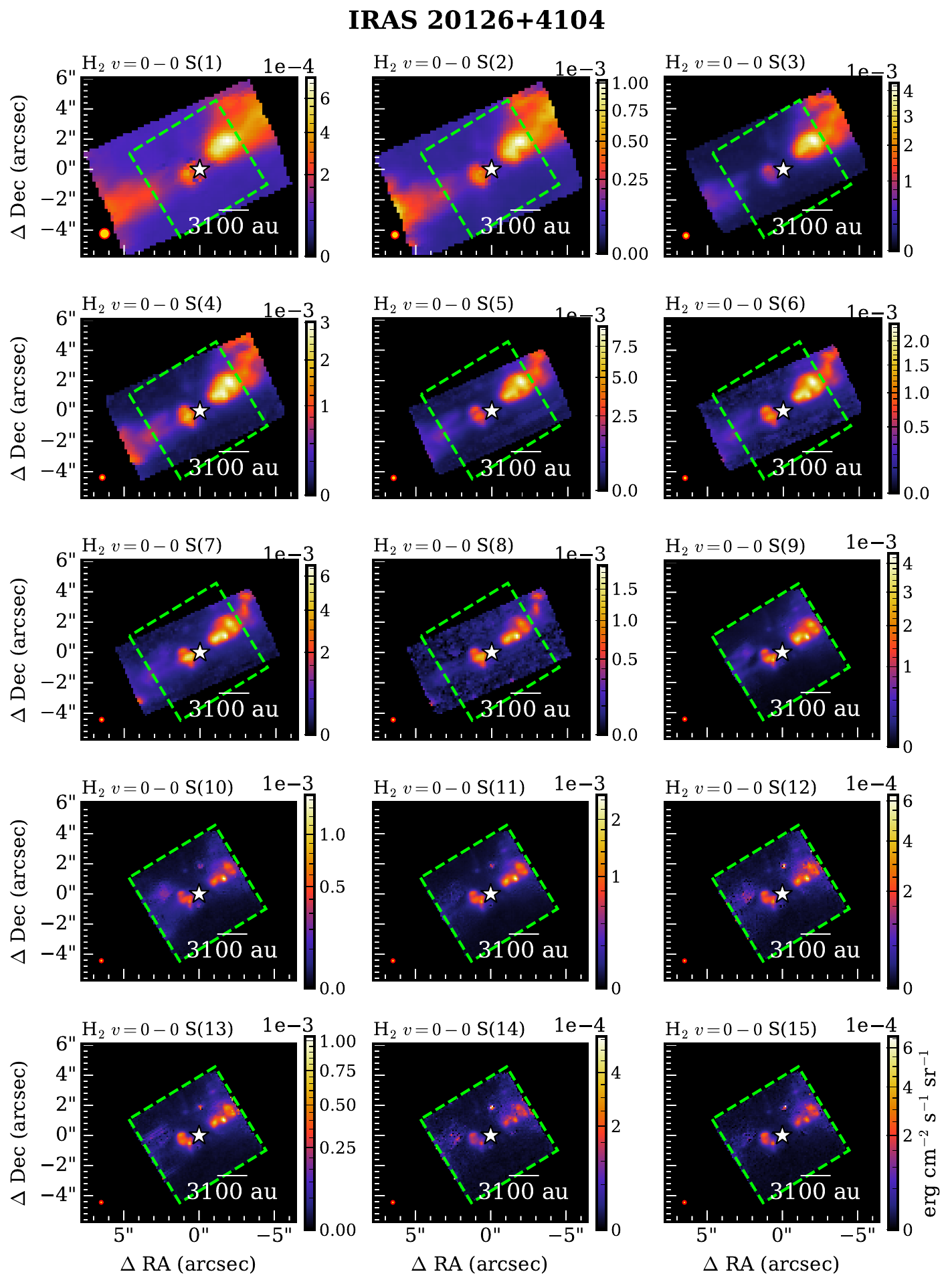}
    \caption{Line Maps of detected \hydrogenmol\ $\nu=0-0$ S($J$) lines (labeled on the top of images) in IRAS 20126. Other labels are same as in Figure \ref{fig:I16253Linemap}.}
    \label{fig:IRAS20126Linemap}
\end{figure}

In this appendix, we present all the line maps of the remaining \hydrogenmol\ lines in Figures \ref{fig:I16253Linemap}, \ref{fig:B335Linemap}, \ref{fig:HOPS153Linemap}, \ref{fig:HOPS370Linemap}, and \ref{fig:IRAS20126Linemap}.

\clearpage
\section{Edge-detection}
\begin{figure}[htbp]
    \centering
    \gridline{
    \fig{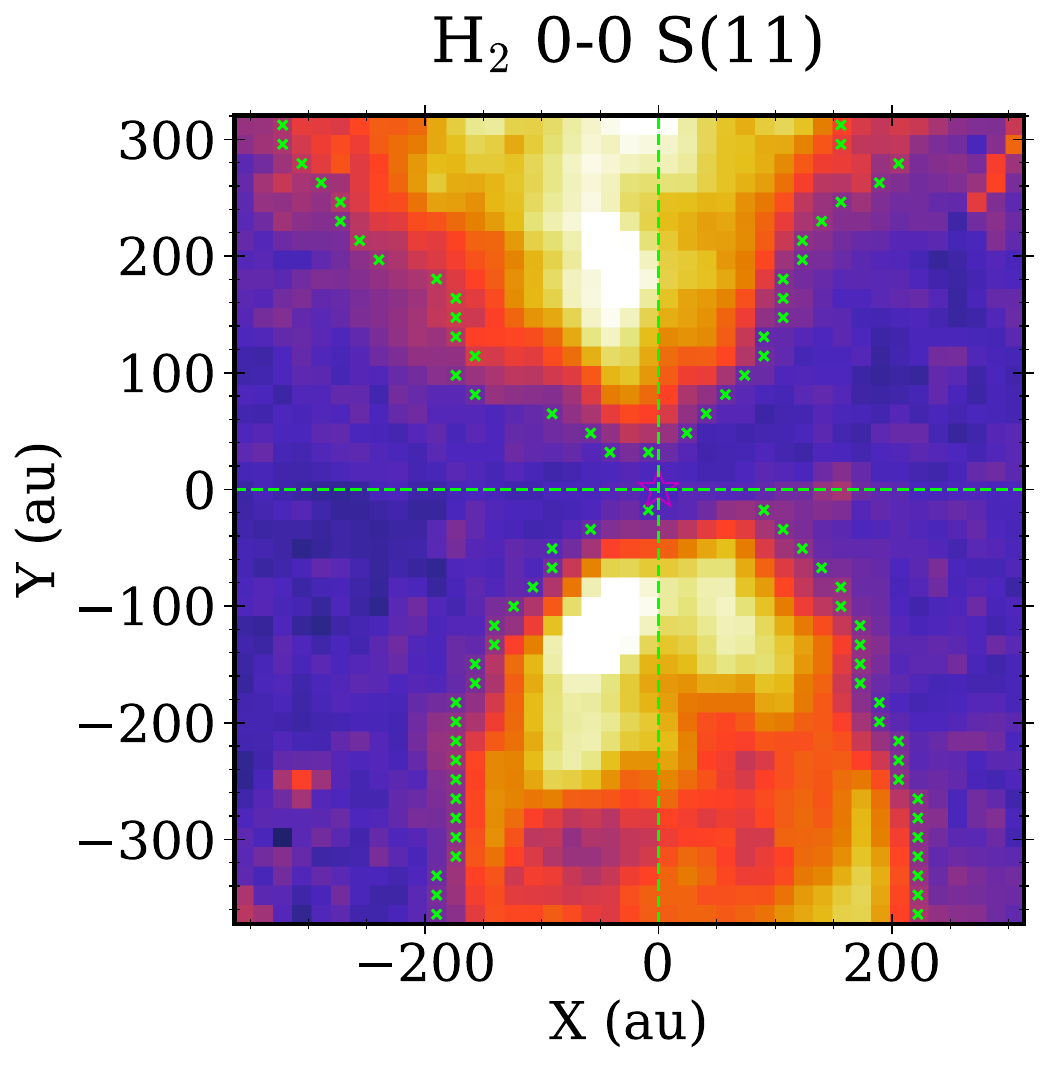}{0.25\textwidth}{}
    \fig{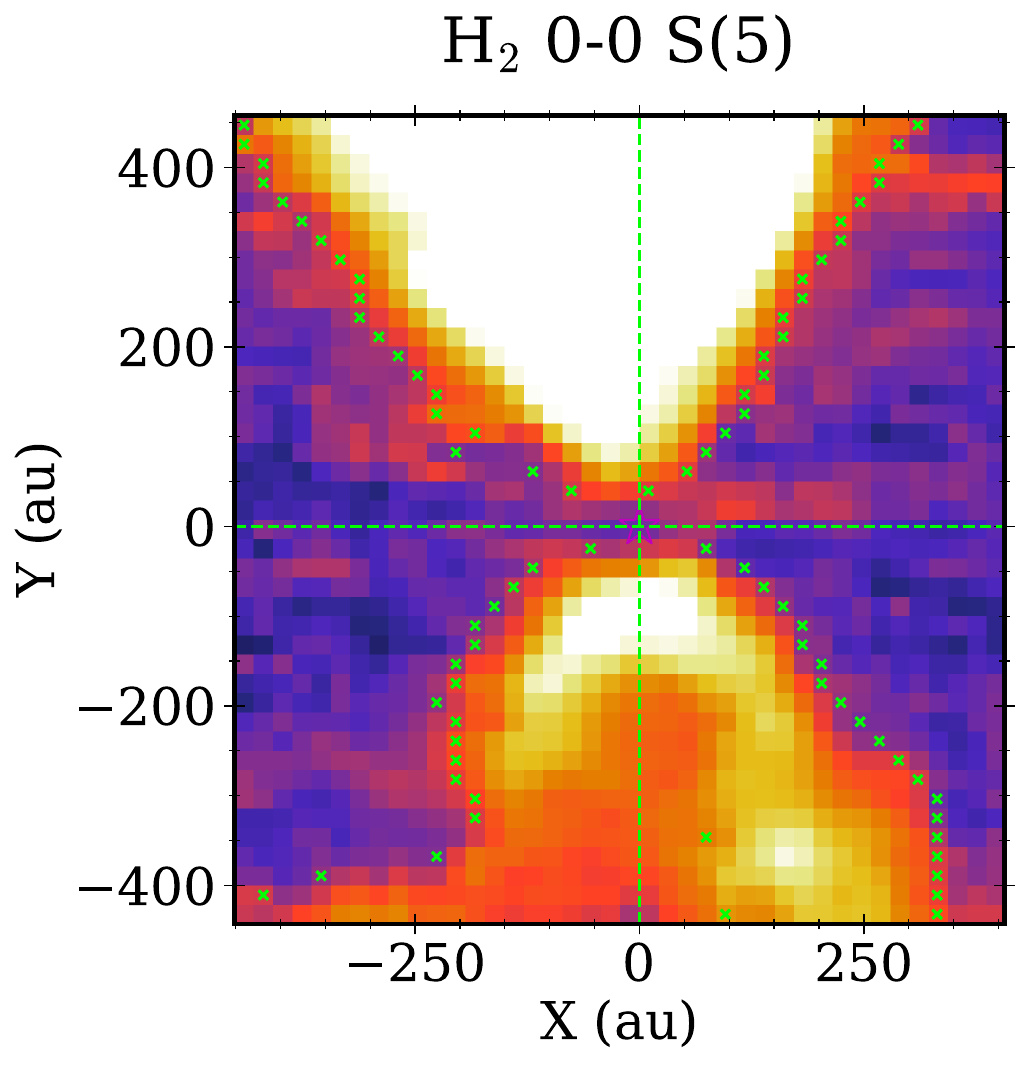}{0.25\textwidth}{(a) B335}
    \fig{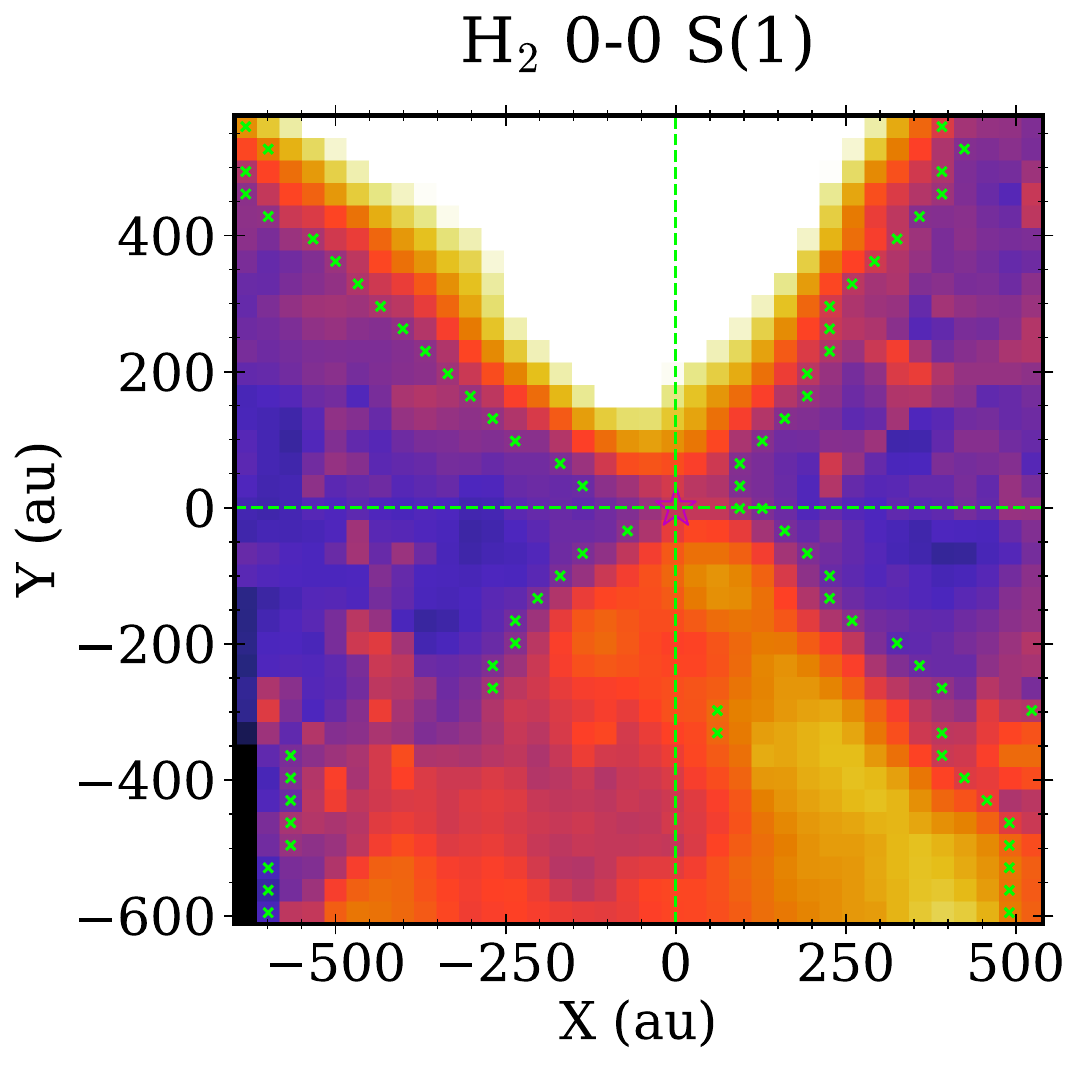}{0.25\textwidth}{}
    }
    \vspace{-0.4cm}
    \gridline{
    \fig{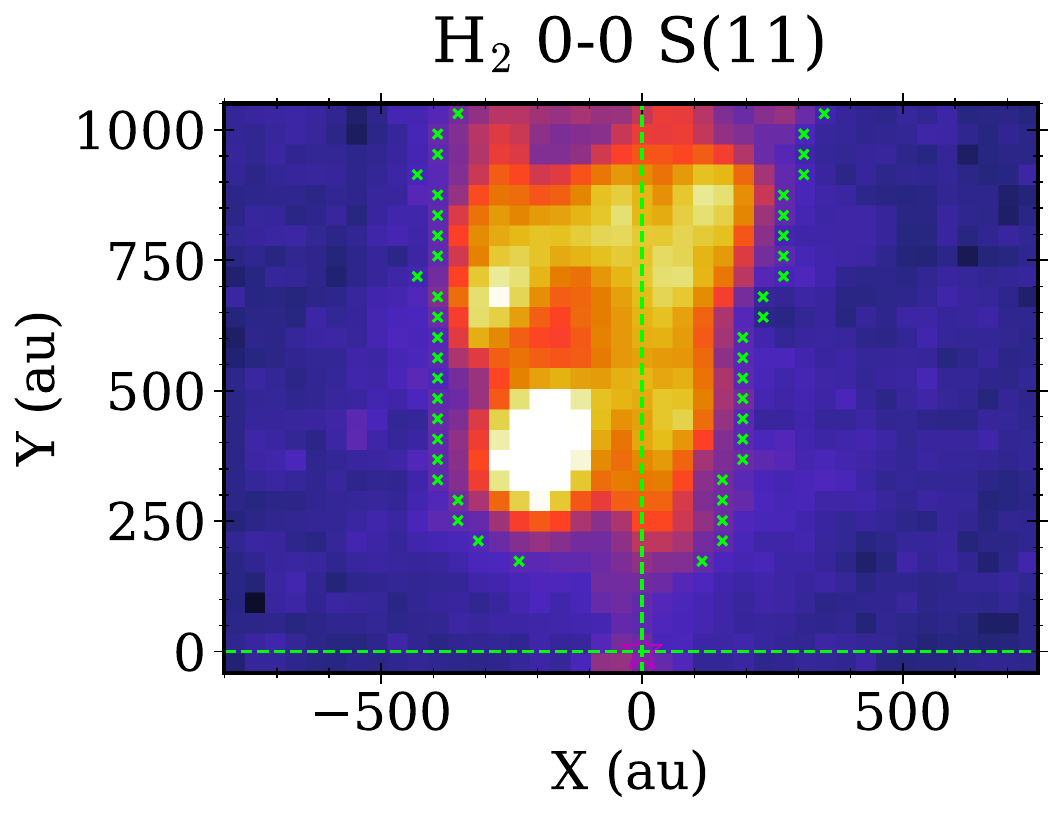}{0.25\textwidth}{}
    \fig{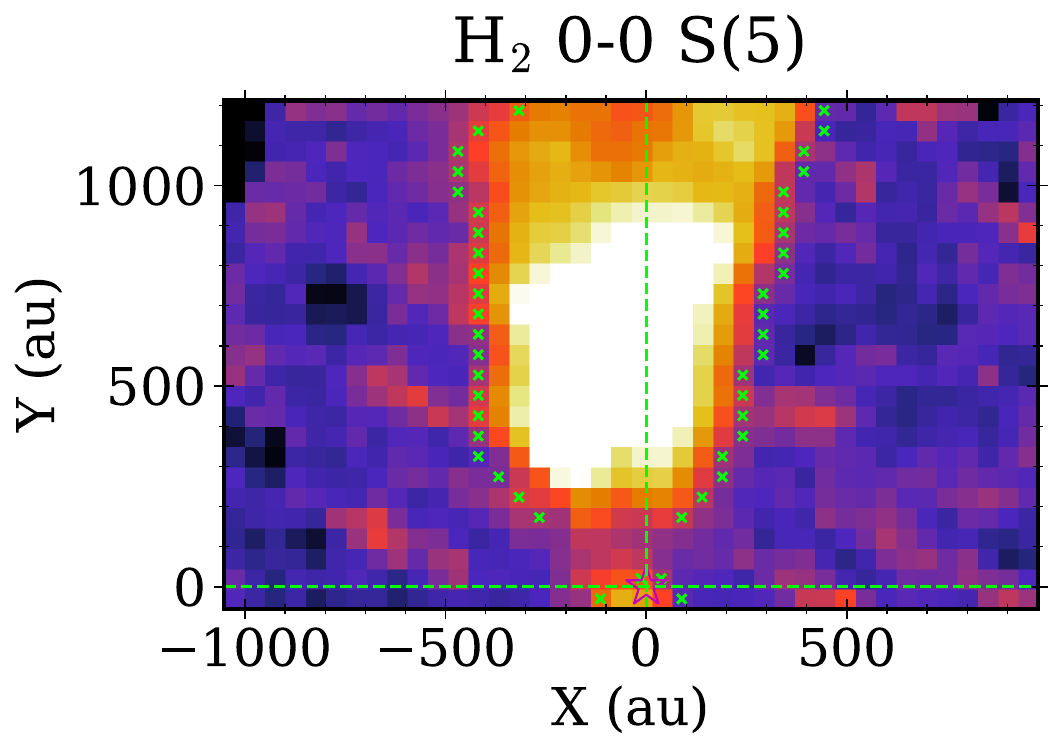}{0.25\textwidth}{(b) HOPS 153}
    \fig{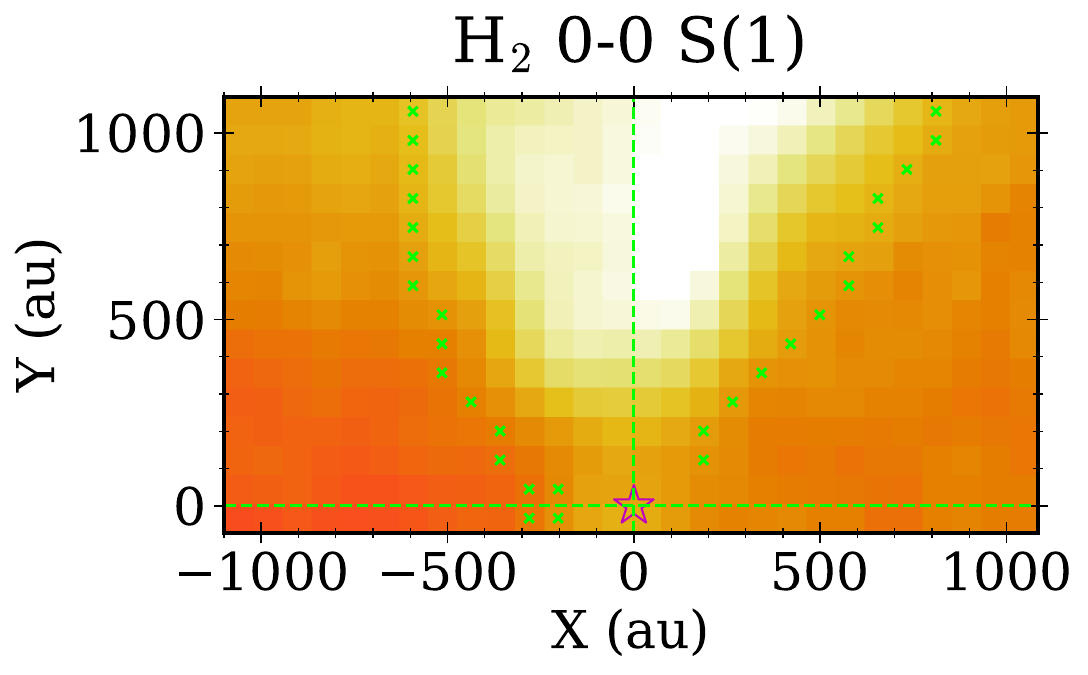}{0.25\textwidth}{}
    }
    \gridline{
    \fig{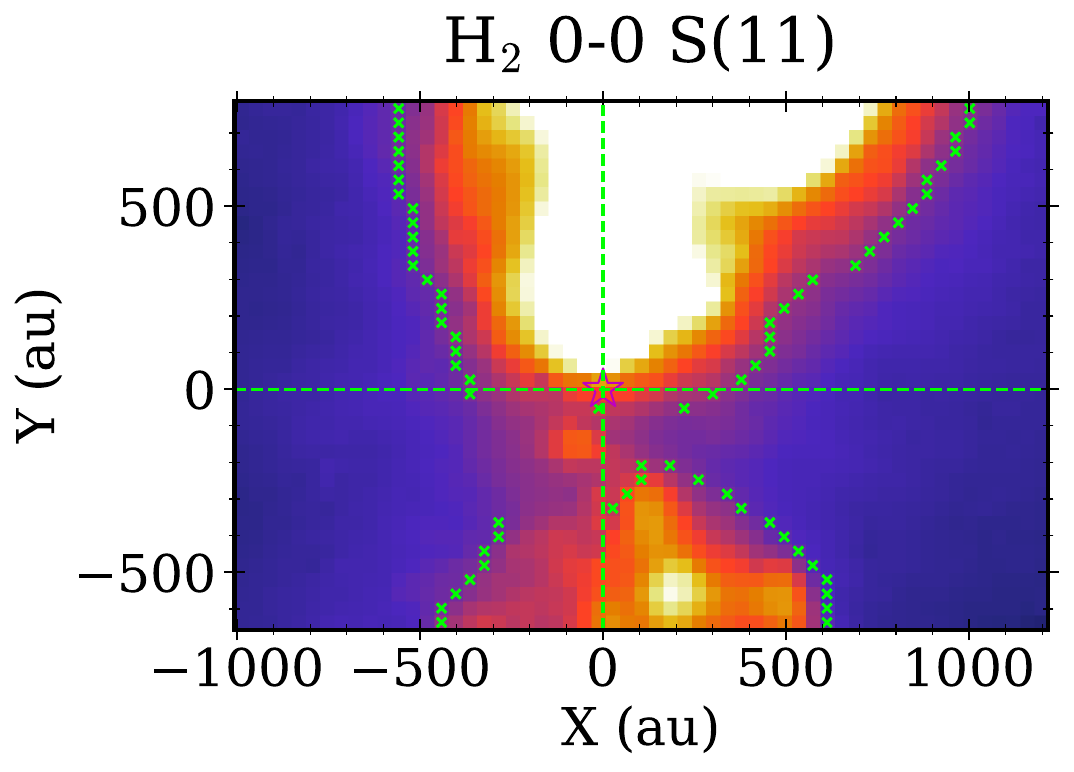}{0.25\textwidth}{}
    \fig{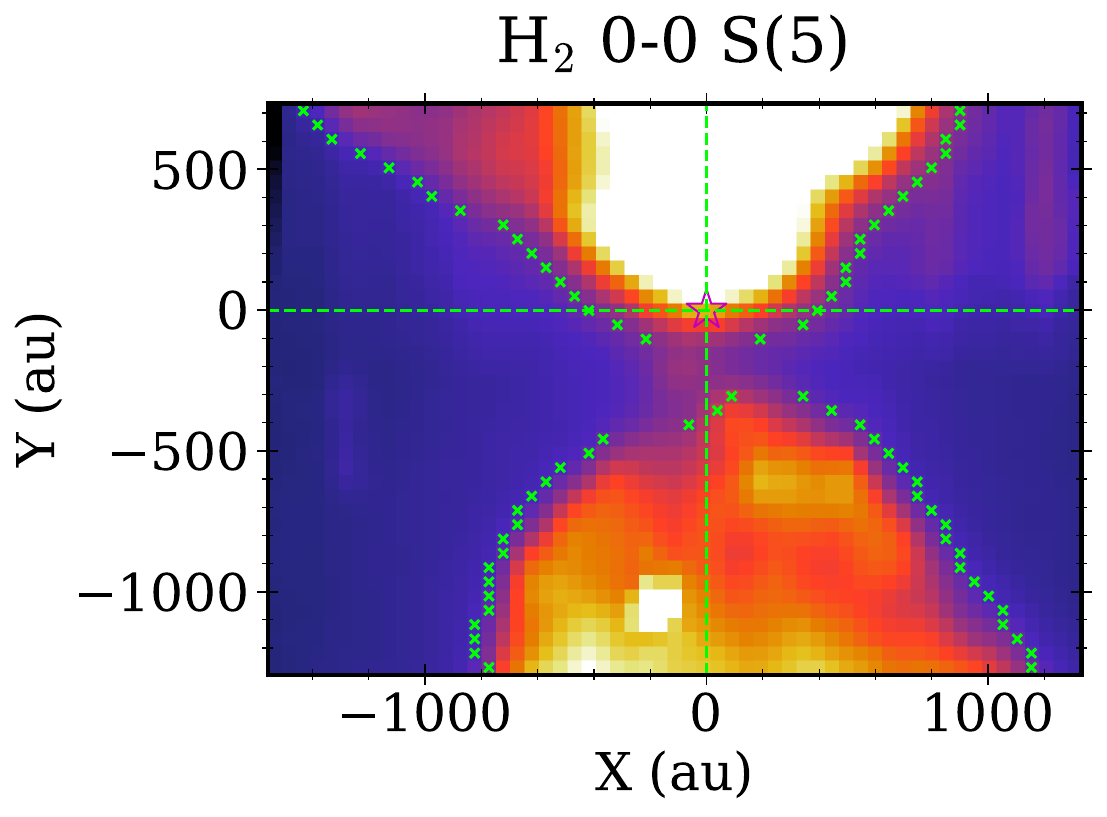}{0.25\textwidth}{(c) HOPS 370}
    \fig{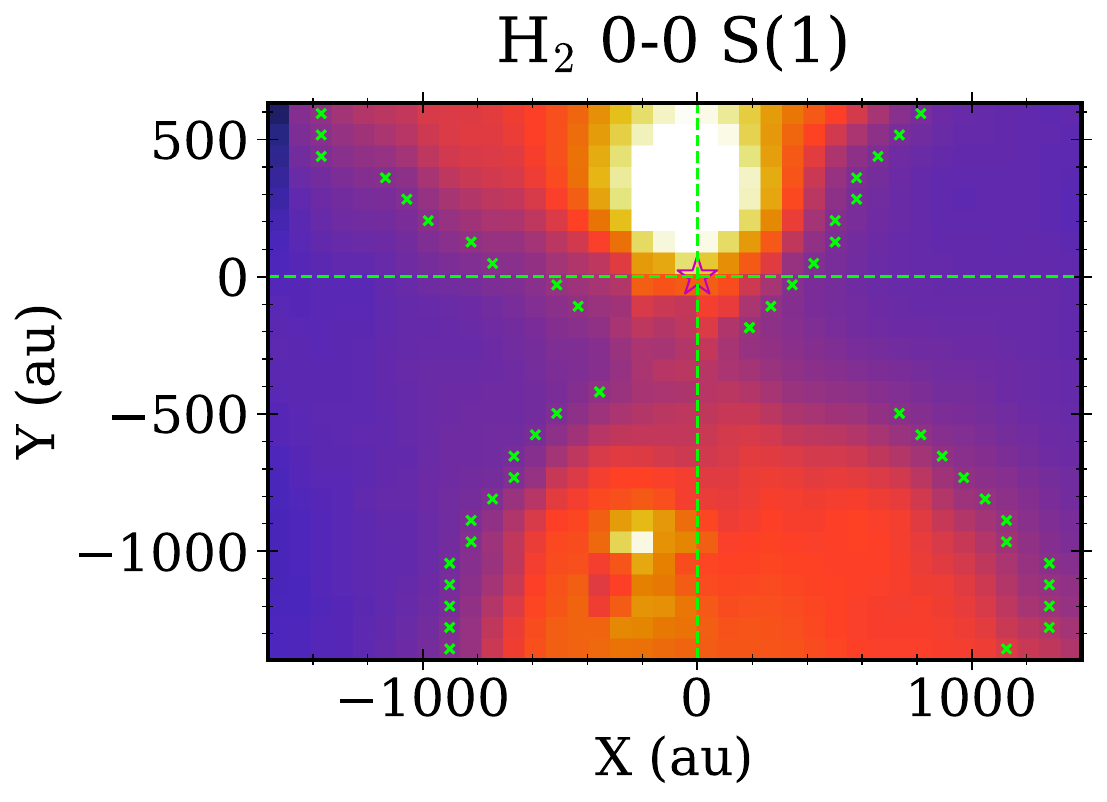}{0.25\textwidth}{}
    }
    \vspace{-0.4cm}
    \gridline{
    \fig{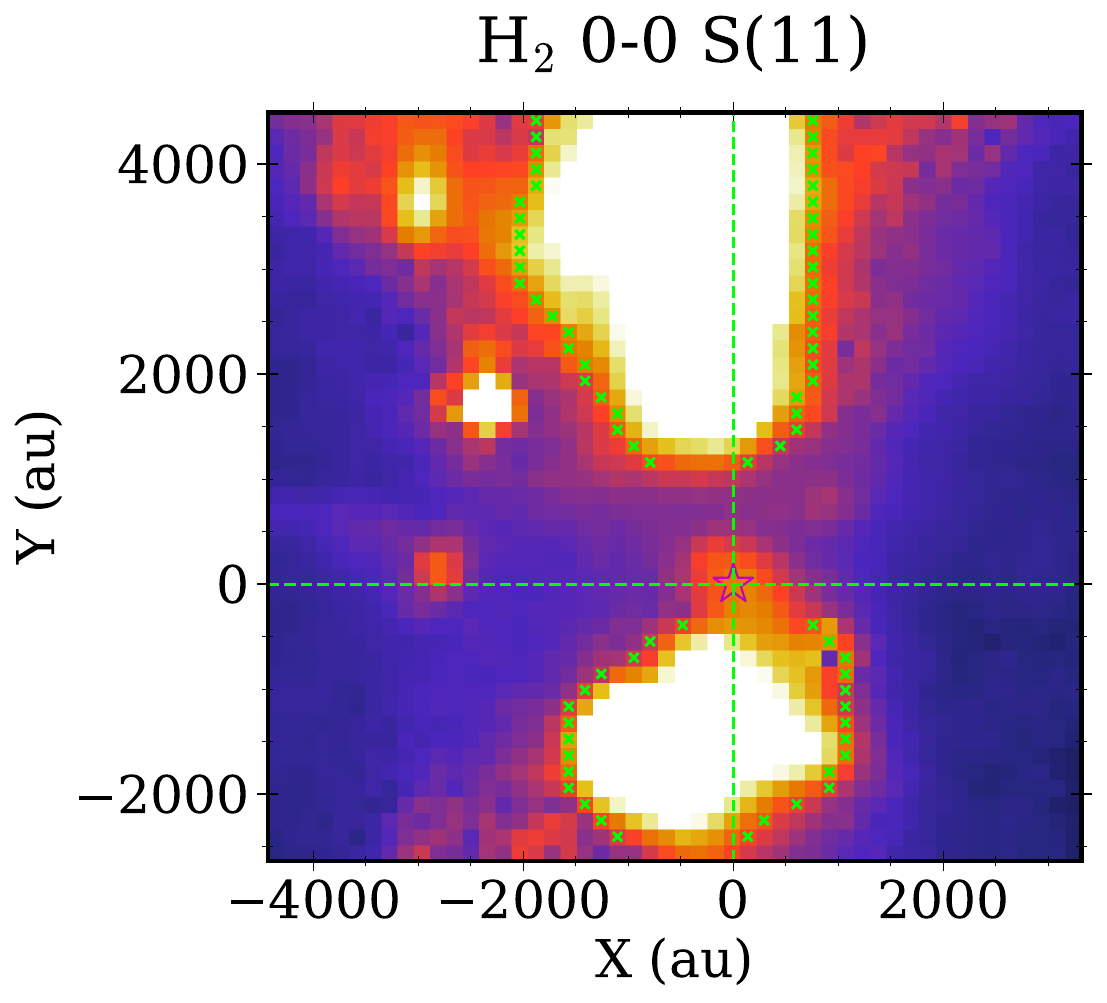}{0.25\textwidth}{}
    \fig{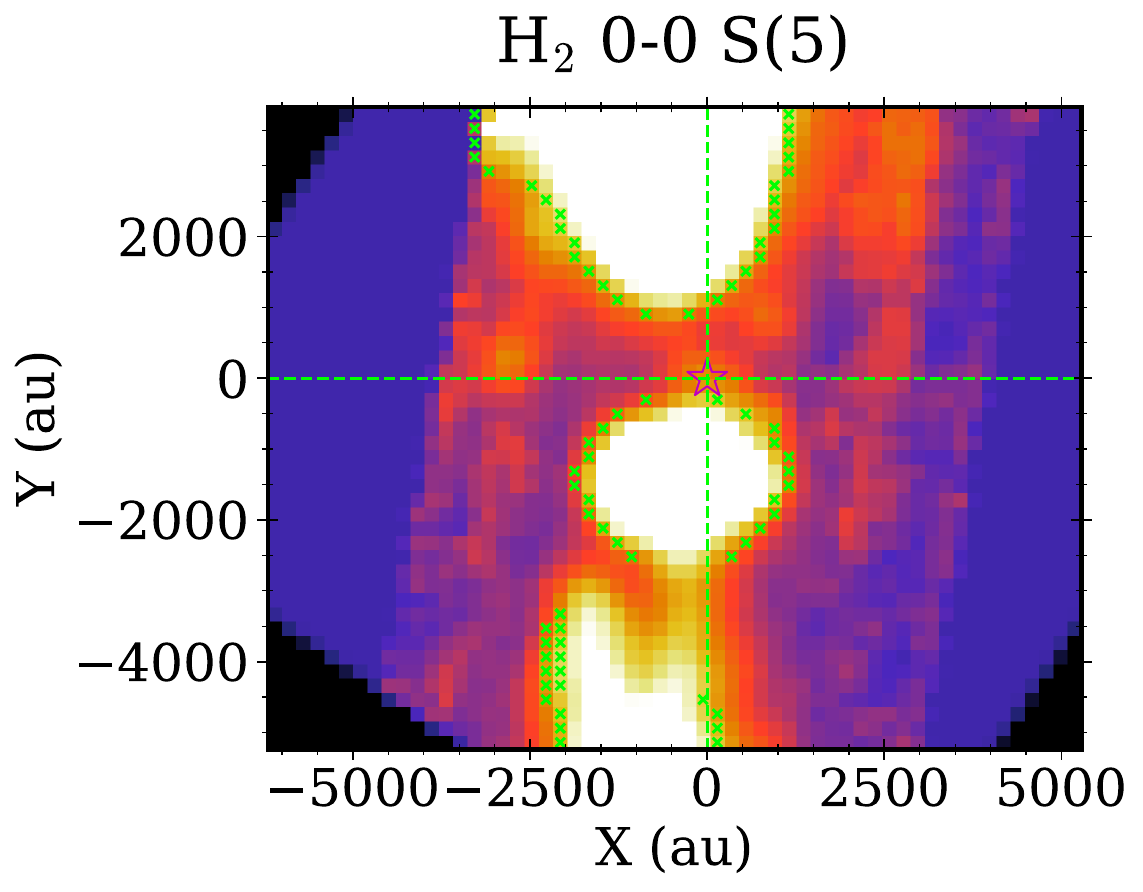}{0.25\textwidth}{(d) IRAS 20126}
    \fig{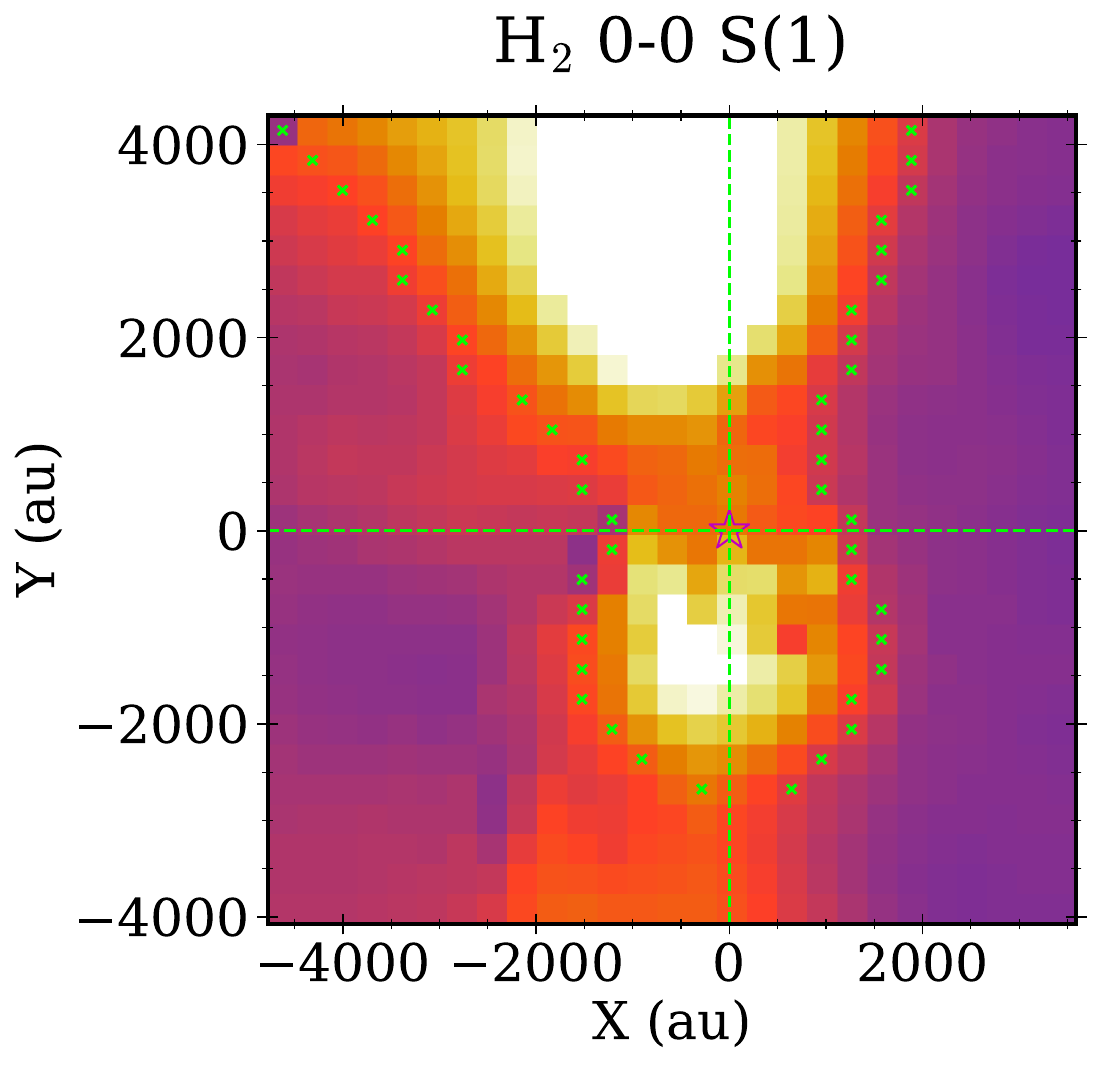}{0.25\textwidth}{}
    }
    
    \caption{Example of edge detection in B335, HOPS 153, HOPS 370, and IRAS 20126. The color scale depicts the rotated line maps of {\hhnu(11), S(5), and S(1)} lines. The lime crosses mark the edges of the cavity. The magenta star marks the ALMA 870 \micron\ peak. }
    \label{fig:All-edge-detection}
\end{figure}

\restartappendixnumbering
In this section, we show the detected edges, overplotted on the respective line maps in Figure \ref{fig:All-edge-detection}.

\section{Effect of Beam size on the width measurements \label{Appendix:Deconvolution}}

{Because the pure-rotational \hydrogenmol\ lines are distributed across a wide spectral range ($3.4-17$ \micron), they are observed at significantly different angular resolutions with JWST. To remove the effect of this varying instrumental angular resolutions on the wind width, we derived the intrinsic \hydrogenmol\ wind widths by subtracting the Gaussian beam FWHM in quadrature from the observed widths as follows:
\begin{equation}
    \theta_{\rm Deconvolved} = \sqrt{\theta_{\rm Observed}^2-\theta_{\rm PSF}^2}
\end{equation}
Where $\theta_{\rm PSF} = 0.033 (\lambda/\mu m) + 0.106$ (in arcsec; \citealt{Law2023AJ....166...45L}).}

\begin{figure}
    \centering
    \includegraphics[width=0.7\linewidth]{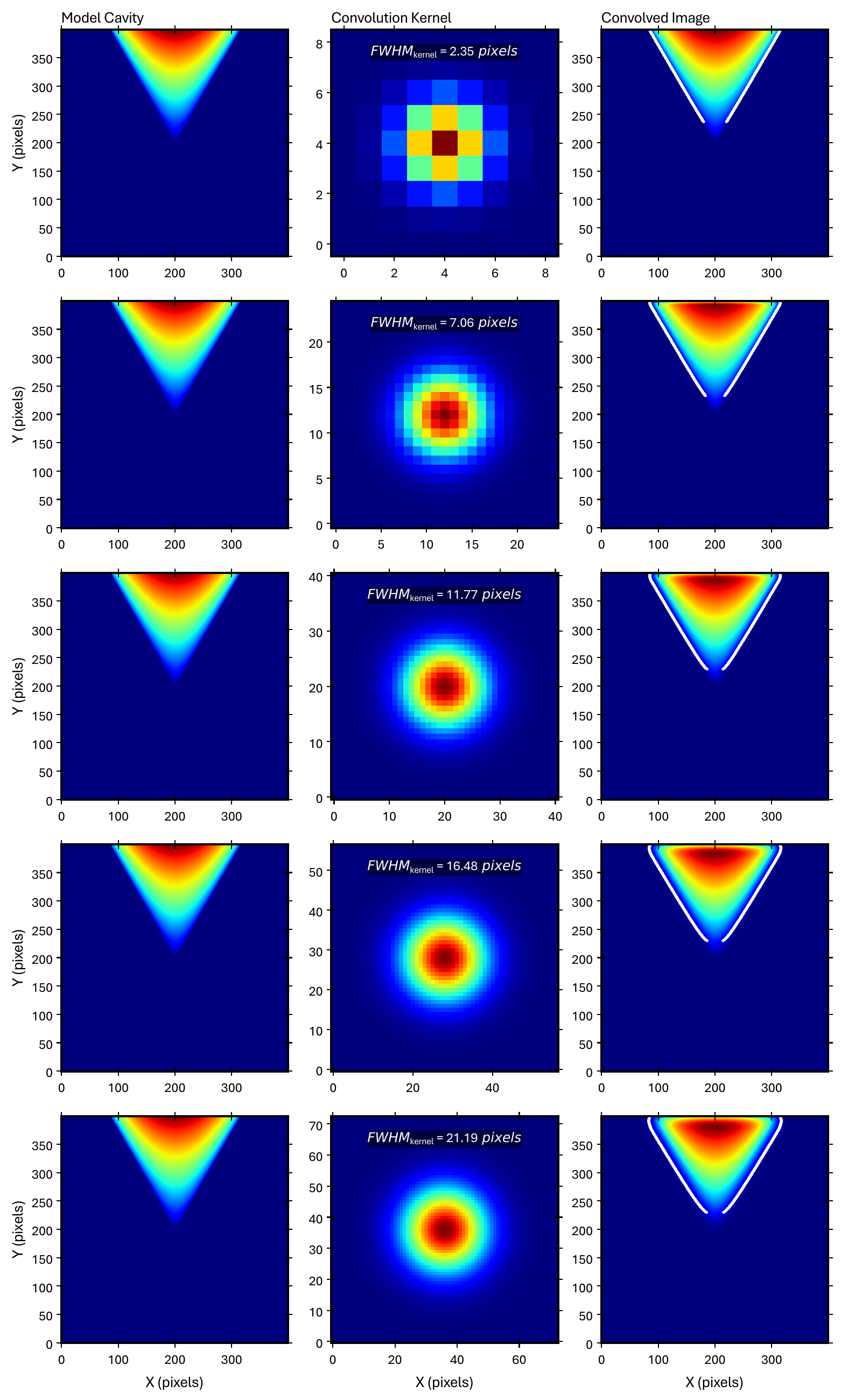}
    \caption{{The intrinsic synthetic conical model (\textit{left column}), the applied Gaussian convolution kernels (\textit{middle column}), and the resulting convolved model images (\textit{right column}). The edges identified by our algorithm are overlaid on the convolved images in white colored lines.}}
    \label{fig:convolved_Images}
\end{figure}

\begin{figure}
    \centering
    \includegraphics[width=\linewidth]{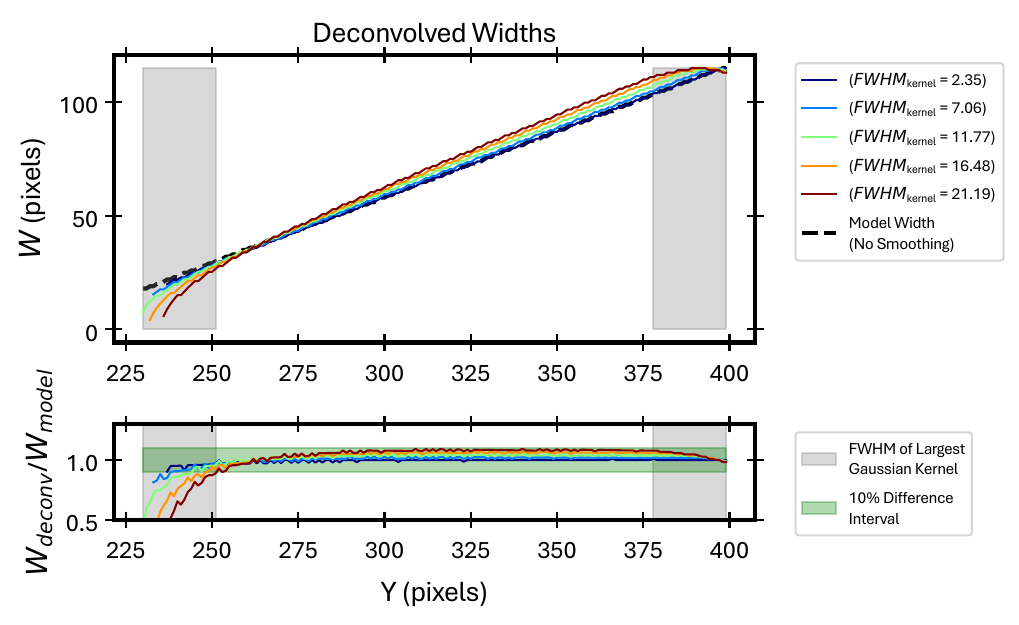}
    \caption{{
    \textit{Top:} Deconvolved widths measured from their respective convolved models are plotted in solid colored lines as a function of the Y-axis. The intrinsic width of the synthetic model is shown by the thick black dashed line. 
    \textit{Bottom:} Ratio of the deconvolved widths to the intrinsic width. The green shaded region highlights the 10\% difference interval.
    In both panels, the gray shaded regions mark distances within one FWHM of the largest Gaussian kernel from the center and the edges of the model image.}}
    \label{fig:deconv_width_appendix}
\end{figure}

{To test the accuracy of this deconvolution approach, we generated a synthetic conical model and convolved it with 2D Gaussian kernels with $\sigma=1, 3, 5, 7,$ and $9$ pixels. The intrinsic synthetic model, convolution kernel, and convolved model are shown in Figure \ref{fig:convolved_Images}. We then ran our edge-detection algorithm on both the intrinsic and convolved models to measure their respective widths. 
Subsequently, to measure the deconvolved widths, the FWHMs of their respective Gaussian kernels were subtracted in quadrature from the measured widths of the convolved models. These deconvolved widths as a function of the $y$-axis are displayed in Figure \ref{fig:deconv_width_appendix} (top panel), with the intrinsic width of the synthetic model marked by a thick dashed black line. To estimate the accuracy of our method, we show the ratio of the deconvolved widths to the intrinsic width in the bottom panel of Figure \ref{fig:deconv_width_appendix}.
We find that outside of the unresolved central region (distances $\gtrsim 1$ FWHM from the central pixel) and away from the extreme edges of the model, the deconvolved widths recover the intrinsic width to within 10\%.}

\end{document}